\documentclass[aps,twocolumn,floatfix,titlepage,balancelastpage,raggedbottom,amsfonts,amssymb,amsmath,showkeys]{revtex4}  
\usepackage{epsfig}
\usepackage{bm}
\usepackage{times}
\usepackage{mathptmx}
\usepackage{amsmath}
\usepackage{graphicx}
\usepackage{graphics}
\usepackage{amssymb}
\usepackage{bm}
\usepackage[english]{babel}
\begin{document}
\title{Lattice Boltzmann simulations on the role of channel structure for 
reactive capillary infiltration}
\author{Danilo Sergi}
\email[Corresponding author. E-mail: ]{danilo.sergi@icimsi.ch}
\affiliation{University of Applied Sciences SUPSI, 
The iCIMSI Research Institute, 
Galleria 2, CH-6928 Manno, Switzerland}
\author{Loris Grossi}
\affiliation{University of Applied Sciences SUPSI, 
The iCIMSI Research Institute, 
Galleria 2, CH-6928 Manno, Switzerland}
\author{Tiziano Leidi}
\affiliation{University of Applied Sciences SUPSI, 
The iCIMSI Research Institute, 
Galleria 2, CH-6928 Manno, Switzerland}
\author{Alberto Ortona}
\affiliation{University of Applied Sciences SUPSI, 
The iCIMSI Research Institute, 
Galleria 2, CH-6928 Manno, Switzerland}

\date{\today}


\begin{abstract}
It is widely recognized that the structure of porous media is of relevance for a variety of mechanical and physical 
phenomena. The focus of the present work is on capillarity, a pore-scale process occurring at the micron scale. 
We attempt to characterize the influence of pore shape for capillary infiltration by means of Lattice Boltzmann simulations
in 2D with reactive boundaries leading to surface growth and ultimately to pore closure. The systems under investigation consist of single 
channels with different simplified morphologies: namely, periodic profiles with sinusoidal, step-shaped and zig-zag walls, as
well as constrictions and expansions with rectangular, convex and concave steps. This is a useful way to decompose the complexity
of typical porous media into basic structures. The simulations show that the minimum radius alone fails to
characterize properly the infiltration dynamics. The structure of the channels
emerge as the dominant property controlling the process. A factor responsible
for this behavior is identified as being the occurrence of pinning of the contact line.
It turns out that the optimal configuration for the pore structure arises from the packing of large particles with round shapes.
In this case, the probability to have flow paths wide and straight is higher. Faceted surfaces presenting sharp edges should be
avoided because of the phenomenon of pinning near narrow-to-wide parts. This study is motivated by the infiltration of molten
metals into carbon preforms. This is a manufacturing technique for ceramic components devised to advanced applications. Guidelines
for experimental work are discussed.
\end{abstract}
\maketitle

\begin{figure*}[t]
\includegraphics[width=12cm]{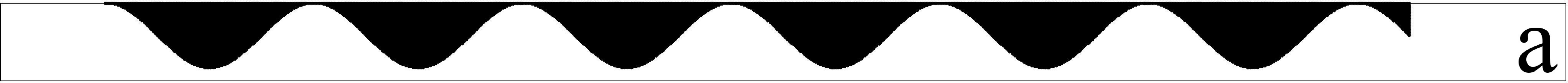}\\
\vspace{0.1cm}
\includegraphics[width=12cm]{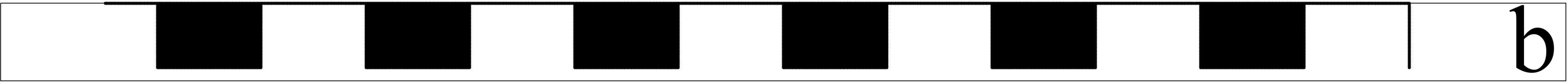}\\
\vspace{0.1cm}
\includegraphics[width=12cm]{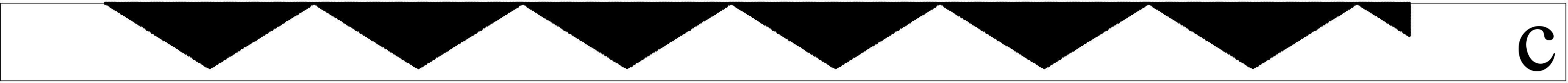}\\
\vspace{0.1cm}
\includegraphics[width=12cm]{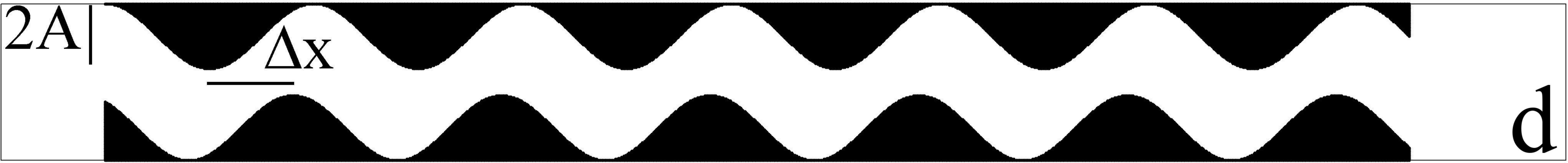}
\caption{\label{fig:profile1}
Periodic profiles for the upper wall: that is, sinusoidal profile, rectangular step and zig-zag profile.
The last figure shows the case of a sinusoidal capillary with the lower wall misaligned with respect
to the upper one. $A$ is the amplitude of the profiles and $\Delta x$ denotes the displacement 
inducing misalignment. Similar definitions apply also for the other geometries.}
\end{figure*}
\begin{figure*}[t]
\includegraphics[width=12cm]{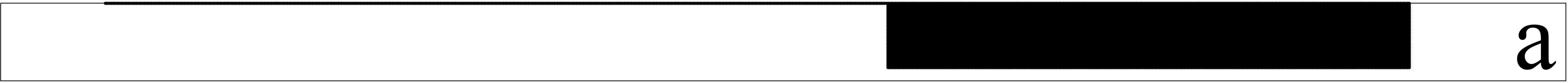}\\
\vspace{0.1cm}
\includegraphics[width=12cm]{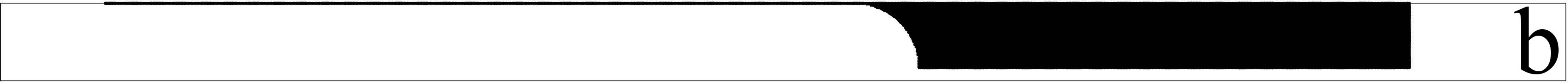}\\
\vspace{0.1cm}
\includegraphics[width=12cm]{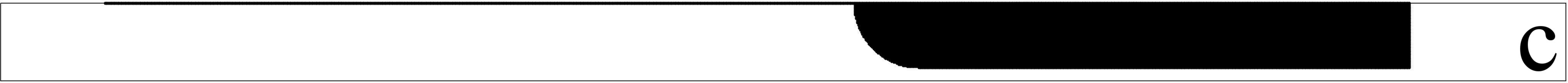}\\
\vspace{0.1cm}
\includegraphics[width=12cm]{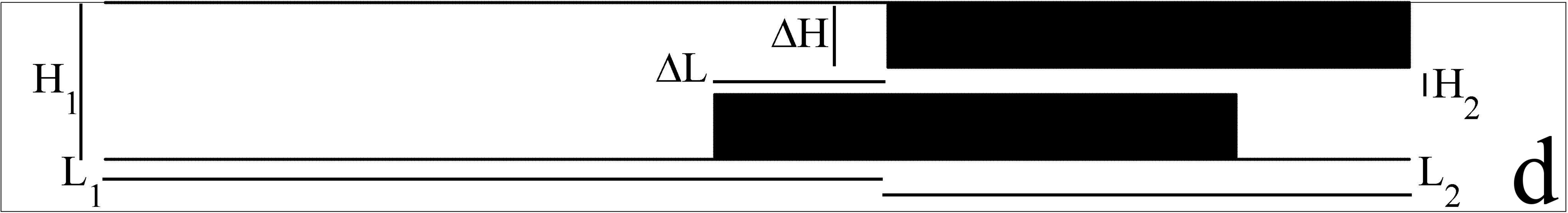}
\caption{\label{fig:profile2}
Constricted channel with rectangular, convex and concave junction (upper wall). In the last panel, $L=L_{1}+L_{2}$ is
the overall length of the interstice, $H_{1}$ is the radius of the first segment, $H_{2}$ that of the second one, 
$\Delta H=(H_{1}-H_{2})/2$ the width of the solid phase and $\Delta L$ the displacement introducing misalingnment 
between the upper and lower walls. Similar definitions hold also for the capillaries with expansion.}
\end{figure*}

\section*{1.~~~INTRODUCTION}

A porous medium is a material presenting some void content in the solid matrix 
(Dullien, 1992). The fraction of void space defines the porosity $\varepsilon$. 
Common examples  of porous materials include wood (Jeje and Zimmermann, 1979; Ota et al., 1995; Studart et al., 2006)
and building materials (Martys and Hagedorn, 2002). For the latters the porosity is exploited in order to trap air with 
the purpose to achieve better thermal insulation (Dullien, 1992). Soil science is another fruitful research area
in the field (Pan et al., 2004; Schaap et al., 2007; Sukop and Or, 2004). In materials 
engineering, the porosity can also be a key factor in order to meet the highest demands in a variety of applications 
(Clyne et al., 2006; Furler et al., 2012a and 2012b; Goodall and Mortensen, 2013; Roberts and Garboczi, 2000; 
Trimis et al., 2006; Zeschky et al., 2005). In general, the pore structure is recognized to have a prominent role for 
the operating behavior of the material. 

In the present work we are concerned with the capillary
properties of porous materials. Capillarity is the phenomenon responsible
for the spontaneous infiltration of a liquid into a porous medium. It is a 
pore-scale process arising from the adhesive forces between the liquid
and solid phases. Computational fluid dynamics (CFD) relying on the finite element method (FEM)
allows to  address such problem (Chibbaro et al., 2009c; Svihla and Xu, 2006;
Xu and Guetari, 2004).
Our study is based on simulations using the Lattice Boltzmann 
(LB) method in 2D (Benzi et al., 1992; Chen and Doolen, 1998; Succi, 2009; 
Sukop and Thorne, 2010; Wolf-Gladrow, 2005). This approach has been 
gaining consideration for handling hydrodynamic systems out of equilibrium 
involving complex boundaries and interfacial phenomena. More generally, many
applications can be figured out in the incompressible limit, i.e.~at low Mach
numbers (Ghosh et al., 2012; Guiet et al., 2011; Haghani et al., 2013). 

The motivating problem for our investigation is the reactive infiltration of molten silicon (Si) into carbon (C) preforms 
(Bougiouri et al., 2006; Dezellus and Eustathopoulos, 2010; Dezellus et al., 2003; Einset, 1996 and 1998; 
Eustathopoulos et al., 1999; Hillig et al., 1975; Israel et al., 2010; Liu et al., 2010; Messner and Chiang, 1990;
Mortensen et al., 1997; Voytovych et al., 2008). 
This process is of special relevance for the industrial practice dependent on the 
processing of carbon and graphite materials (Gadow, 2000; Gadow and Speicher, 2000; 
Krenkel, 2005; Paik et al., 2002; Salamone et al., 2008). Specifically, ceramization 
through liquid Si infiltration, or impregnation, is necessary for high-temperature 
applications such as tribology (Gadow, 2000). In this process, the reaction between Si and C 
to form silicon carbide (SiC) has been the subject of intense research activity 
since it coincides with a wetting transition 
(Bougiouri et al., 2006; Dezellus and Eustathopoulos, 2010; Dezellus et al., 2003; Einset, 1996 and 1998; 
Eustathopoulos et al., 1999; Hillig et al., 1975; Israel et al., 2010; Liu et al., 2010; Messner and Chiang, 1990;
Mortensen et al., 1997; Voytovych et al., 2008). 
In other words, without this reaction, infiltration would not occur (Bougiouri et al., 2006). 
Importantly, the formation of SiC can lead to the thickening of the surface behind 
the invading front (contact line) (Bougiouri et al., 2006; Einset, 1996 and 1998; Israel et al., 2010). Surface 
growth can result in retardation effects and ultimately hinder the impregnation because of pore 
obstruction (Bougiouri et al., 2006; Einset, 1996 and 1998; Israel et al., 2010; Messner and Chiang, 1990). In the LB 
framework, the surface reaction is treated as a precipitation process
(Kang et al., 2007, 2002b, 2003, 2004; Lu et al., 2009; Sergi et al., 2014). 
In this model, the driving mechanism for surface growth is exhaustively compliant 
with the requirement of mass conservation (Kang et al., 2007). Among the pioneering works
for this subject the article by Miller and Succi (2002) deserves attention.

A porous medium can be regarded as a network of capillary channels. For the 
sake of clarity, we shall still limit ourselves to single capillaries. In general, the analysis 
of more complex geometries is based on the theoretical results for this kind of systems. In a 
previous work (Sergi et al., 2014), we investigated the retardation effects induced by surface growth 
for the linear Washburn law for interstices of uniform radii. The linear time dependence for the infiltration depth
is characteristic of reactive Si infiltration (Israel et al., 2010; Voytovych et al., 2008). 
It is important to note that the employed LB models (Chibbaro, 2008; Chibbaro et
al., 2009b; Diotallevi et al., 2009a and 2009b) only reproduce the macroscopic
behavior observed in experiments without providing any explanation for its origin
(Israel et al., 2010; Voytovych et al., 2008).

Here, attention is paid to simplified channel structures consisting of single capillaries with different 
geometric attributes. Namely, we consider periodic profiles (Gern and Kochend\"{o}rfer, 1997; Patro et al., 2007) 
with varying degrees of angularity and tortuosity (Duda et al., 2011; Matyka and Koza, 2012). 
These patterns could arise for example from the juxtaposition of medium-sized C grains. The patterns resulting 
from the juxtaposition of larger C grains are instead modeled as capillary systems with constrictions 
or expansions (Einset, 1996). Also in this case, the results for different morphologies and degrees of 
tortuosity are analyzed.

Our simulations confirm previous results for a uniform channel (Sergi et al., 2014), pointing out that the 
infiltration velocity affects marginally the process of surface growth and pore closure. The main result of 
the present study consists in a system of rules for the evaluation of the response of microstructures of
different characteristics to capillary infiltration and surface growth. For ceramic products 
(Gadow, 2000; Gadow and Speicher, 2000; Krenkel, 2005; Paik et al., 2002; Salamone et al., 2008), these are 
competing phenomena both important in order to obtain complete densification and optimize the final properties. As 
explained before, the complexity of fully-developed porous systems is decomposed into basic structures. In the 
following, a comparative analysis is proposed in order to understand the role of the various structural features. 
Precisely, we shall address questions such as front dynamics, flow retardation, thickening of the surface, pore 
closure, analysis of characteristic radii and tortuosity. Notably, it arises that the structure of the channels 
can affect significantly the resulting effective radius. There appears that our findings provide useful
inputs for preform preparation. The optimal configuration for the porosity is
that containing the larger number of constrictions since the effect of surface
growth is weaker. They could realize a percolating network allowing to guide the
flow into the porous preforms. Another advantage of this configuration is that
it requires coarse particles and so it should be easier to obtain the desired
arrangement. Of course, these guidelines need to be verified and improved by
more simulation work before to proceed to tests by experiments.


\begin{figure*}[b]
\includegraphics[width=8.5cm]{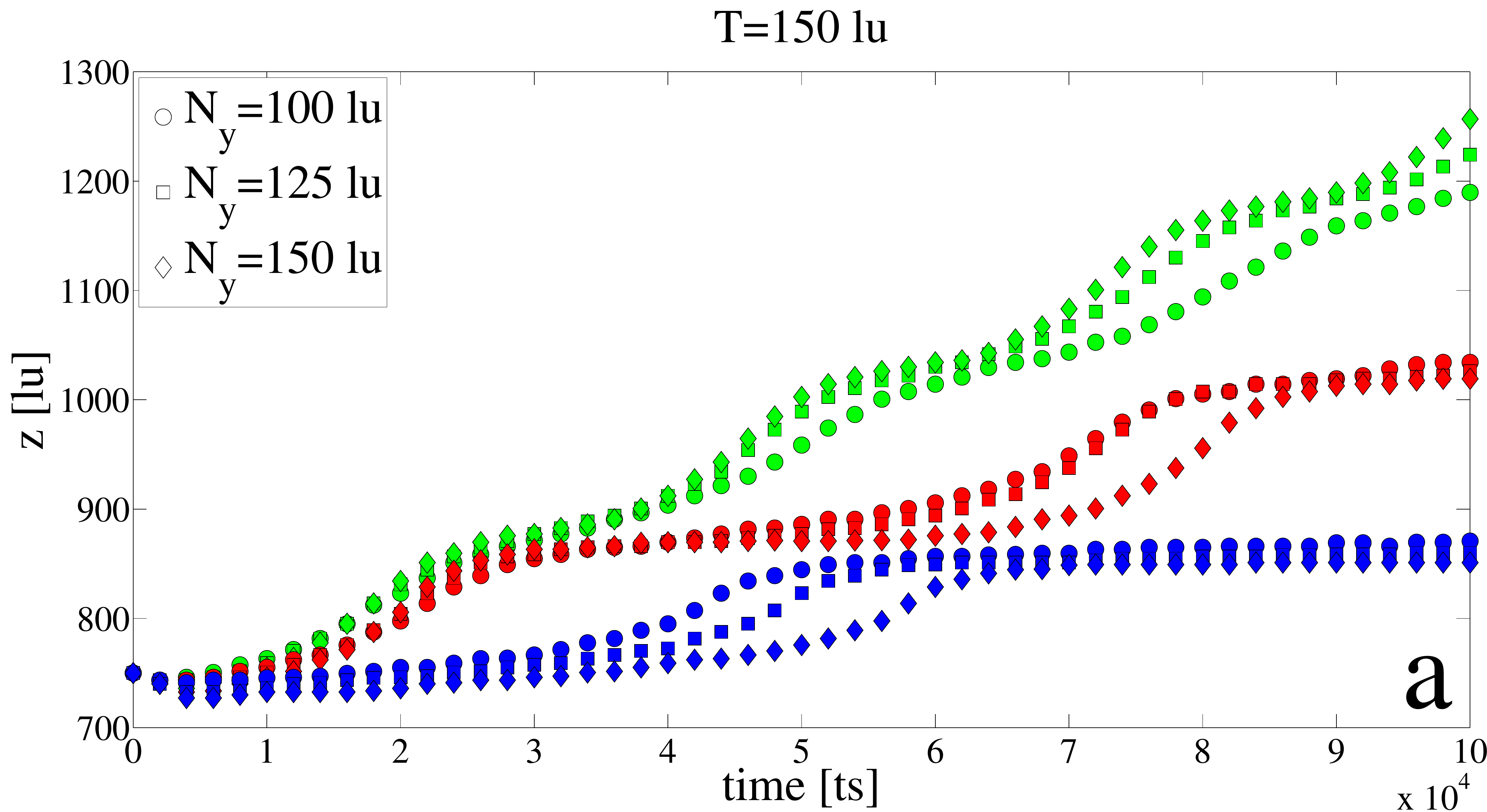}
\includegraphics[width=8.5cm]{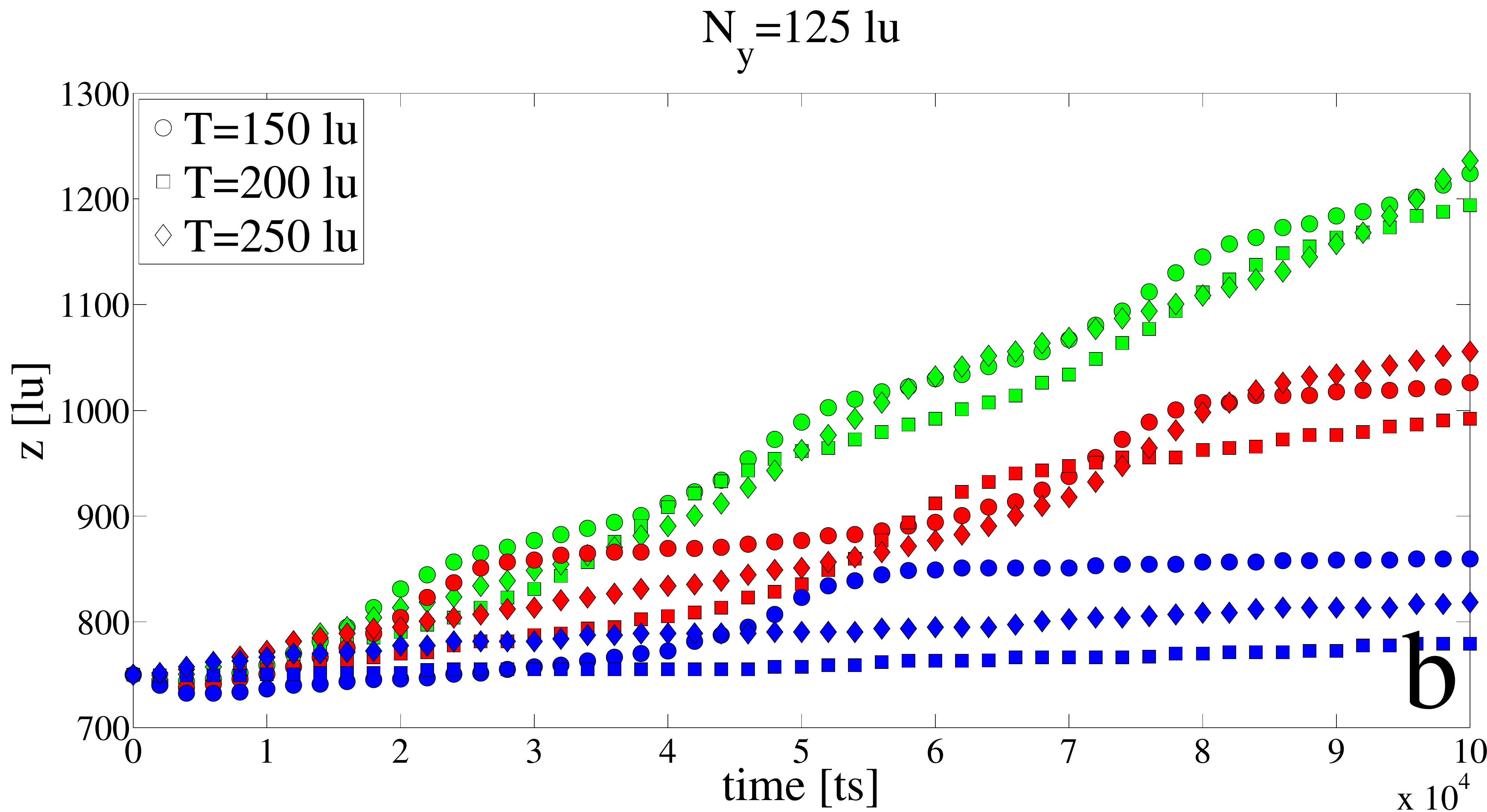}
\caption{\label{fig:sin1}
Centerline position $z$ of the advancing front in the course of time for sinusoidal capillaries of length $L=750$ lu 
without surface reaction. Color code based on the amplitude $A$. We use green, red and blue for increasing 
amplitude $A=i(N_{y}-25)/16$, where $i=2,3,4$.}
\end{figure*}
\begin{figure*}[t]
\includegraphics[width=8.5cm]{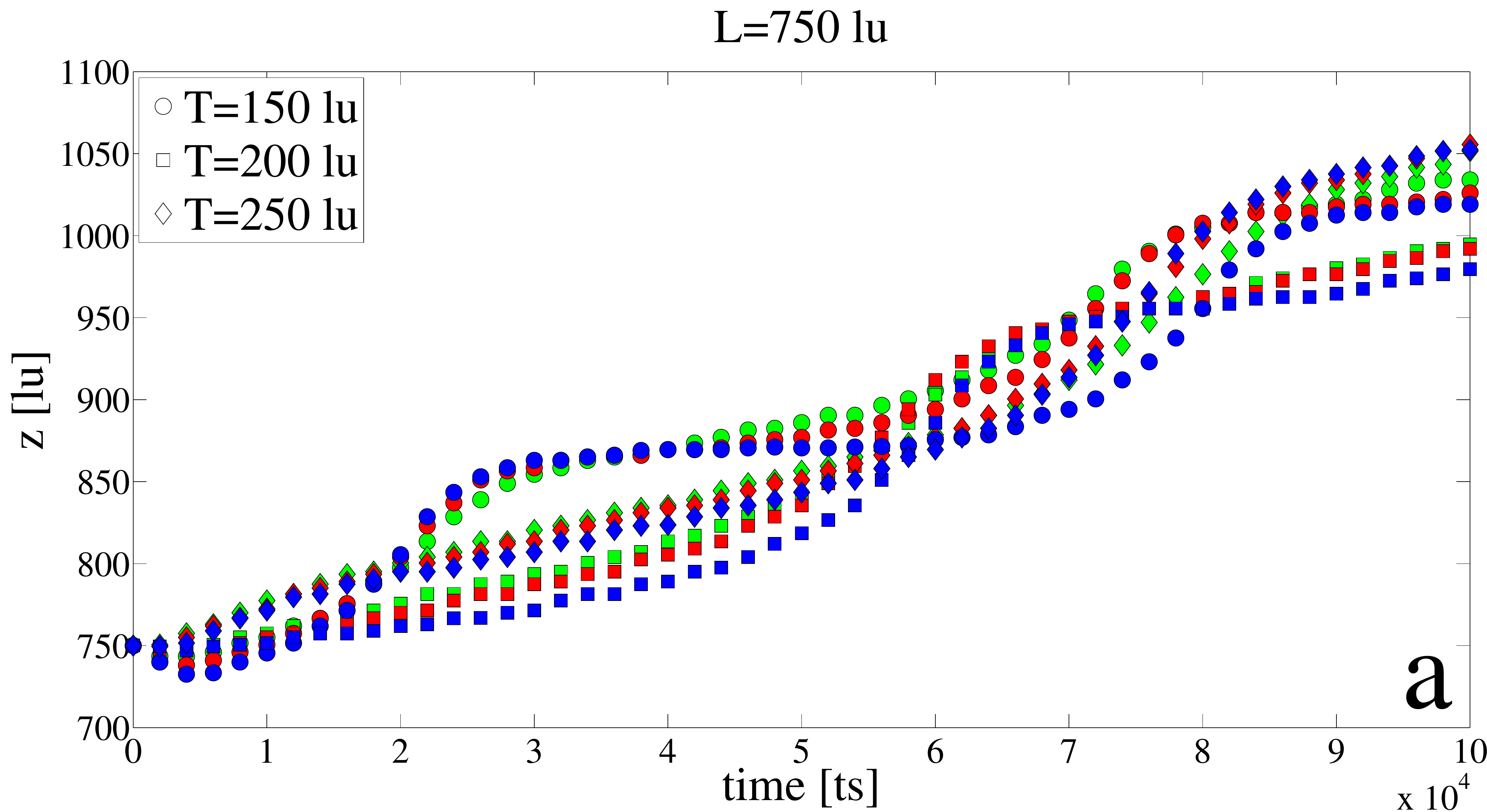}
\includegraphics[width=8.5cm]{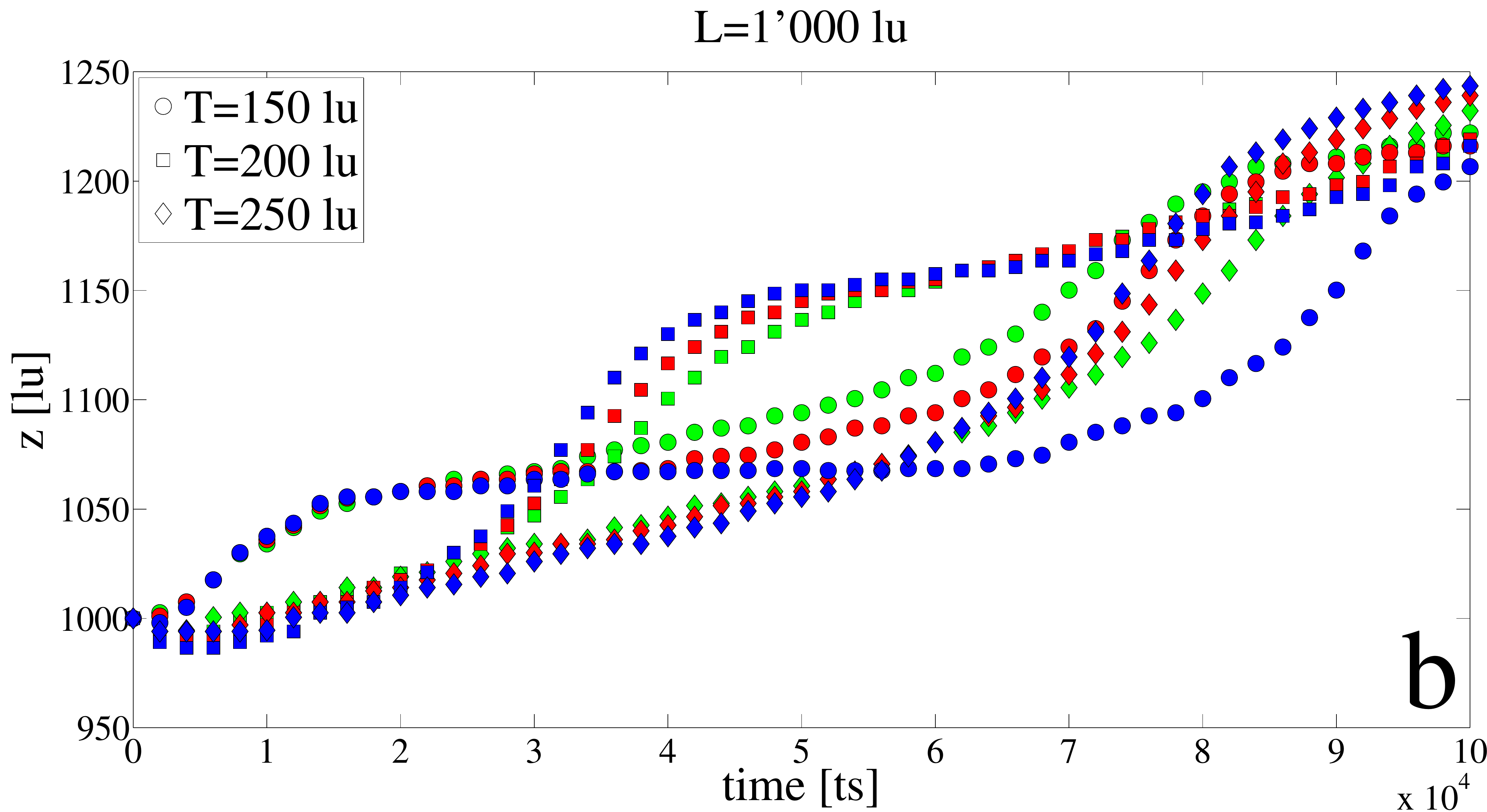}
\caption{\label{fig:sin2}
Time dependence of the centerline position $z$ of the invading front in the absence of reaction for sinusoidal profiles. Color code based on 
the domain width $N_{y}$: green, red and blue for increasing values $N_{y}=100,125,150$ lu. The systems have the average minimum height 
$<H_{\mathrm{min}}>=50$ lu, determined using the amplitude $A=3(N_{y}-25)/16$.}
\end{figure*}
\begin{figure*}[t]
\includegraphics[width=8.5cm]{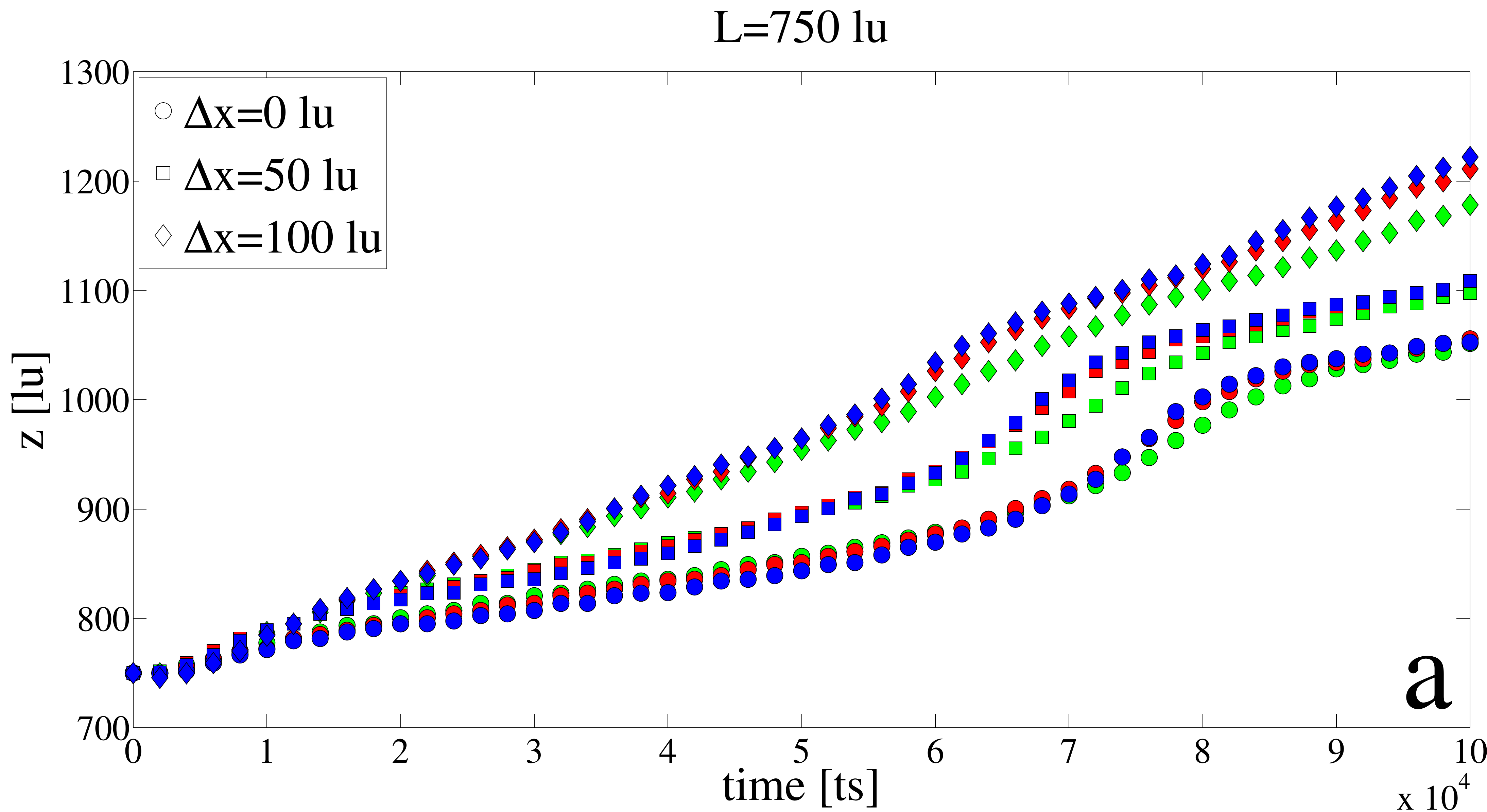}
\includegraphics[width=8.5cm]{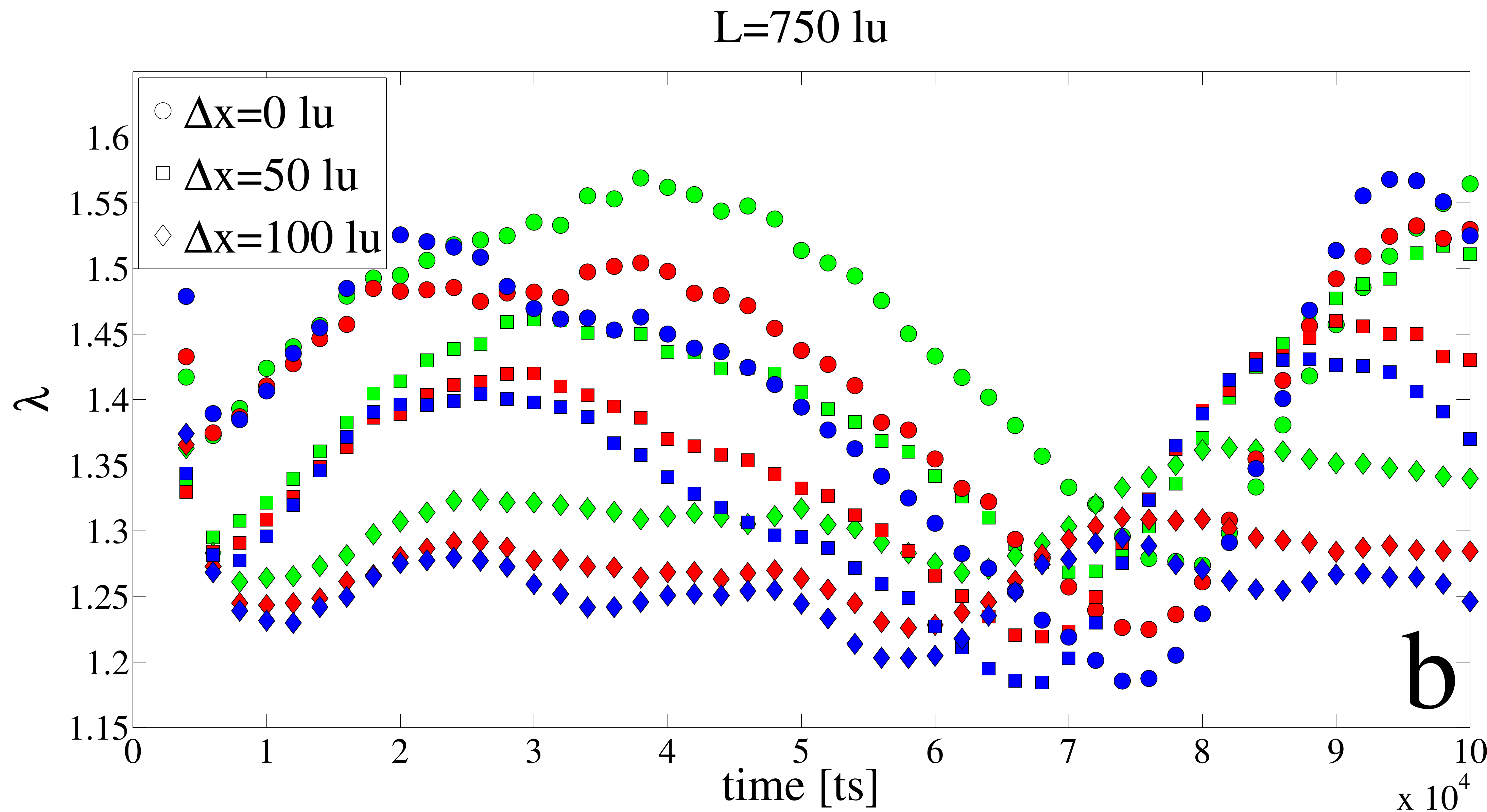}
\caption{\label{fig:sin3}
Sinusoidal capillary systems without reaction in the presence of misalignment between the walls (see Fig.~\ref{fig:profile1}). Color code based 
on the domain width $N_{y}$: green, red and blue for $N_{y}=100,125,150$ lu, respectively. The period is set to $T=250$ lu, while the amplitude is
given by $A=3(N_{y}-25)/16$, providing the average minimum height $<H_{\mathrm{min}}>=50$ lu. (a) Evolution of the centerline position $z$ of 
the meniscus. (b) Evolution of the tortuosity $\lambda$.}
\end{figure*}
\begin{figure*}[t]
\includegraphics[width=8.5cm]{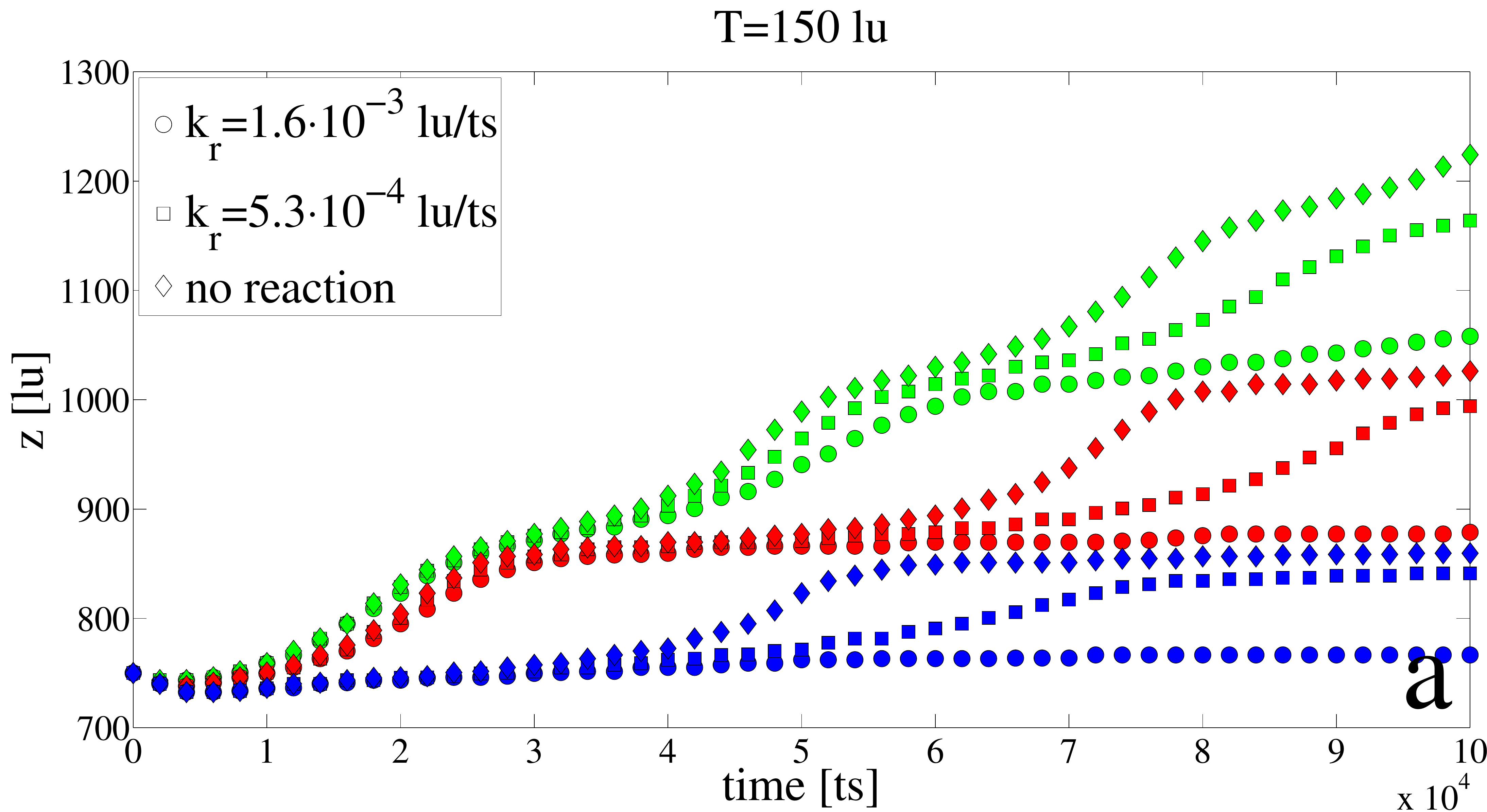}
\includegraphics[width=8.5cm]{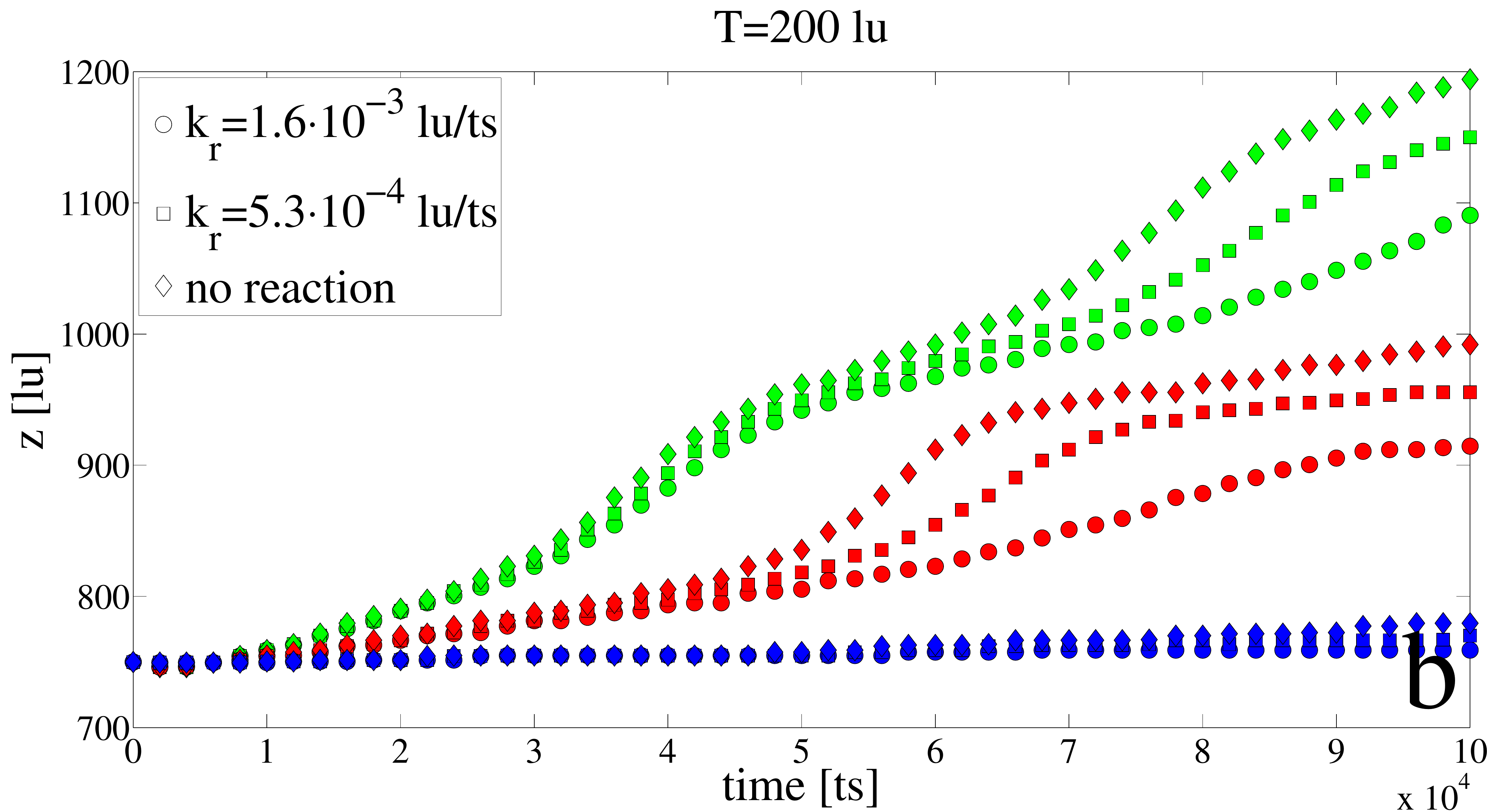}\\
\includegraphics[width=8.5cm]{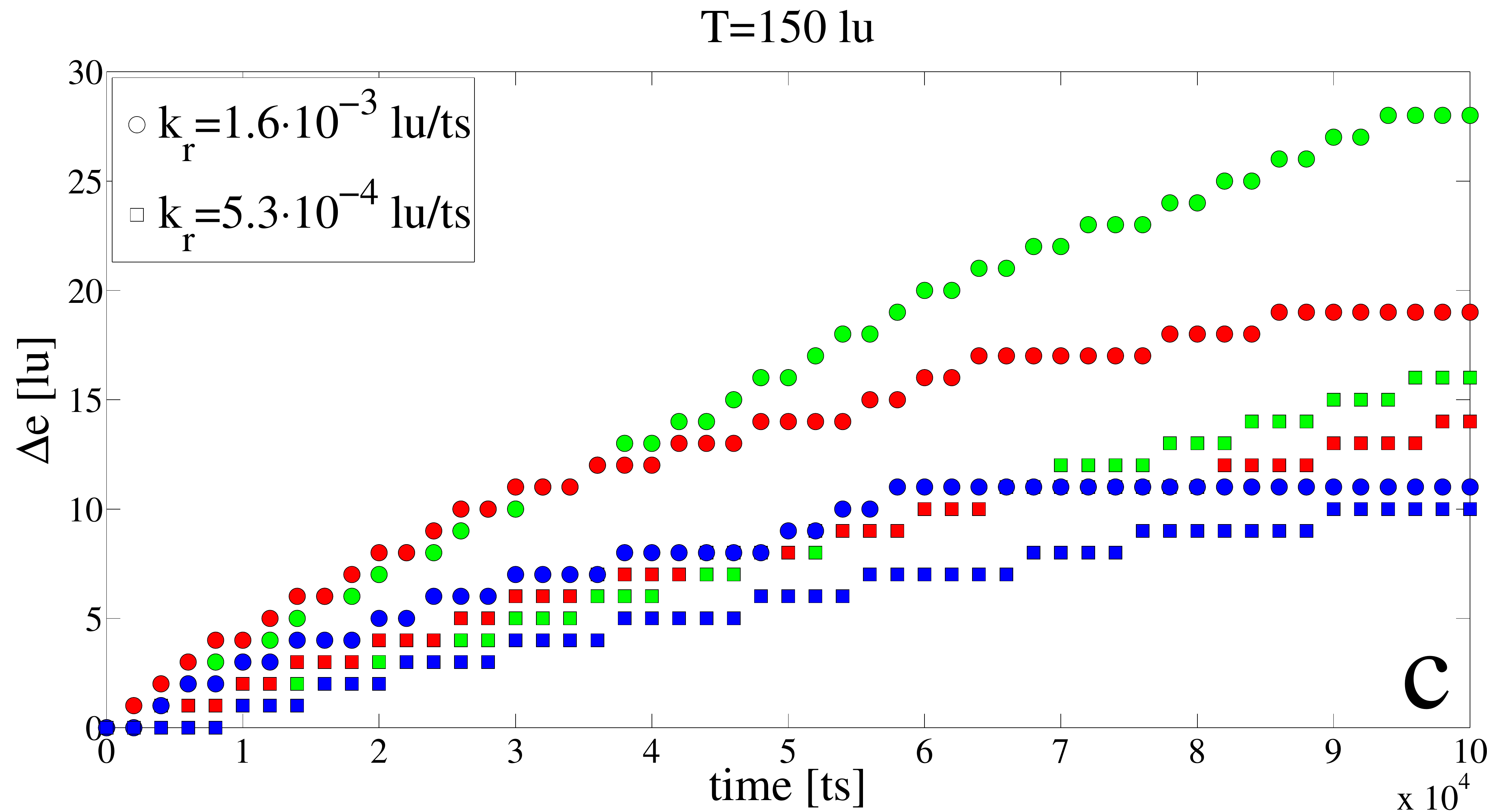}
\includegraphics[width=8.5cm]{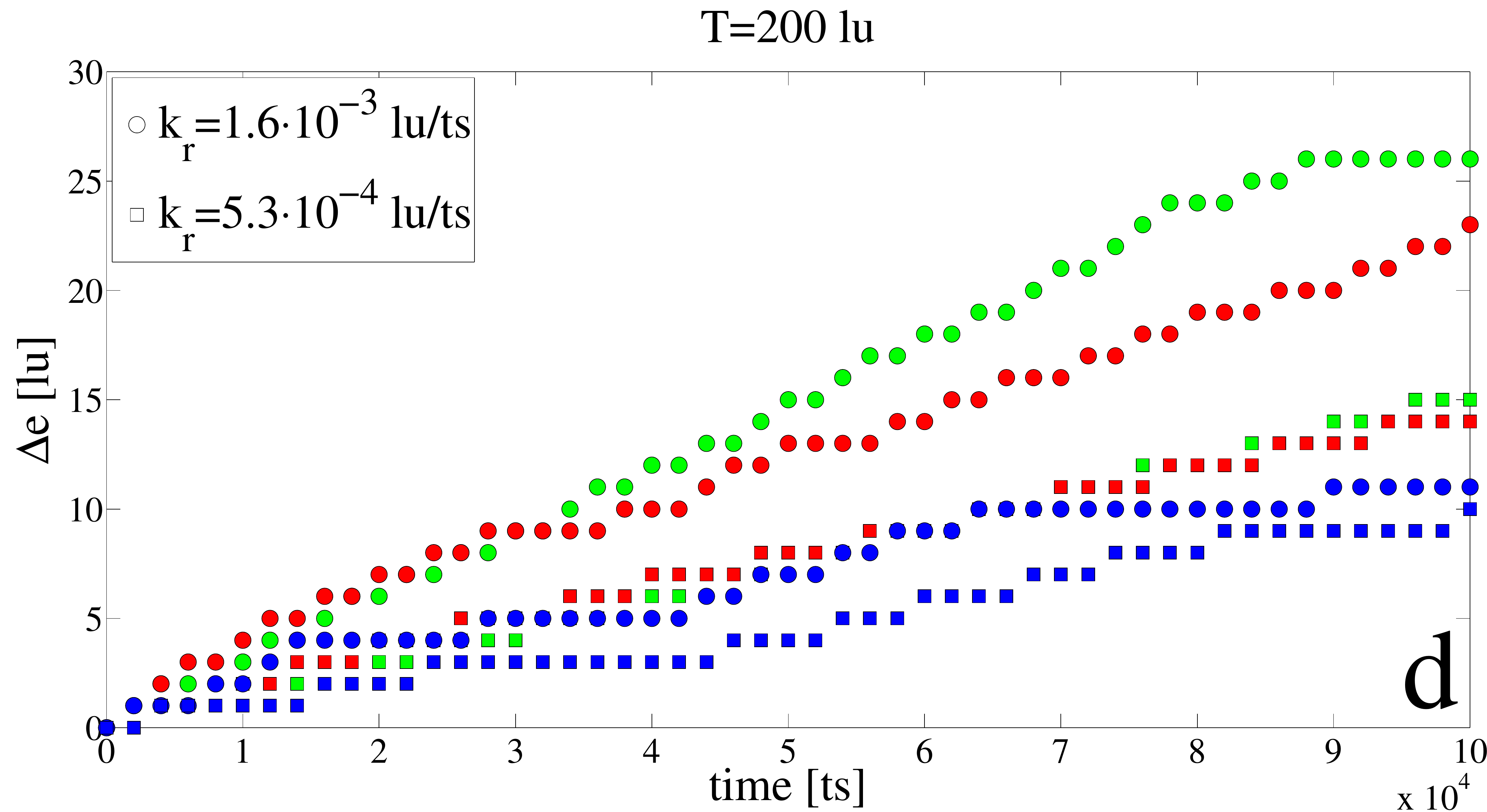}
\caption{\label{fig:sin4}
Results for sinusoidal capillary systems of length $L=750$ lu with $N_{y}=125$ lu in the presence of surface reaction,
controlled by the reaction-rate constant $k_{\mathrm{r}}$. Color code based on the amplitude $A=i(N_{y}-25)/16$ with $i=2,3,4$, for green, red
and blue, respectively. Top: Time dependence of the invading front. Bottom: Maximal thickening $\Delta e$ of the solid surface in the course 
of time.}
\end{figure*}
\begin{figure*}[t]
\includegraphics[width=12cm]{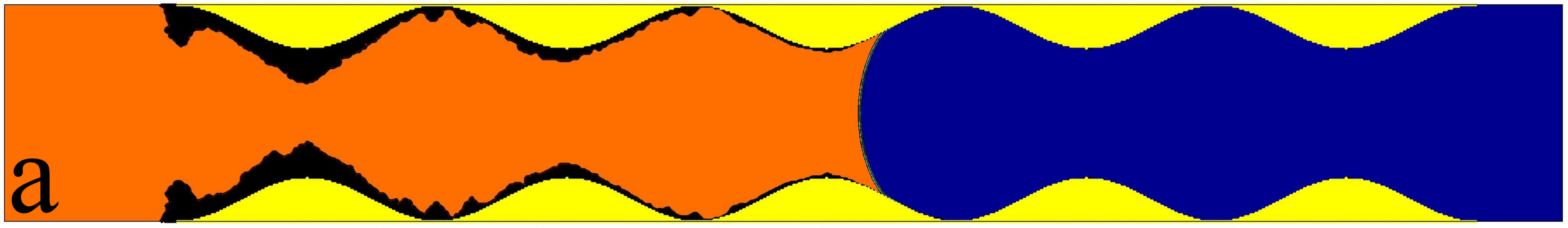}\\
\vspace{0.1cm}
\includegraphics[width=12cm]{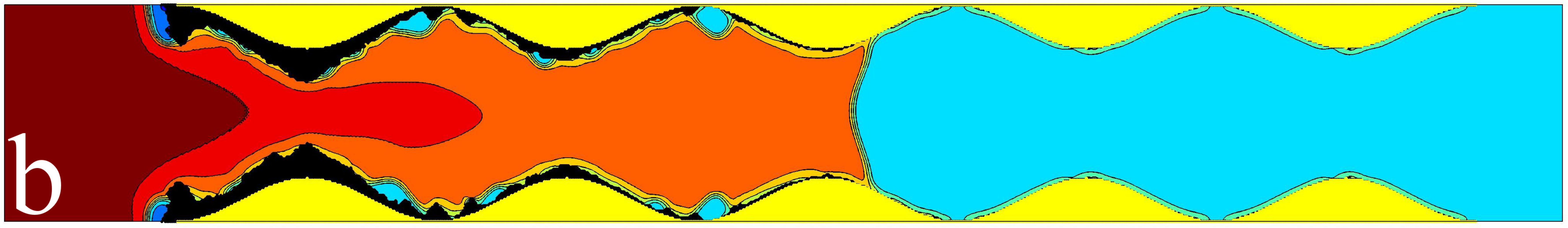}\\
\vspace{0.1cm}
\includegraphics[width=12cm]{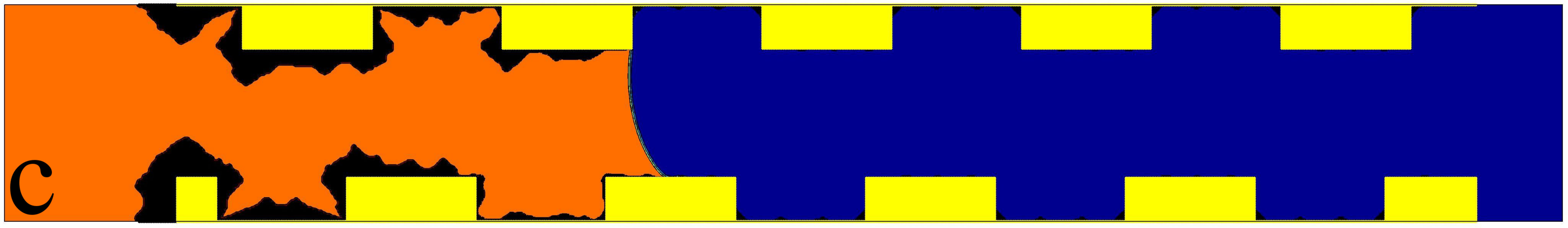}\\
\vspace{0.1cm}
\includegraphics[width=12cm]{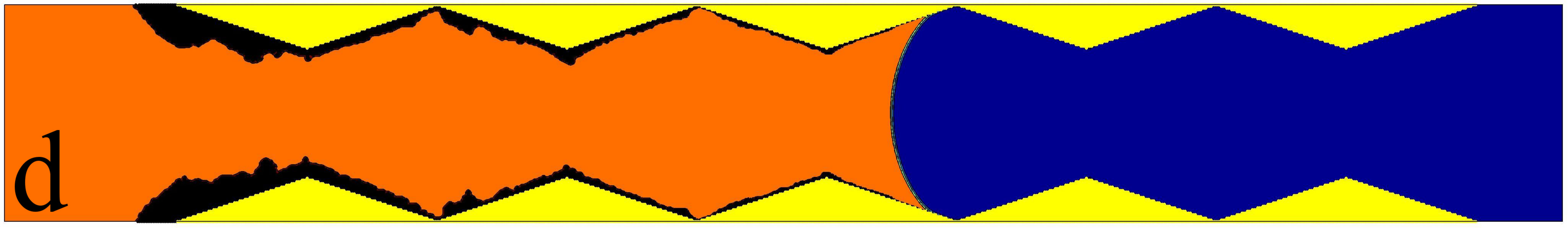}
\caption{\label{fig:system1}
Periodic systems after $60'000$ timesteps with channel width $N_{y}=125$ lu, capillary length $L=750$ lu, period $T=150$ lu, amplitude 
$A=(N_{y}-25)/8$, corresponding to $H_{\mathrm{min}}=75$ lu, and reaction-rate constant $k_{\mathrm{r}}=1.6\cdot10^{-3}$ lu/ts. Unless stated 
otherwise, the figures represent fluid flow with color code based on the density of the wetting component. Dark points for the phase resulting 
from surface growth and yellow points for the initial solid phase. (a) Sinusoidal profile. (b) Sinusoidal profile. Representation of solute 
transport with color code based on the solute concentration. (c) Step-shaped profile with misalignment given by $\Delta x=100$ lu. (d) Zig-zag profile.}
\end{figure*}
\begin{figure*}[b]
\includegraphics[width=8.5cm]{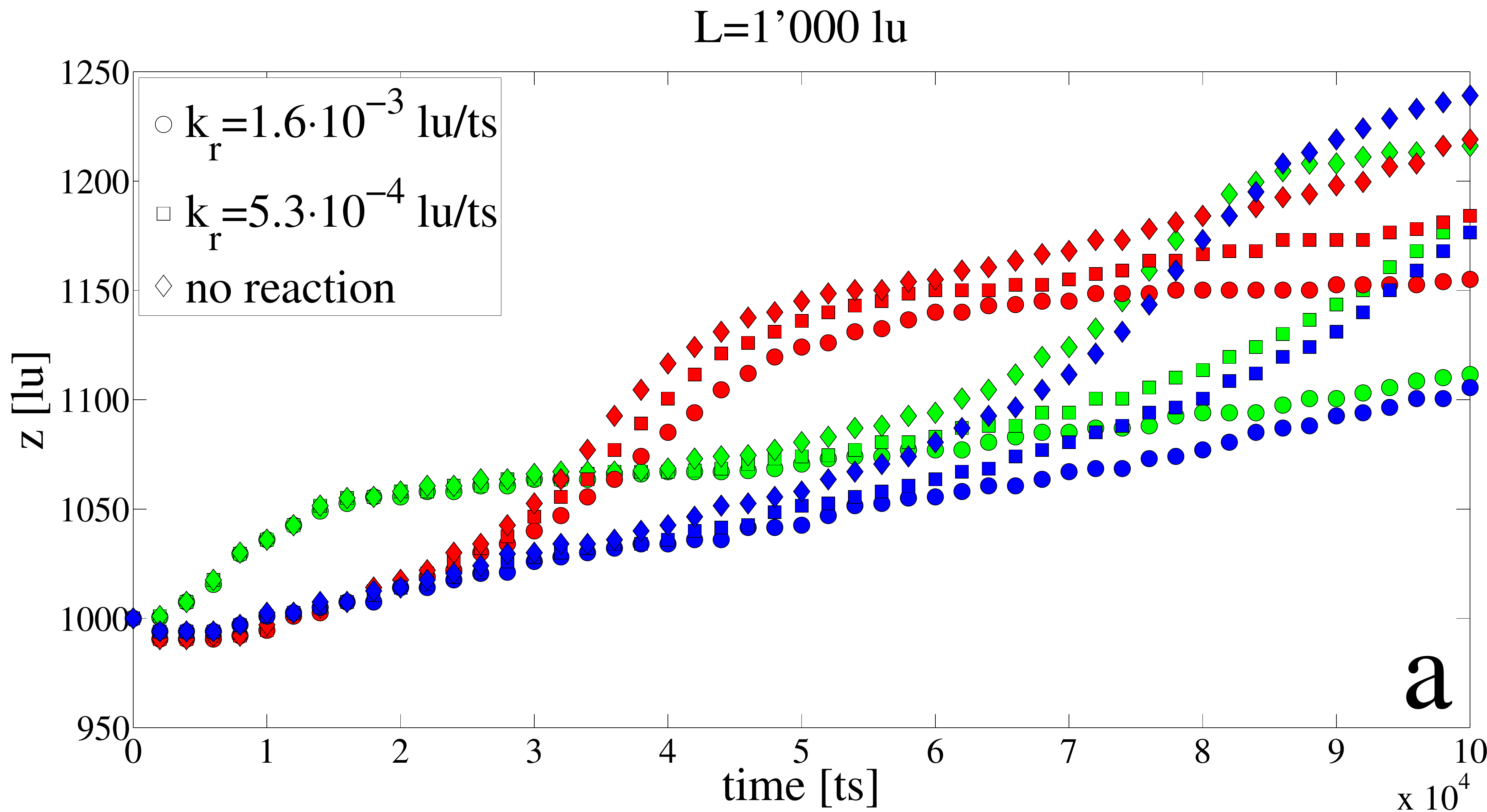}
\includegraphics[width=8.5cm]{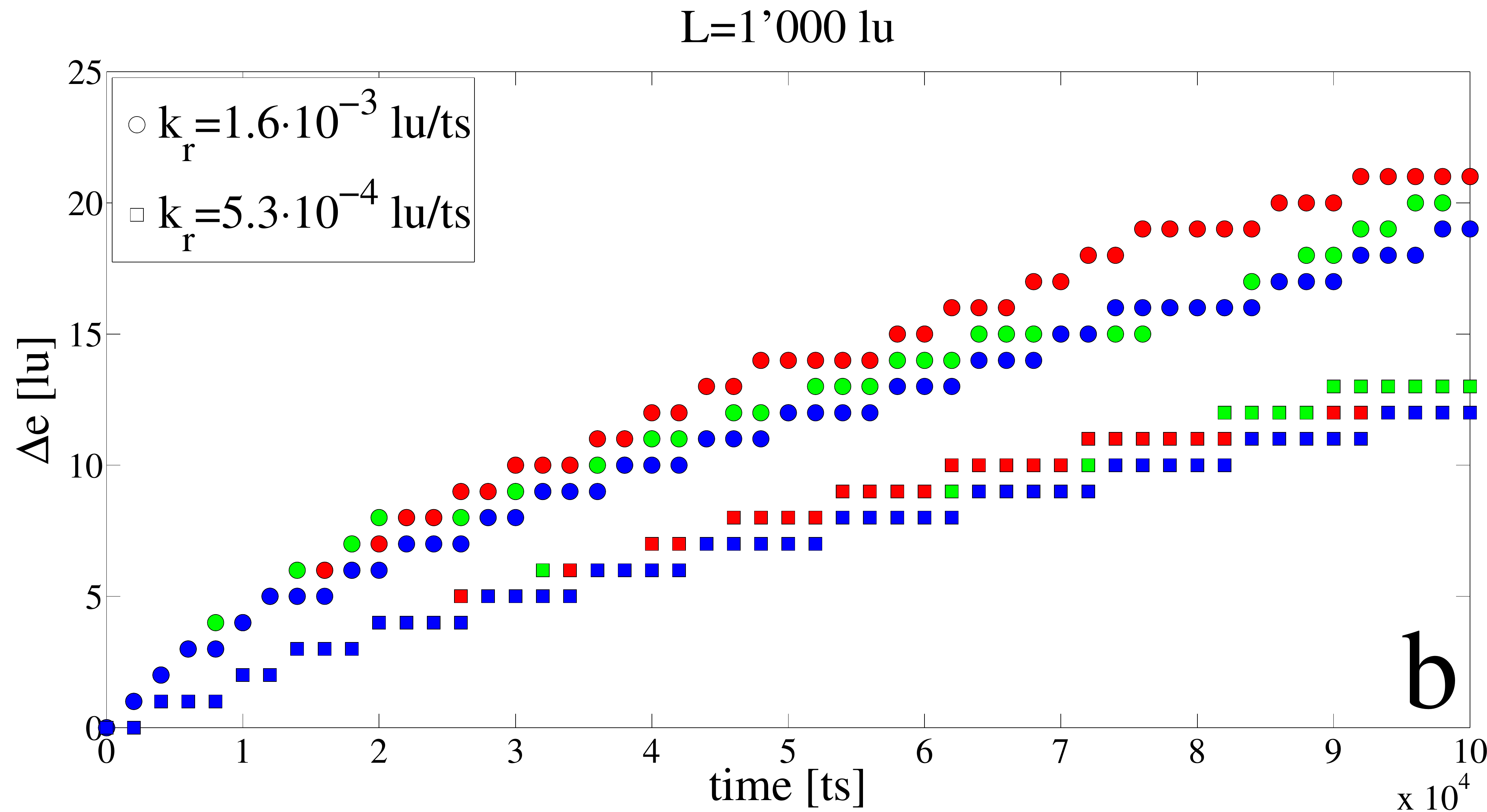}\\
\includegraphics[width=8.5cm]{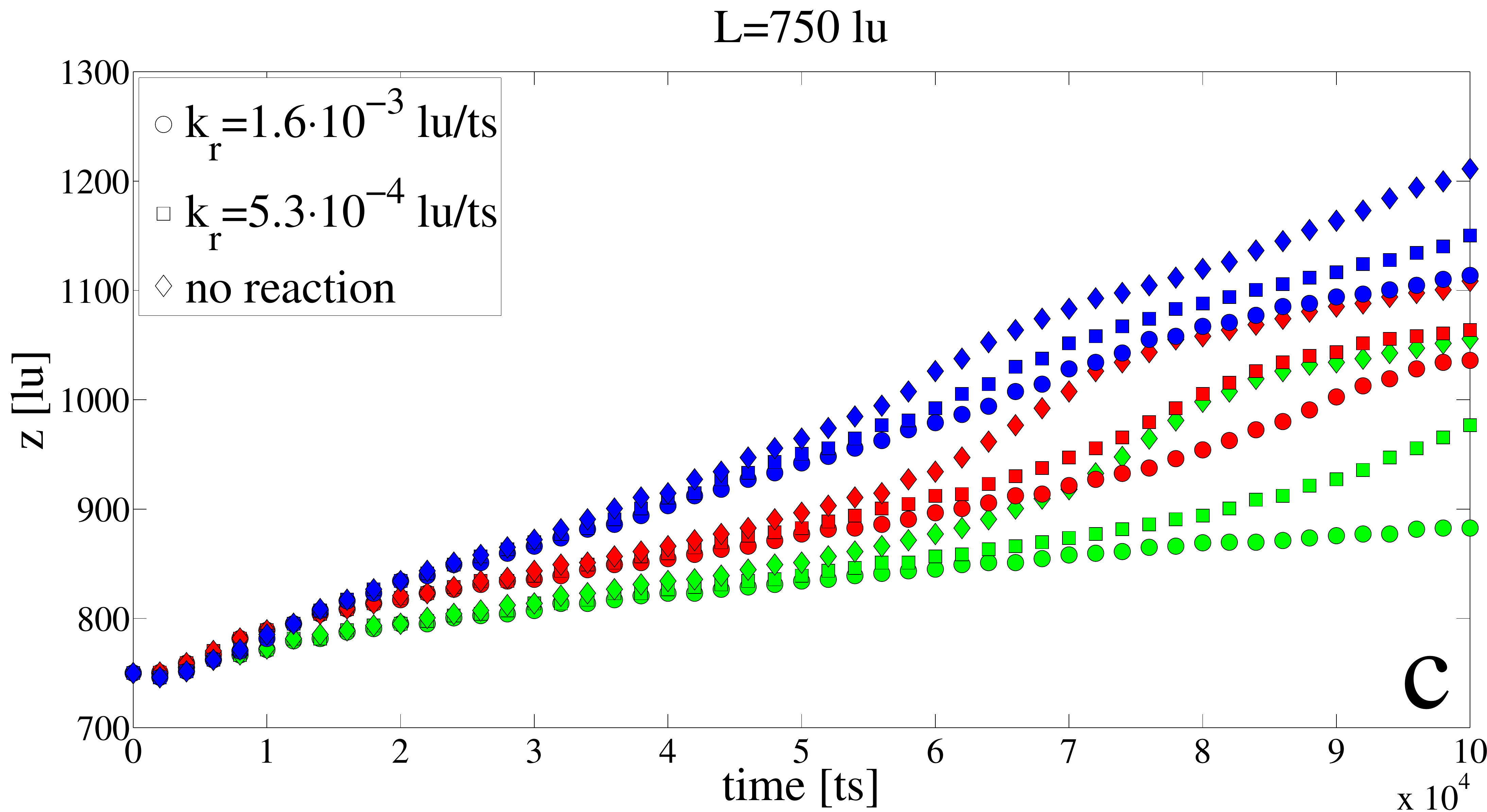}
\includegraphics[width=8.5cm]{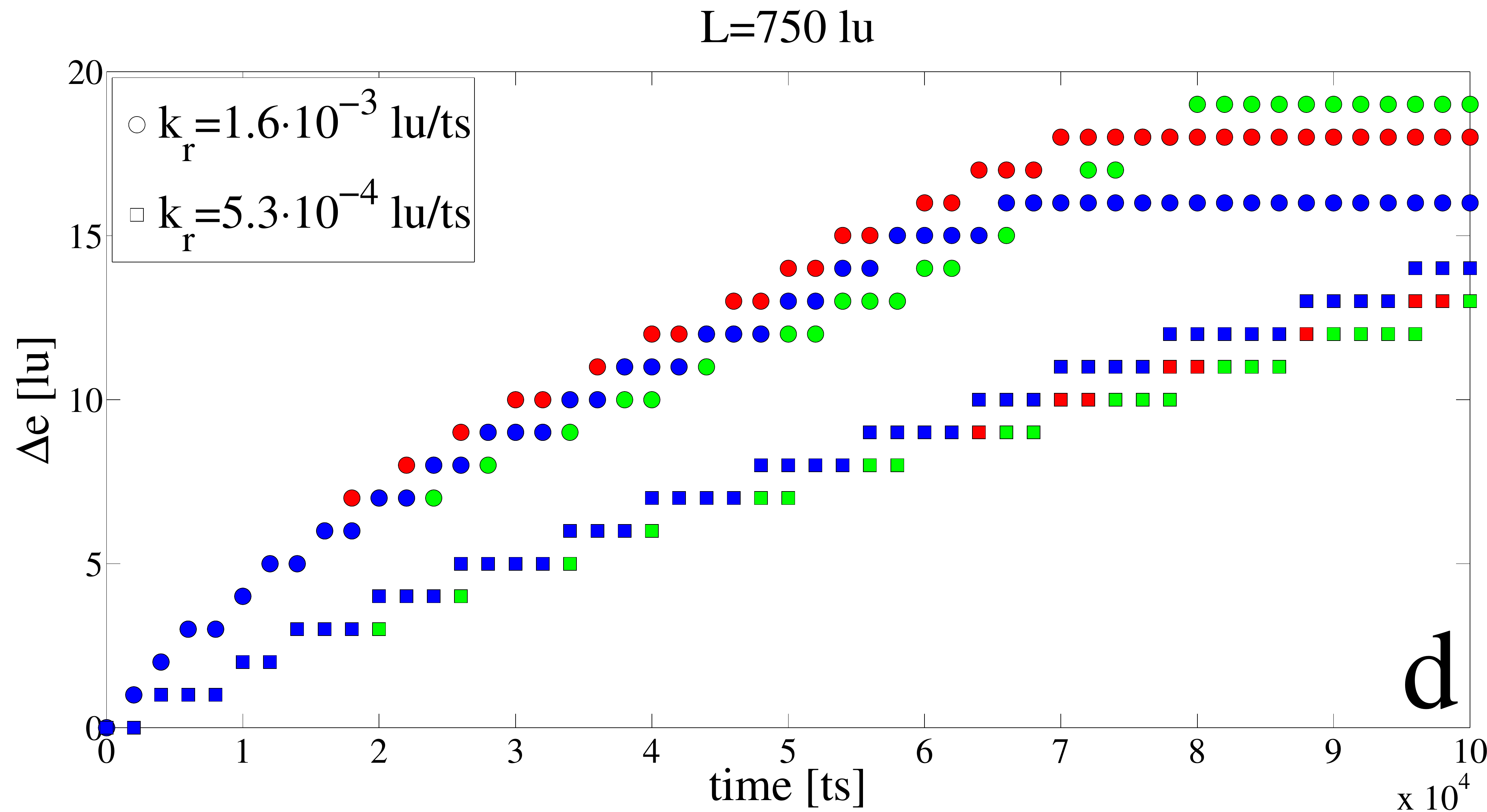}
\caption{\label{fig:sin5}
Sinusoidal capillary systems with reactivity enabled, adjusted via the reaction-rate constant $k_{\mathrm{r}}$. The domain
width is set to $N_{y}=125$ lu and the amplitude is kept fixed at $A=3(N_{y}-25)/16$, yielding the minimum height
$H_{\mathrm{min}}=50$ lu. Top: Color code based on the period $T$: green, red and blue for $T=150,200,250$ lu, 
respectively. Bottom: Color code based on the parameter $\Delta x$ controlling the misalignment between the walls:
green, red and blue for $\Delta x=0,50,100$ lu. The period is $T=250$ lu.}
\end{figure*}
\begin{figure*}[t]
\includegraphics[width=8.5cm]{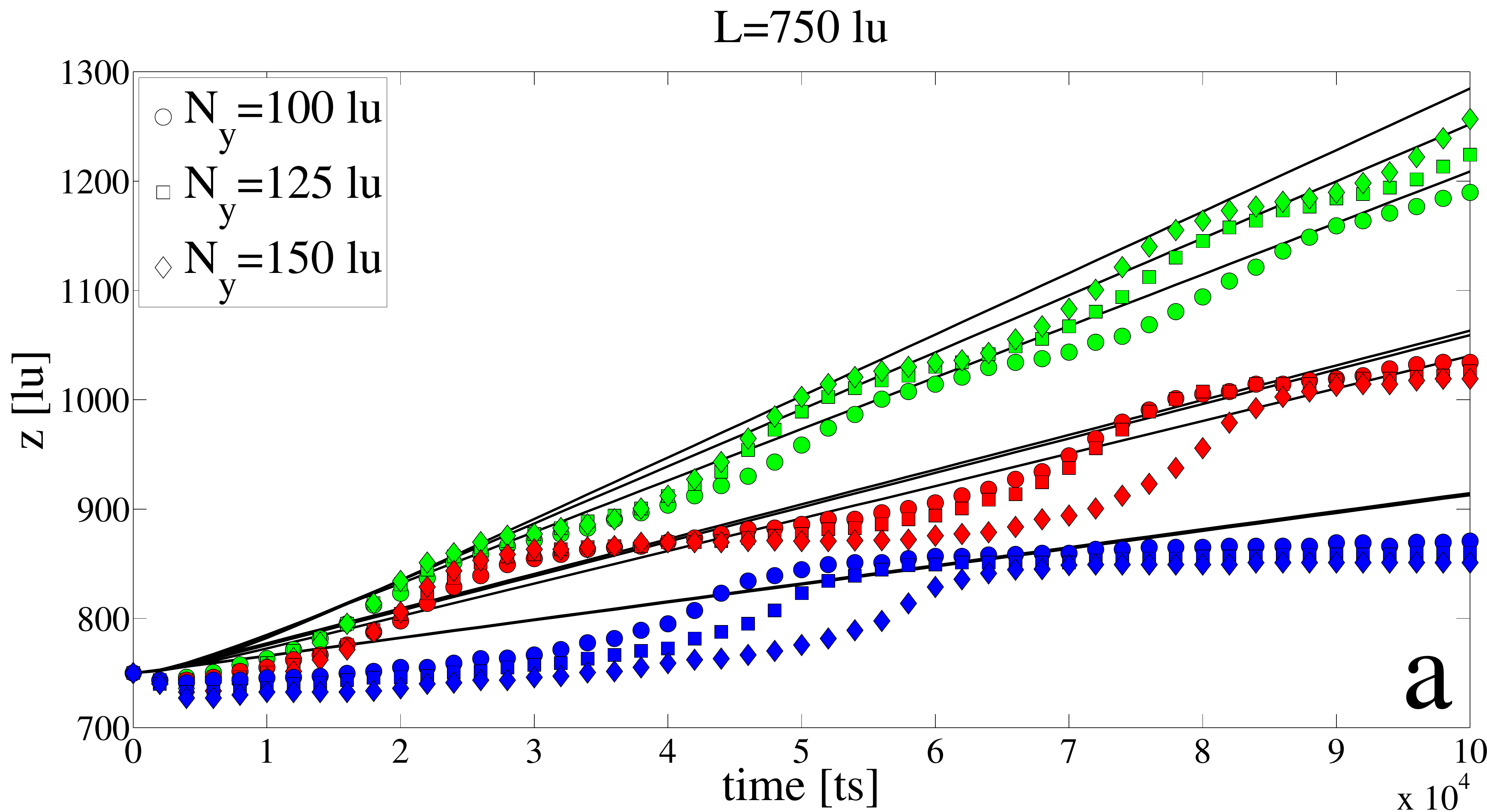}
\includegraphics[width=8.5cm]{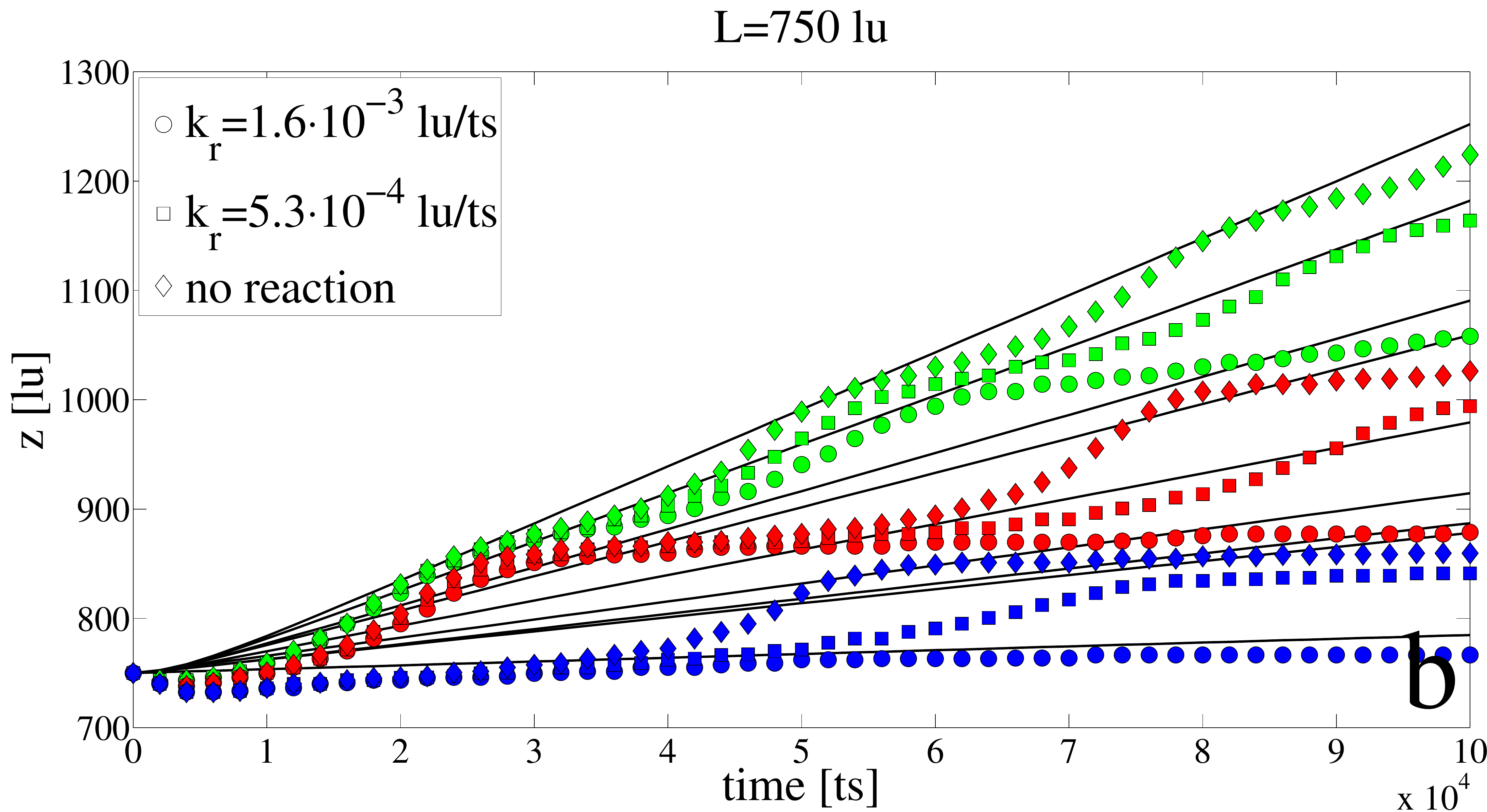}\\
\includegraphics[width=8.5cm]{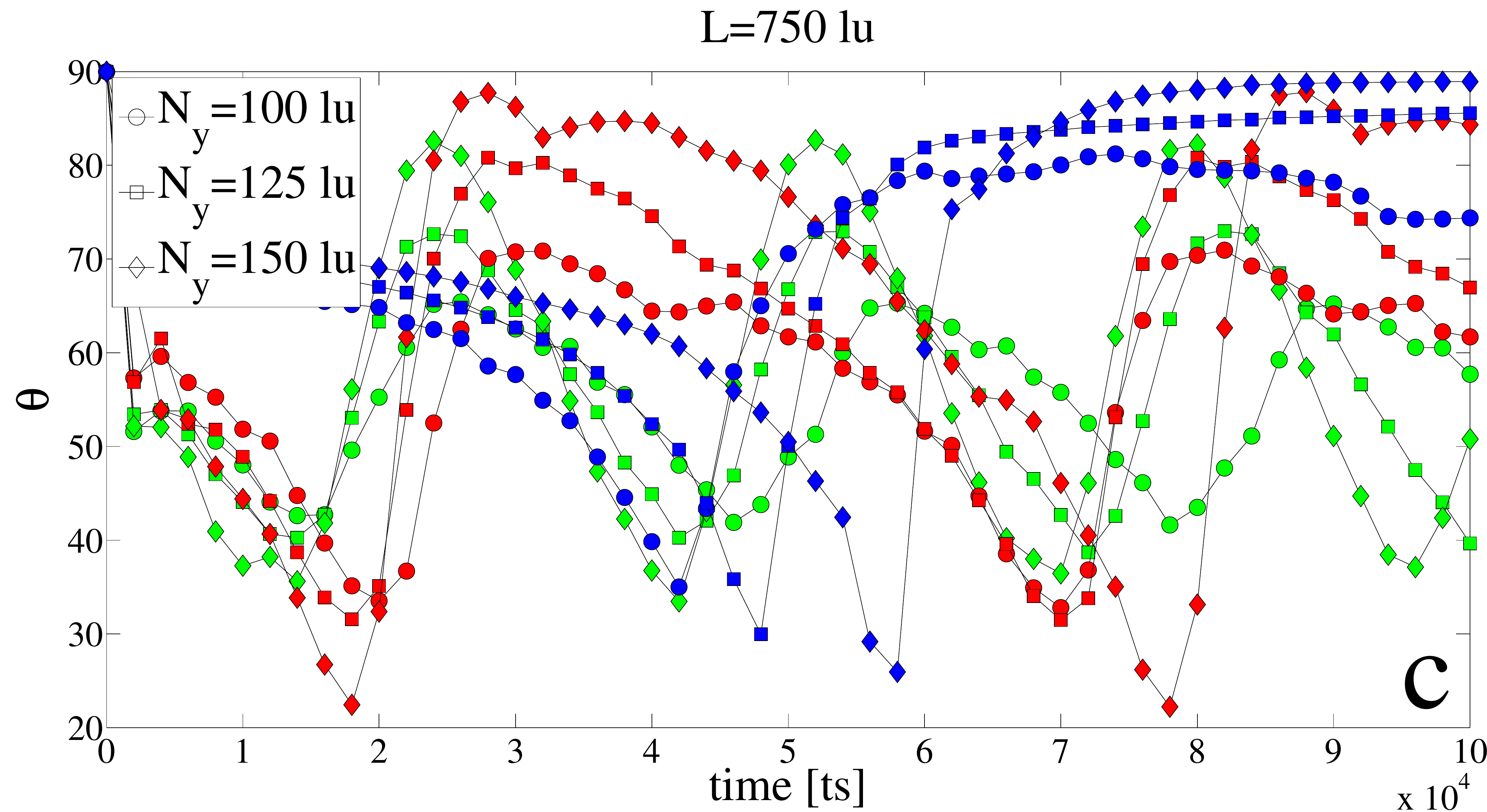}
\includegraphics[width=8.5cm]{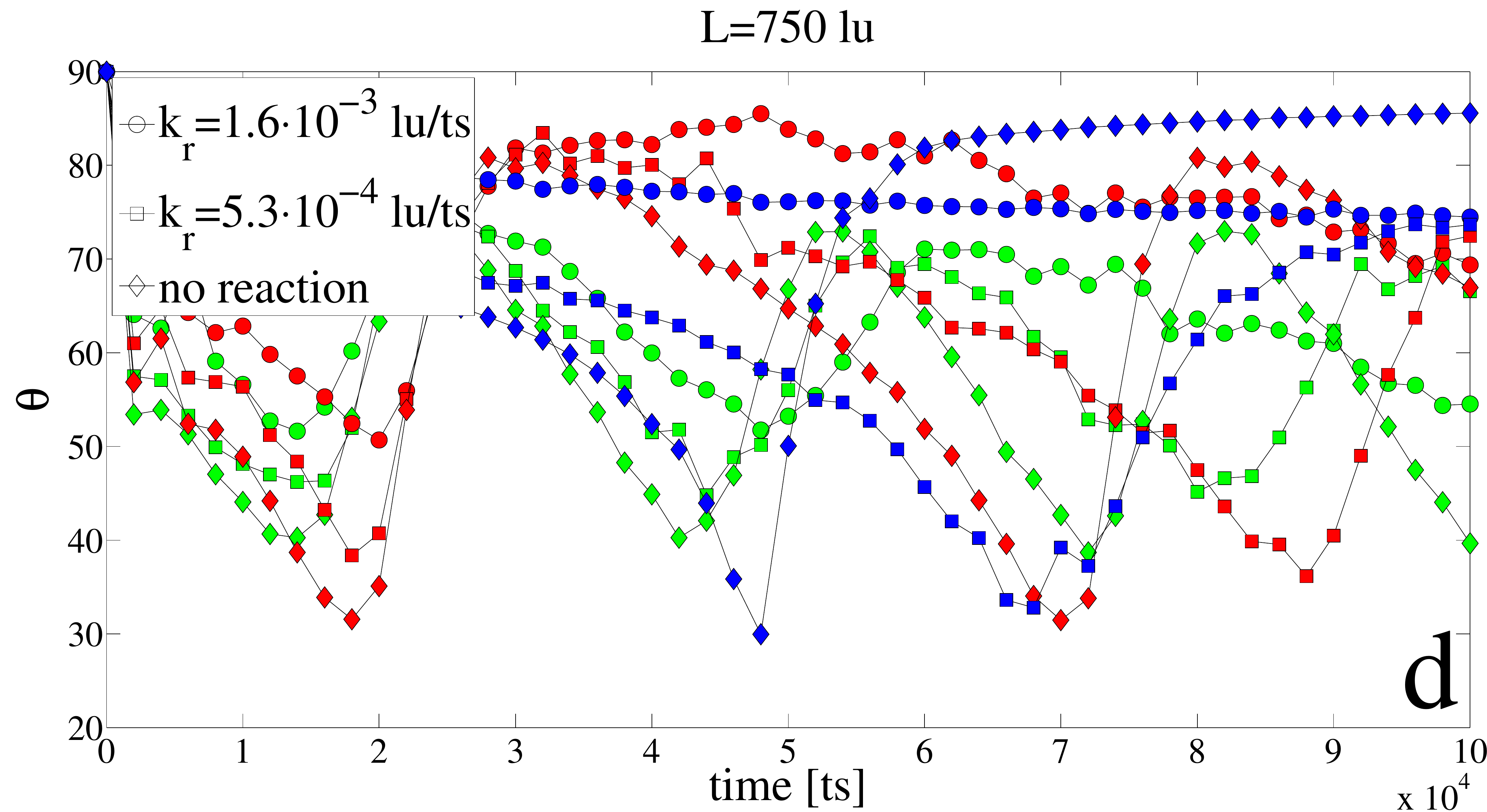}
\caption{\label{fig:sin6}
Sinusoidal capillary systems of period $T=150$ lu. Points for simulation results; the solid lines are determined using
Eq.~\ref{eq:z}. (a) The solid surface is inert. Color code based on the amplitude $A$: green, red and blue for $A=i(N_{y}-25)/16$ with 
$i=2,3,4$, respectively. (b) The surface reaction is controlled via the reaction-rate constant $k_{\mathrm{r}}$. The domain width is 
$N_{y}=125$ lu. Color code based on the amplitude $A=i(N_{y}-25)/16$, with $i=2,3,4$, corresponding to green, red and blue, respectively.
(c) Dynamic contact angle for the conditions of a. (d) Dynamic contact angle for the systems of b.}
\end{figure*}

\section*{2.~~~LB MODELS}

The LB method can simulate hydrodynamic phenomena at a mesoscopic level: streaming
and collision processes are reminiscent of the particulate nature of liquids, demanding
nevertheless for a statistical treatment. More precisely, the velocity space is discretized
and distribution functions account for the particles moving along a given direction 
(Benzi et al., 1992; Chen and Doolen, 1998; Succi, 2009; Sukop and Thorne, 2010; Wolf-Gladrow; 2005). 
In the present work we employ the same models as in Sergi et al.~(2014) with the same notation and
terminology. More details can be found in the related literature. Namely, fluid flow is treated by
means of multicomponent models (Chibbaro, 2008; Chibbaro et al., 2009b; Diotallevi et al., 2009a and
2009b). For solute transport and reactive boundaries we follow closely Kang et al.~(2007). For similar
works the interested reader is addressed to the articles by Kang et al.~(2002b, 2003, 2004) and Lu et
al.~(2009).

In the sequel, we shall express the simulation results in model units. The basic units are those of mass, length
and time; in symbols, mu, lu and ts, respectively (Sukop and Thorne, 2010). Ordinary units are obtained after suitable 
transformations (Gross et al., 2010; Lu et al., 2009). Of course, comparative analyses with experiments based on 
dimensionless parameters (Reynolds, Damkohler, Peclet, capillary, Bond, Weber numbers, etc.) are preferable.
These numbers allow to establish equivalences between systems (Landau and Lifshitz, 2008) and have the same value in 
the common systems for units like the SI.


\begin{figure*}[t]
\includegraphics[width=8.5cm]{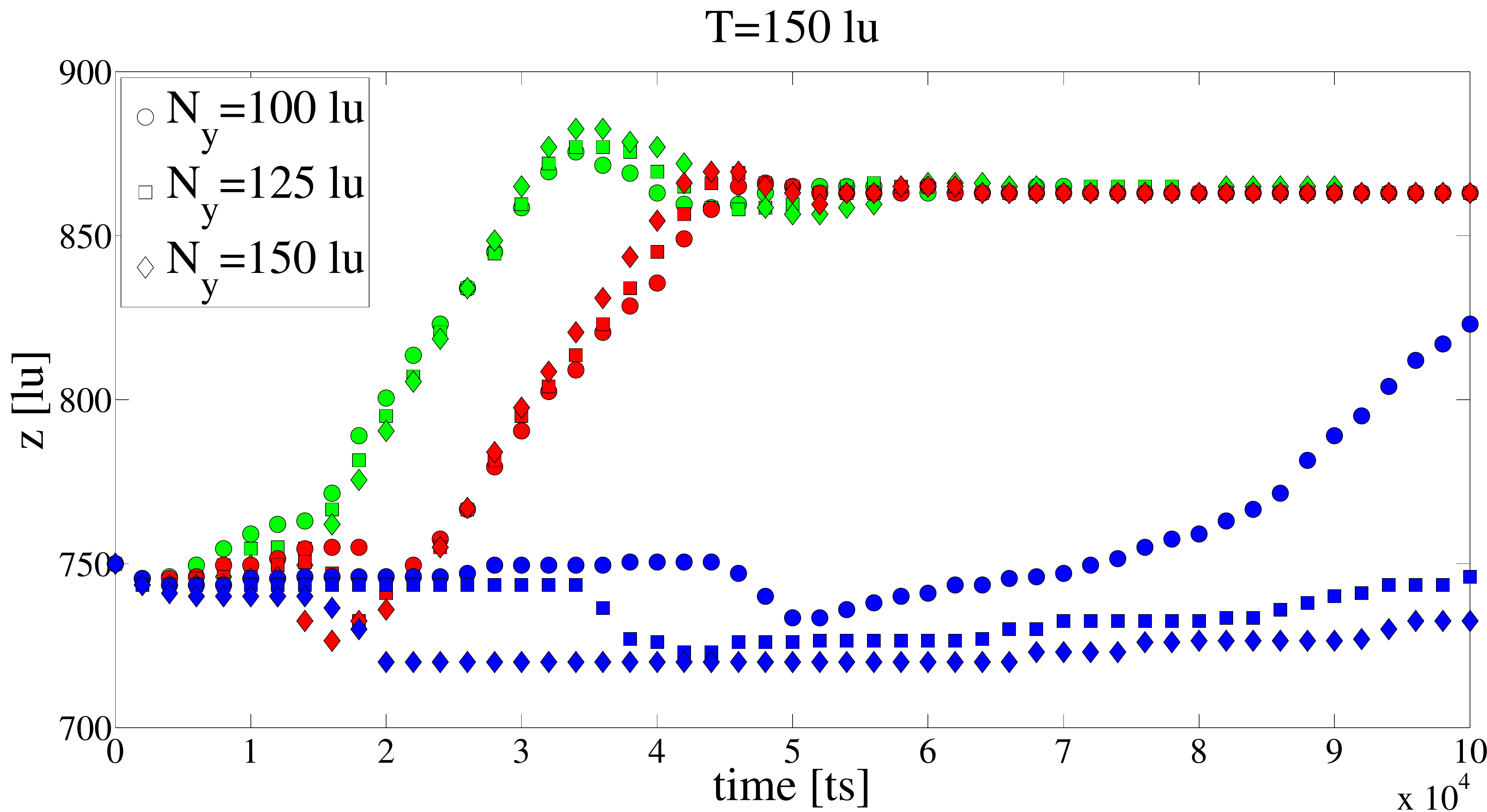}
\caption{\label{fig:step1}
Front displacement in the course of time for step-walled capillaries without reaction.
The length of the capillary is $L=750$ lu. Color code based on the amplitude $A$. It is given by $A=i(N_{y}-25)/16$ 
with $i=2,3,4$, corresponding to the colors green, red and blue, respectively.}
\end{figure*}
\begin{figure*}[t]
\includegraphics[width=8.5cm]{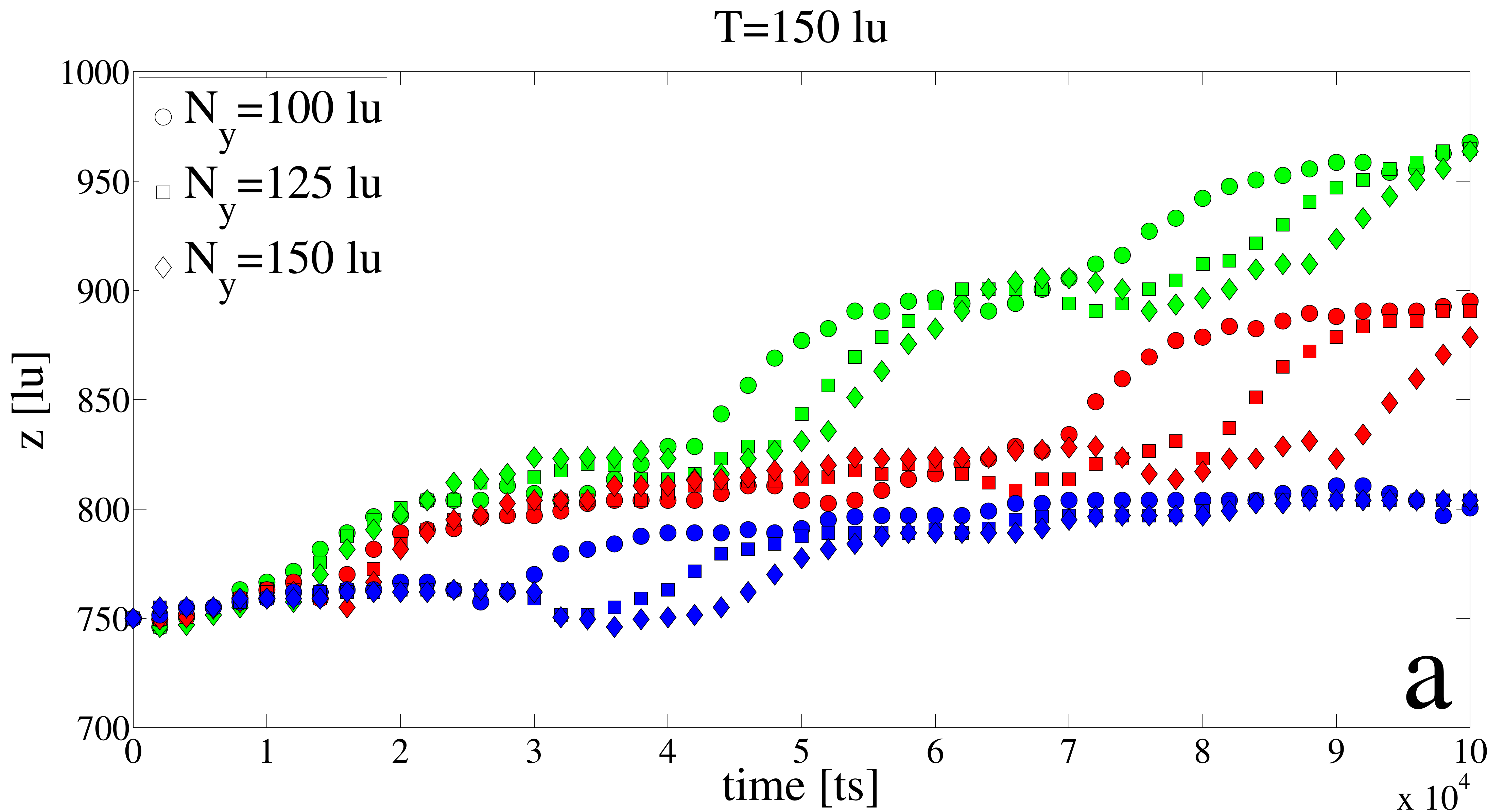}
\includegraphics[width=8.5cm]{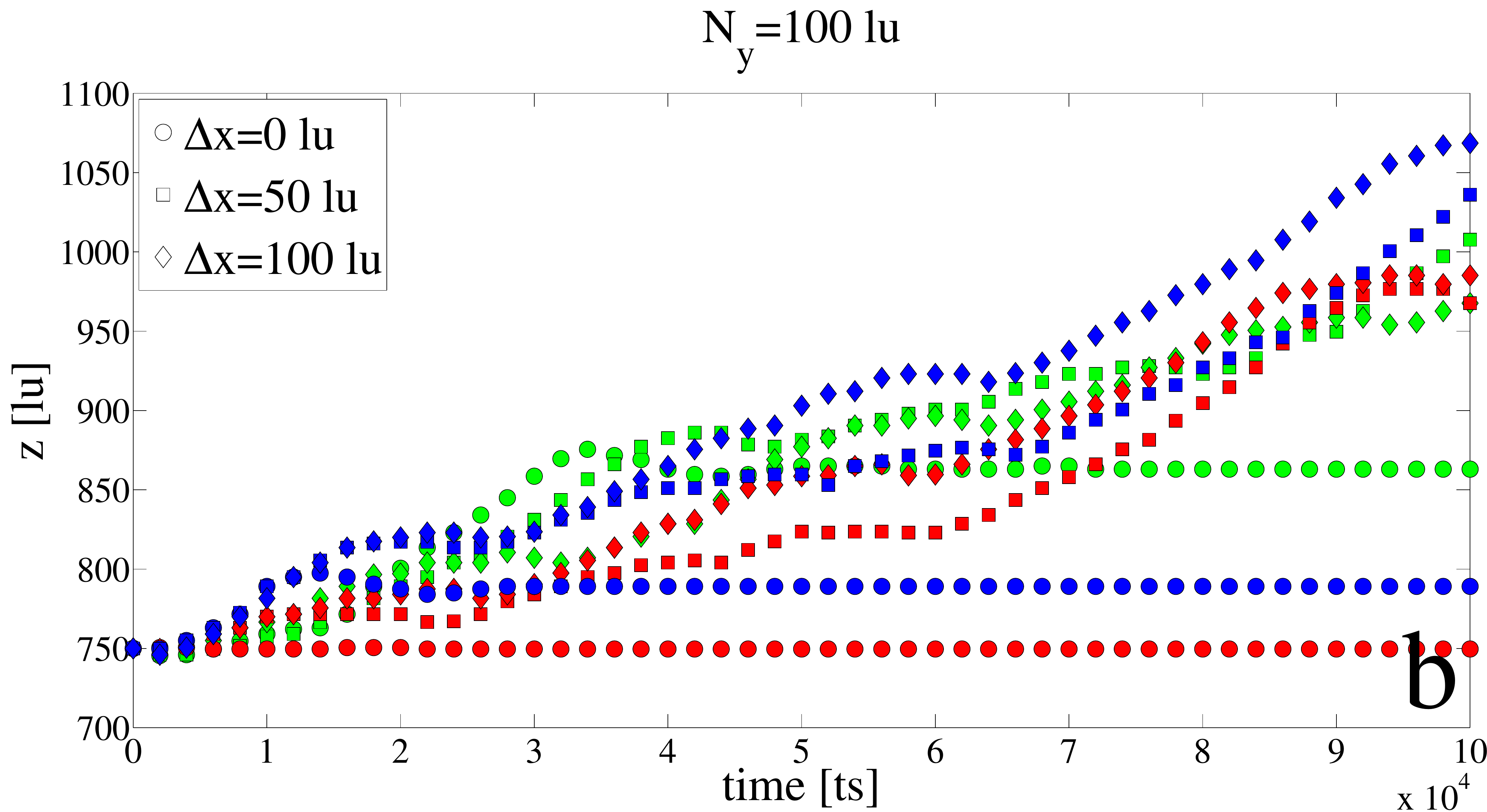}
\caption{\label{fig:step2}
Centerline position of the invading front in the course of time for capillary systems with periodic step-shaped walls in the 
absence of surface reaction. The capillaries have length $L=750$ lu. (a) The 
misalignment between the walls is $\Delta x=100$ lu. Color code based on the amplitude $A=i(N_{y}-25)/16$ with $i=2,3,4$ for 
green, red and blue, respectively. (b) The amplitude is fixed to $A=(N_{y}-25)/8$, determining the minimum height 
$H_{\mathrm{min}}=62$ lu. Color code based on the period: $T=150,200,250$ lu for green, red and blue.}
\end{figure*}
\begin{figure*}[t]
\includegraphics[width=8.5cm]{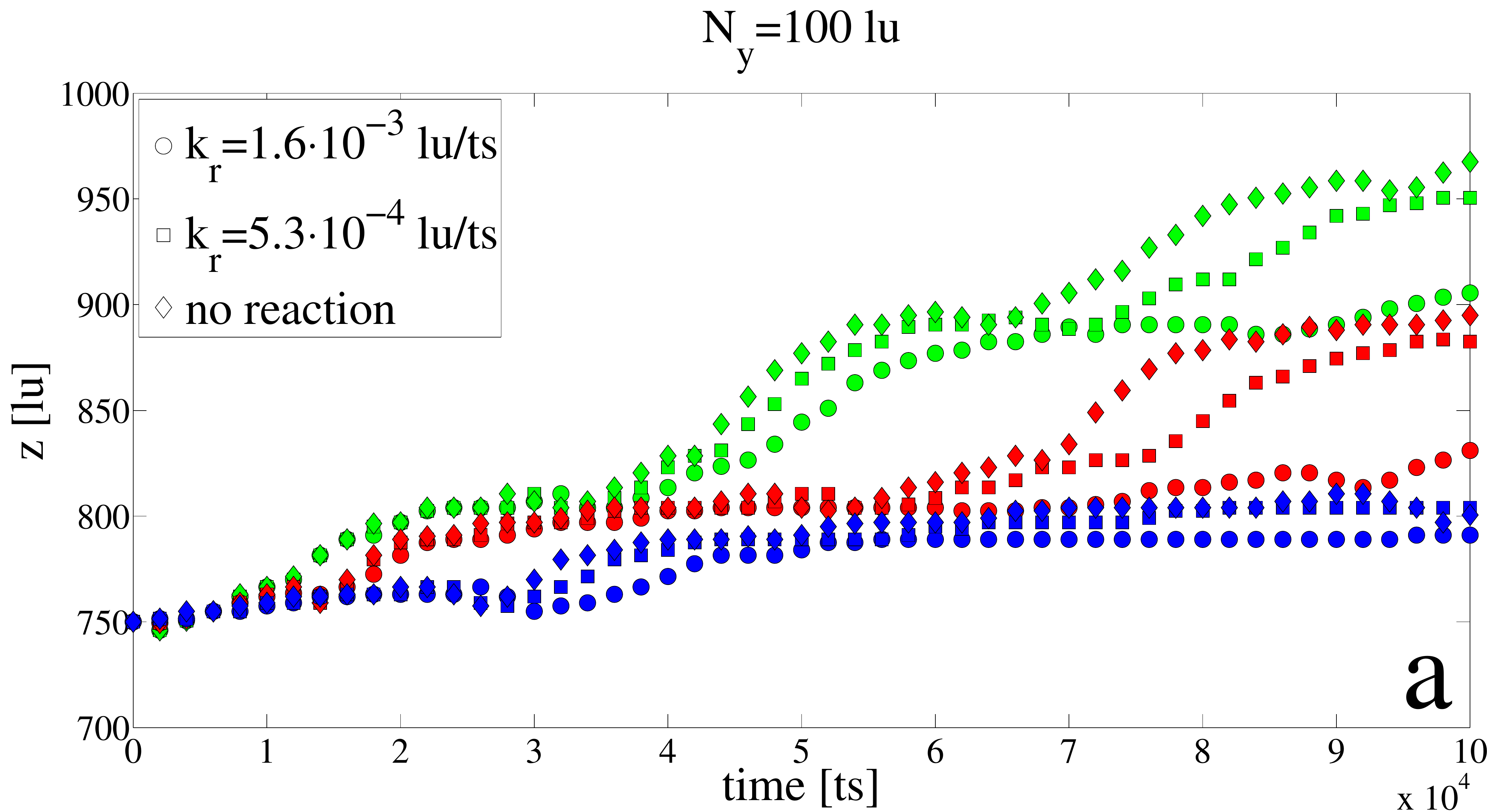}
\includegraphics[width=8.5cm]{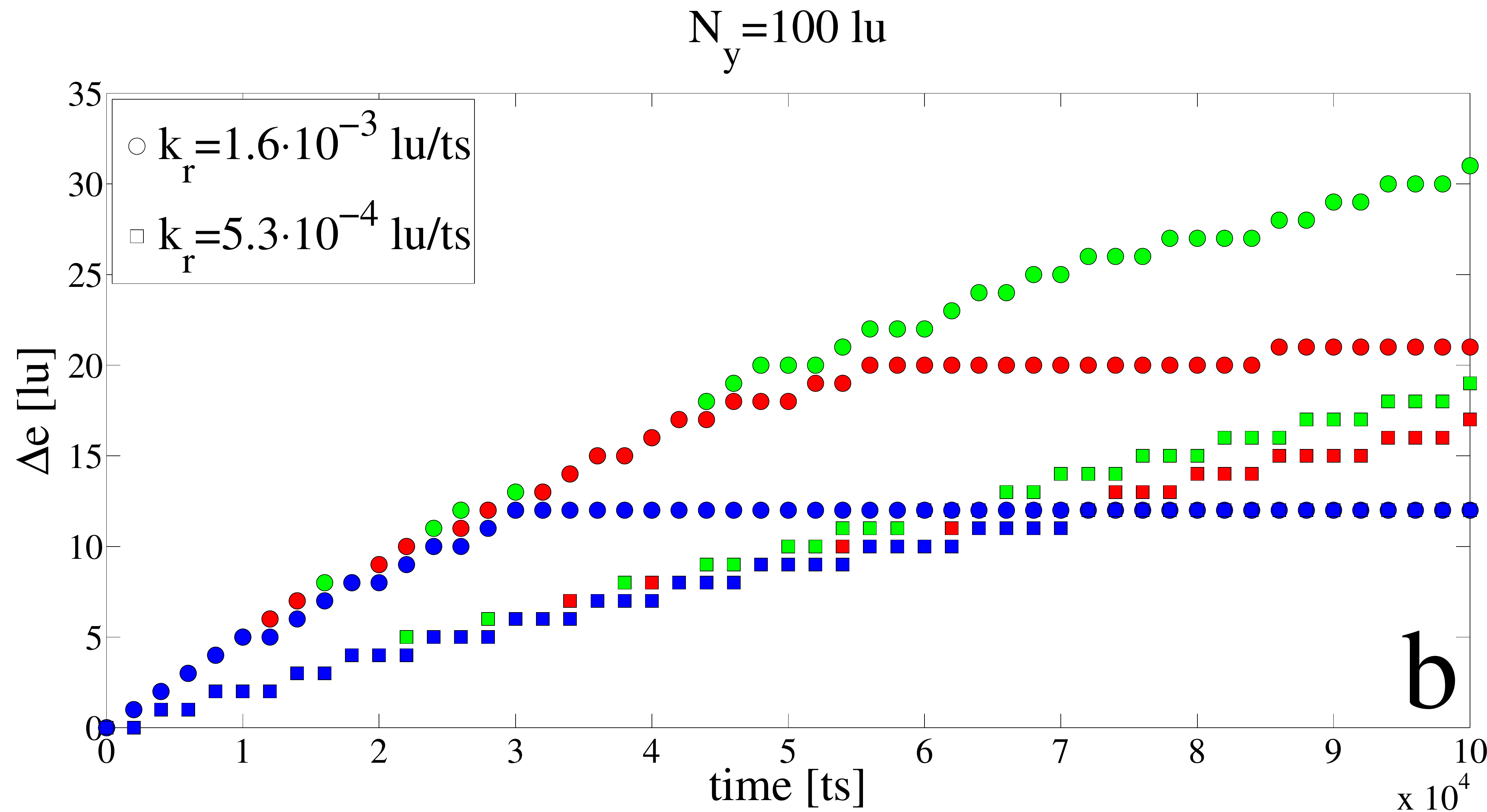}
\caption{\label{fig:step3}
Capillary systems with periodic step-shaped walls in the presence of surface reaction. The period is fixed to $T=150$ lu and
the misalignment is set to $\Delta x=100$ lu. Color code based on the amplitude: green, red and blue for $A=i(N_{y}-25)/16$ 
when $i=2,3,4$. (a) Front displacement in the course of time. (b) Evolution of the maximal thickening of the solid surface.}
\end{figure*}
\begin{figure*}[t]
\includegraphics[width=8.5cm]{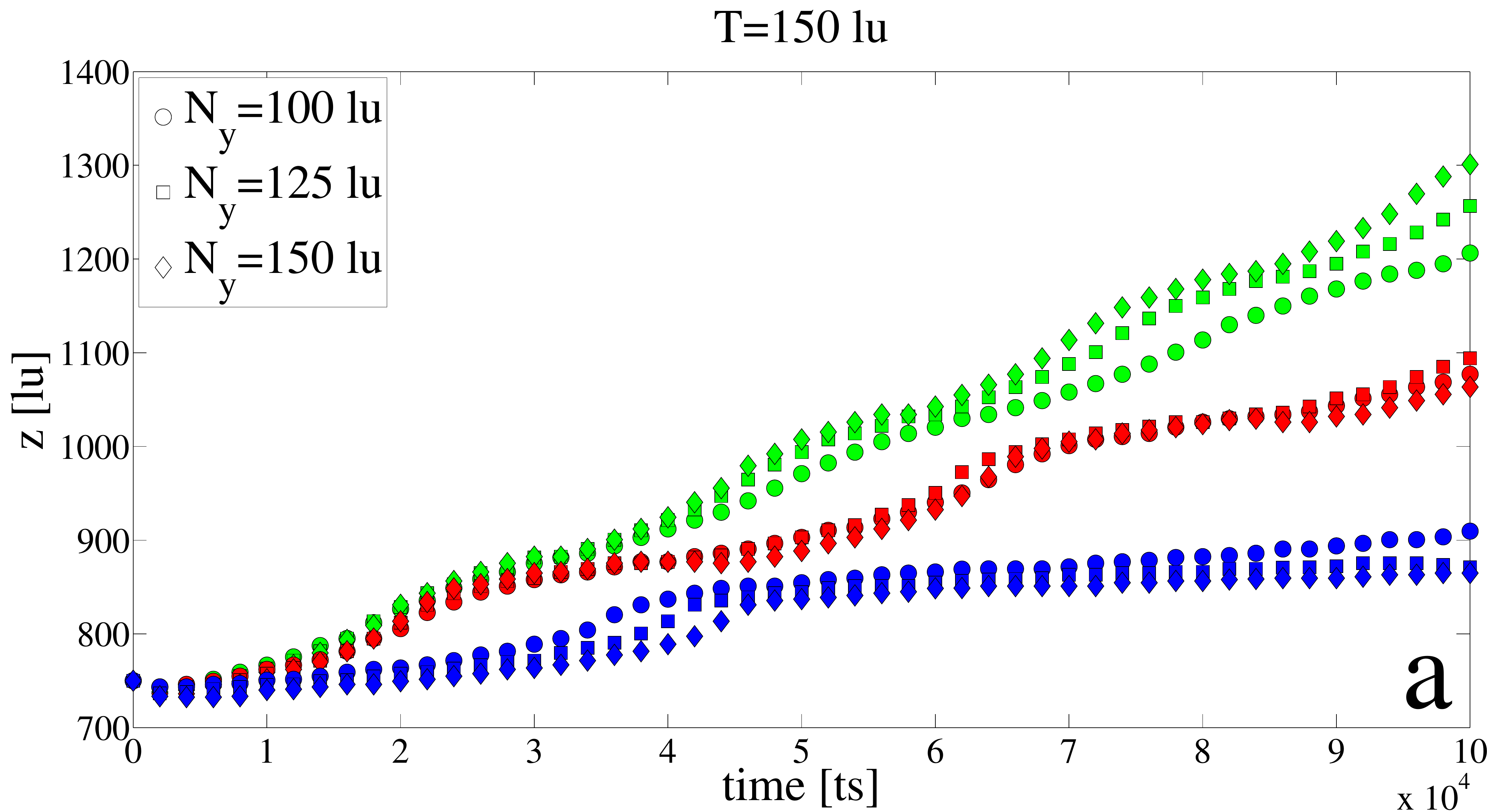}
\includegraphics[width=8.5cm]{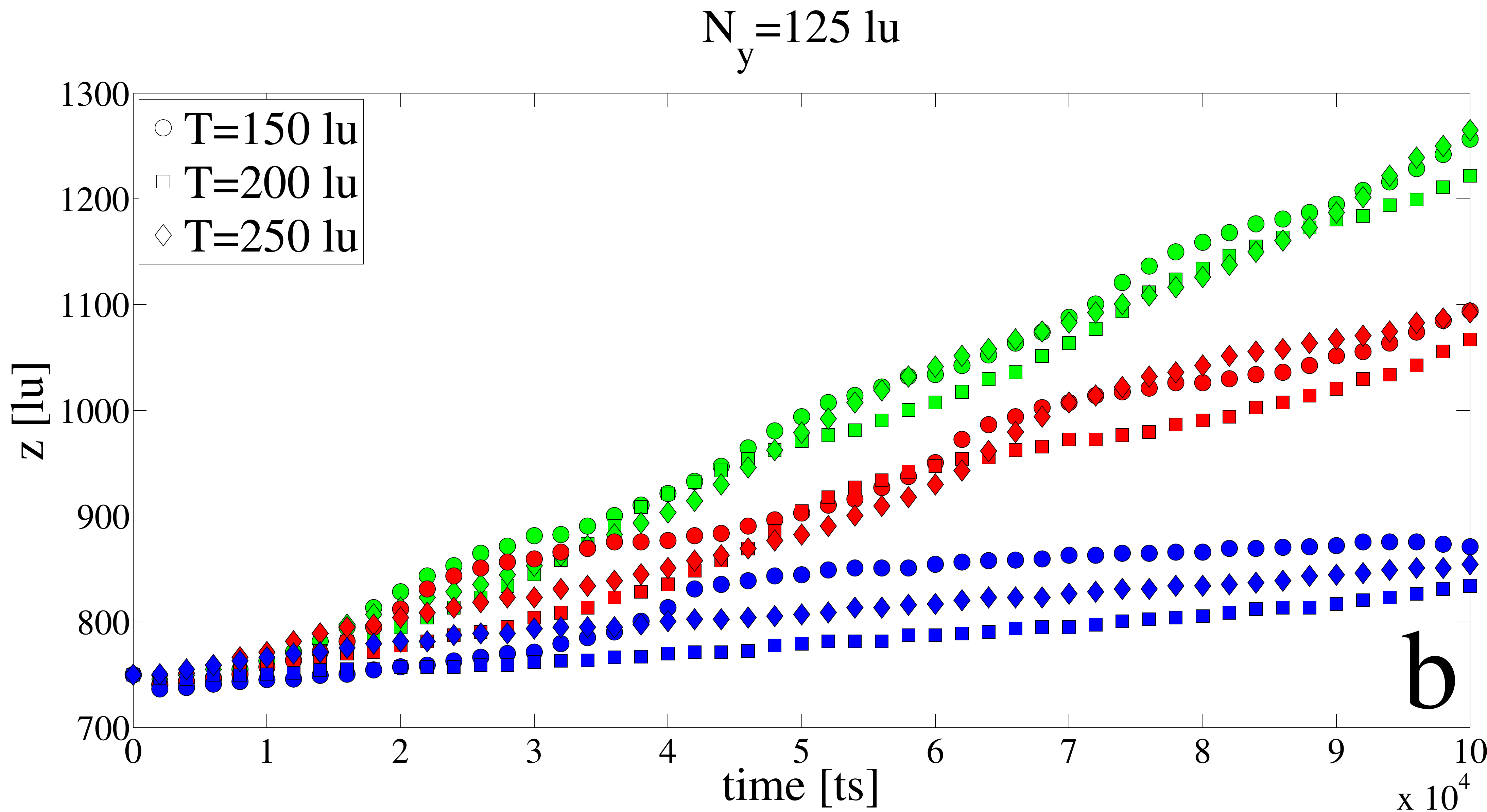}
\caption{\label{fig:zigzag1}
Front dynamics for zig-zag walls in the absence of reactivity. The length of the capillary is 
$L=750$ lu. Color code based on the amplitude $A$: green, red and blue are associated with $A=i(N_{y}-25)/16$ where $i=2,3,4$.}
\end{figure*}
\begin{figure*}[t]
\includegraphics[width=8.5cm]{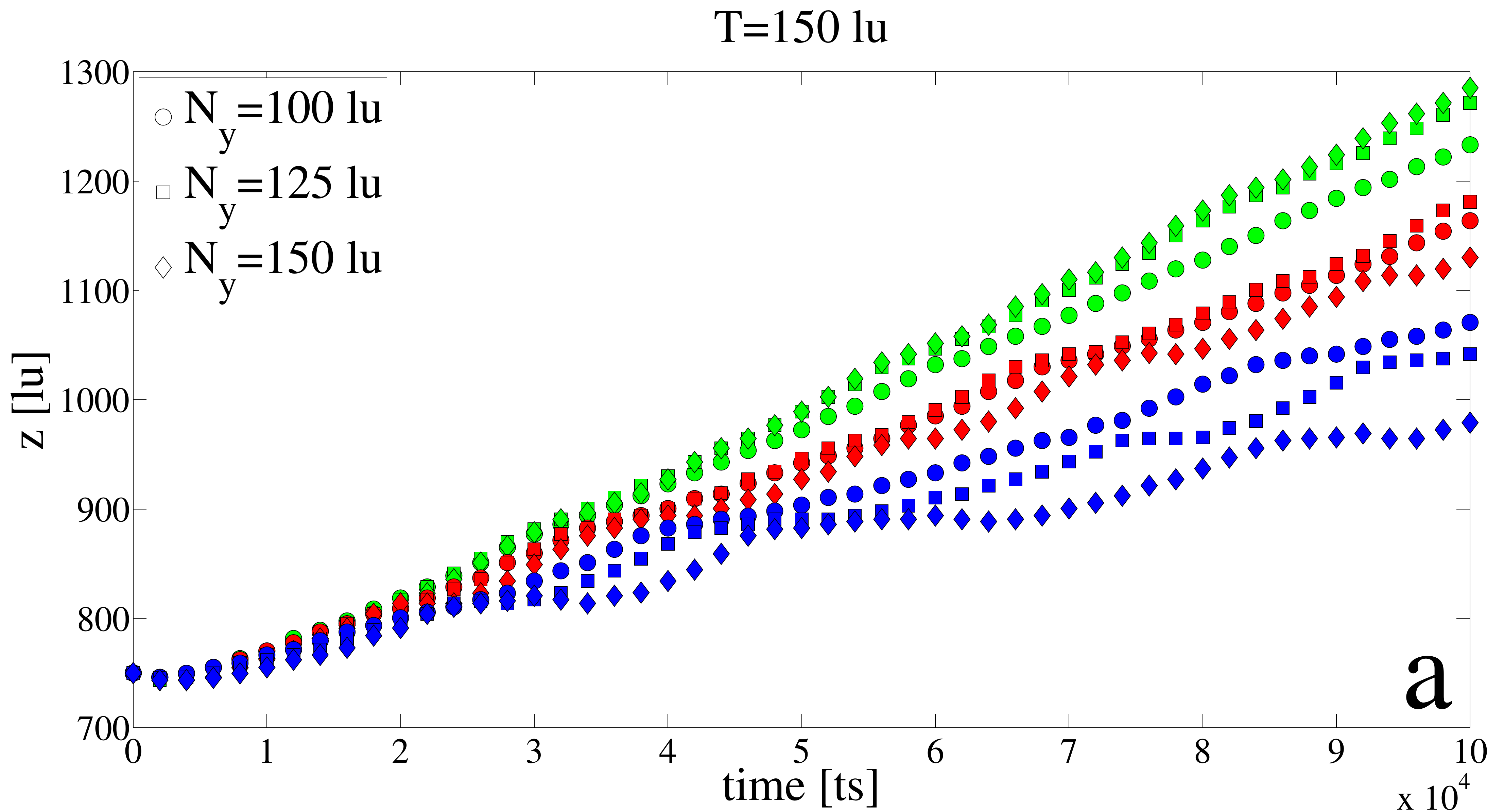}
\includegraphics[width=8.5cm]{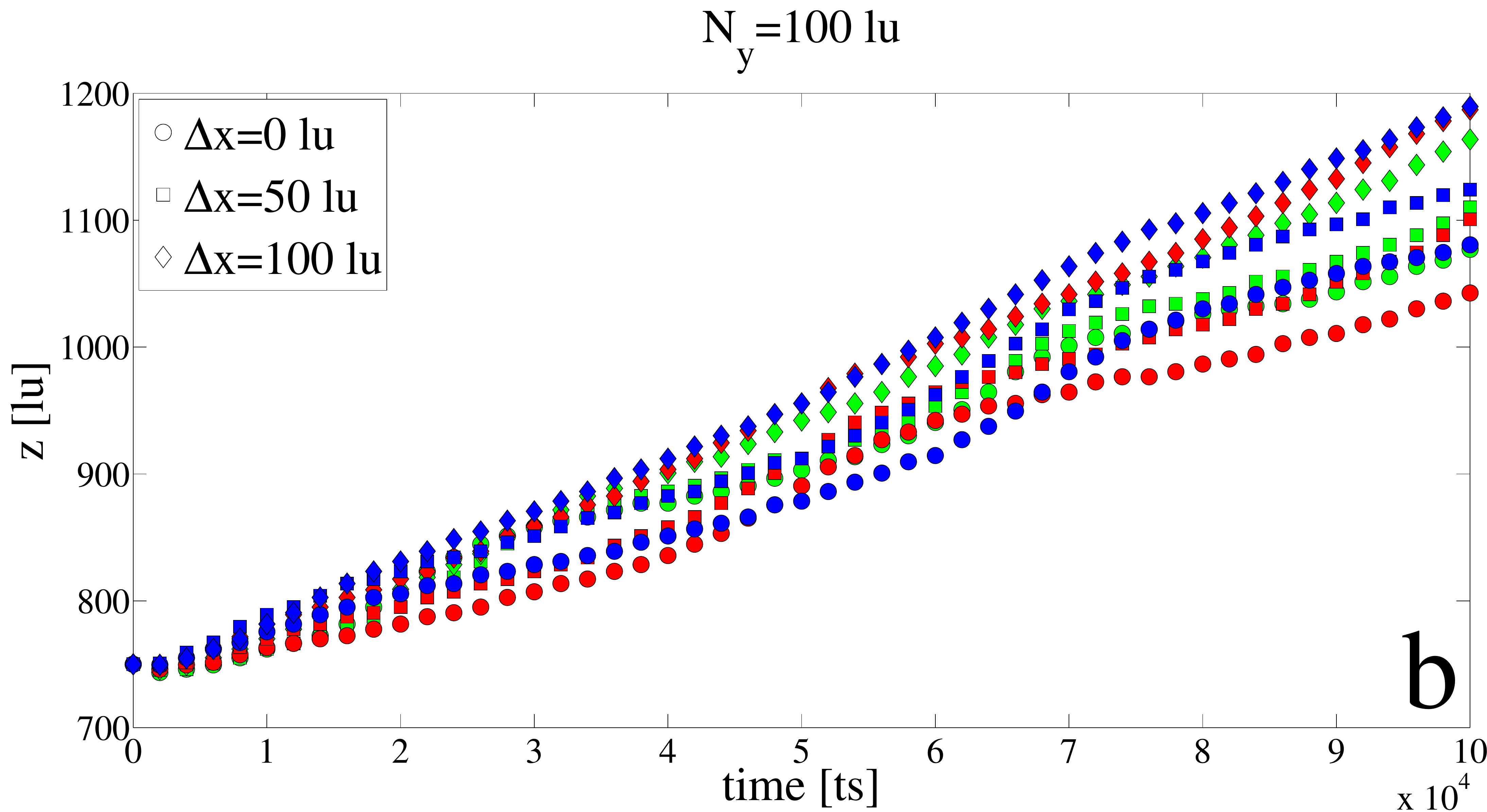}
\caption{\label{fig:zigzag2}
Motion of the invading front for zig-zag walls without reactivity. The capillaries have length $L=750$ lu. (a) The misalignment between the walls is set 
to $\Delta x=100$ lu. Color code based on the amplitude $A=i(N_{y}-25)/16$ with $i=2,3,4$ for green, red and blue. (b) The amplitude is fixed to 
$A=3(N_{y}-25)/16$. Color code based on the period $T=150,200,250$ lu for green, red and blue.}
\end{figure*}
\begin{figure*}[t]
\includegraphics[width=8.5cm]{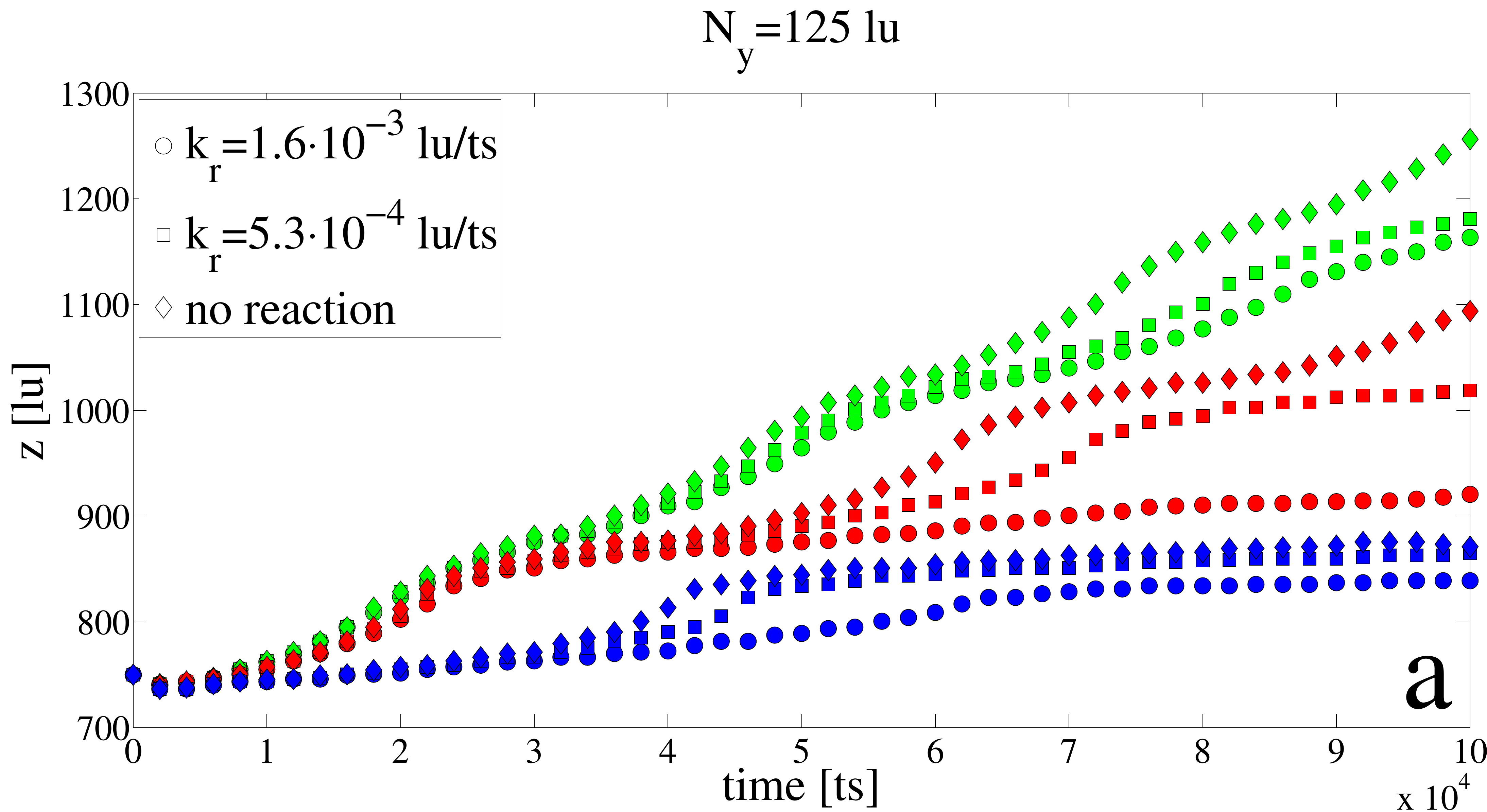}
\includegraphics[width=8.5cm]{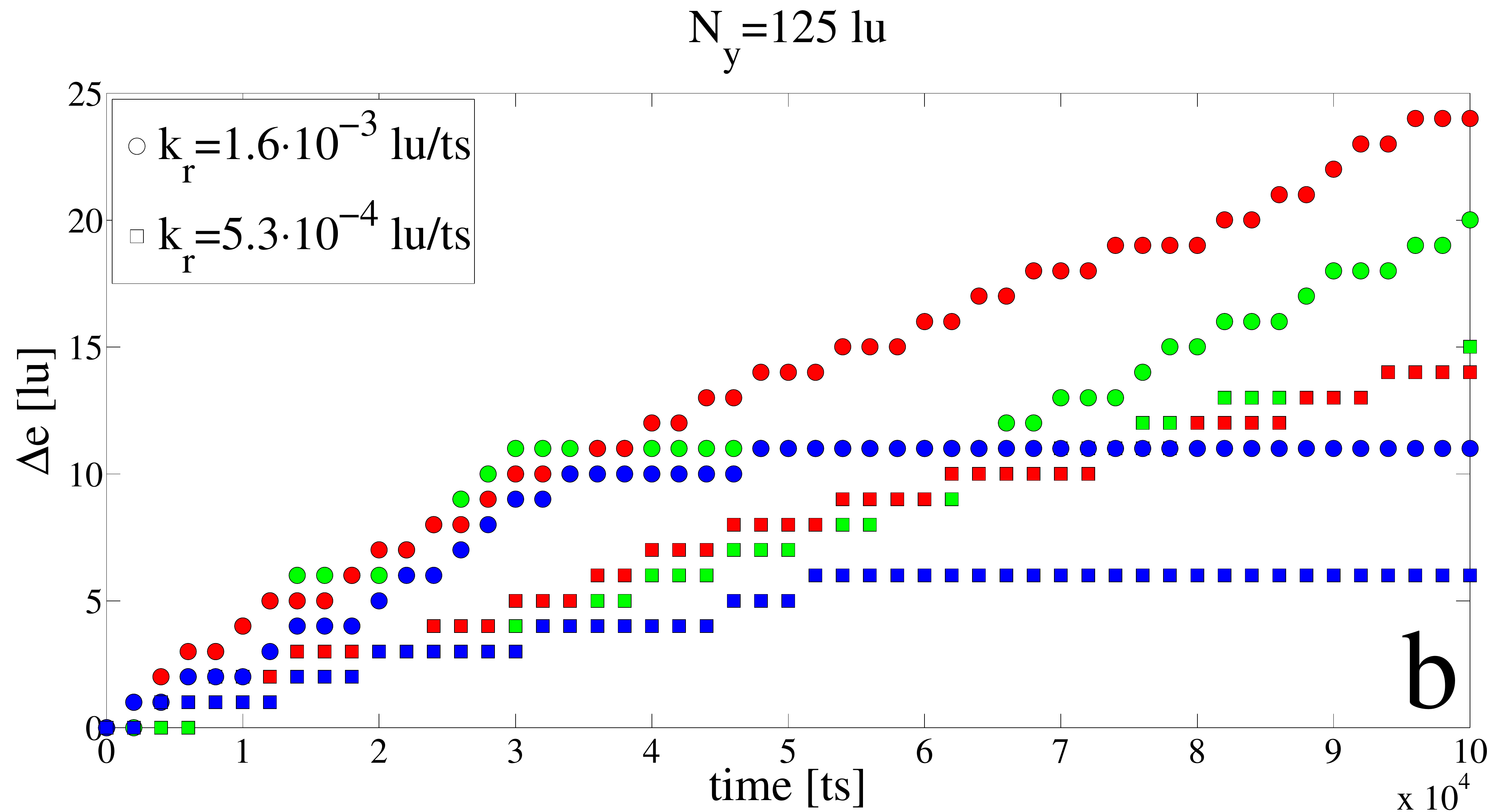}
\caption{\label{fig:zigzag3}
Capillary systems with zig-zag walls in the presence of surface reaction. The length of the capillaries is $L=750$ lu and the period is set
to $T=150$ lu. Color code based on the amplitude $A=i(N_{y}-25)/16$ with $i=2,3,4$ for green, red and blue. (a) Time dependence of the
position of the invading front. (b) Evolution of the maximal thickening of the growing surface.}
\end{figure*}
\begin{figure*}[t]
\includegraphics[width=8.5cm]{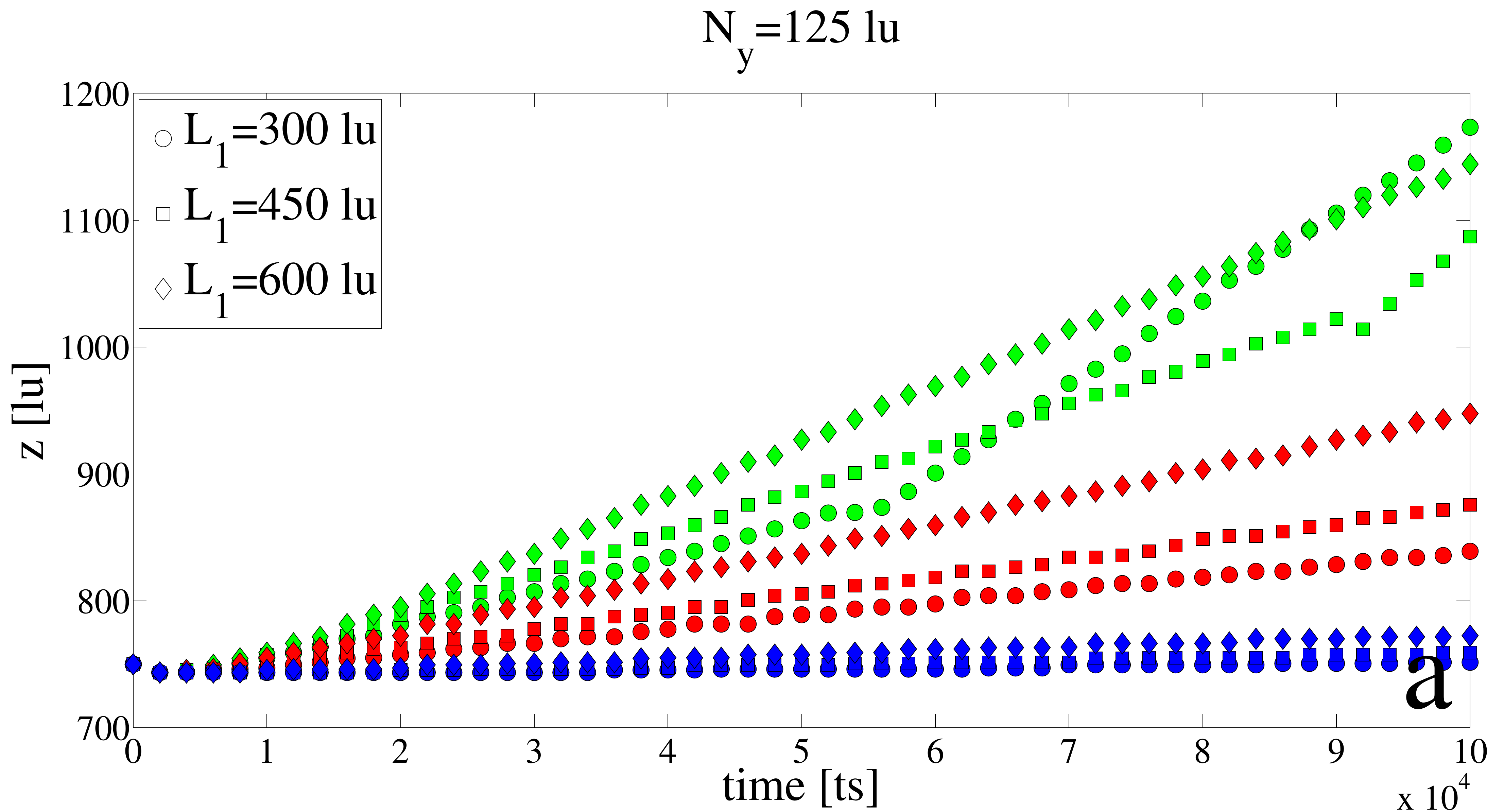}
\includegraphics[width=8.5cm]{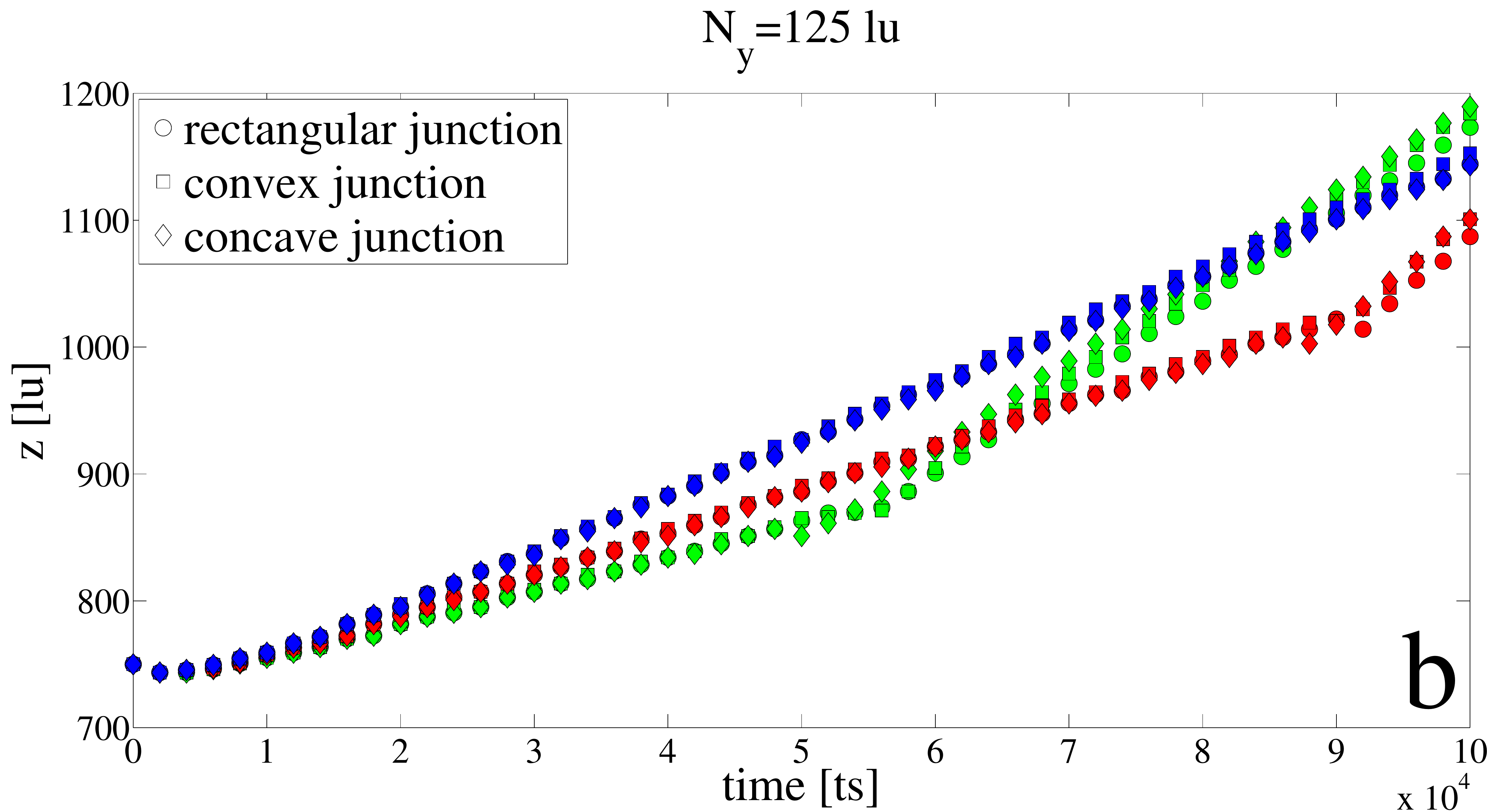}
\caption{\label{fig:constriction1}
Time dependence of the front position for constrictions without reactivity. The length of the capillary is $L=750$ lu. (a)
The junction has a rectangular shape. Color code based on the width $\Delta H=25,38,50$ lu associated with green,
red and blue, respectively. The corresponding minimum heights are $H_{\mathrm{min}}=75,50,25$ lu. (b) The width of the step is 
set to $\Delta H=25$ lu. Green, red and blue for the length of the wider segment given by $L_{1}=300,450,600$ lu.}
\end{figure*}
\begin{figure*}[t]
\includegraphics[width=8.5cm]{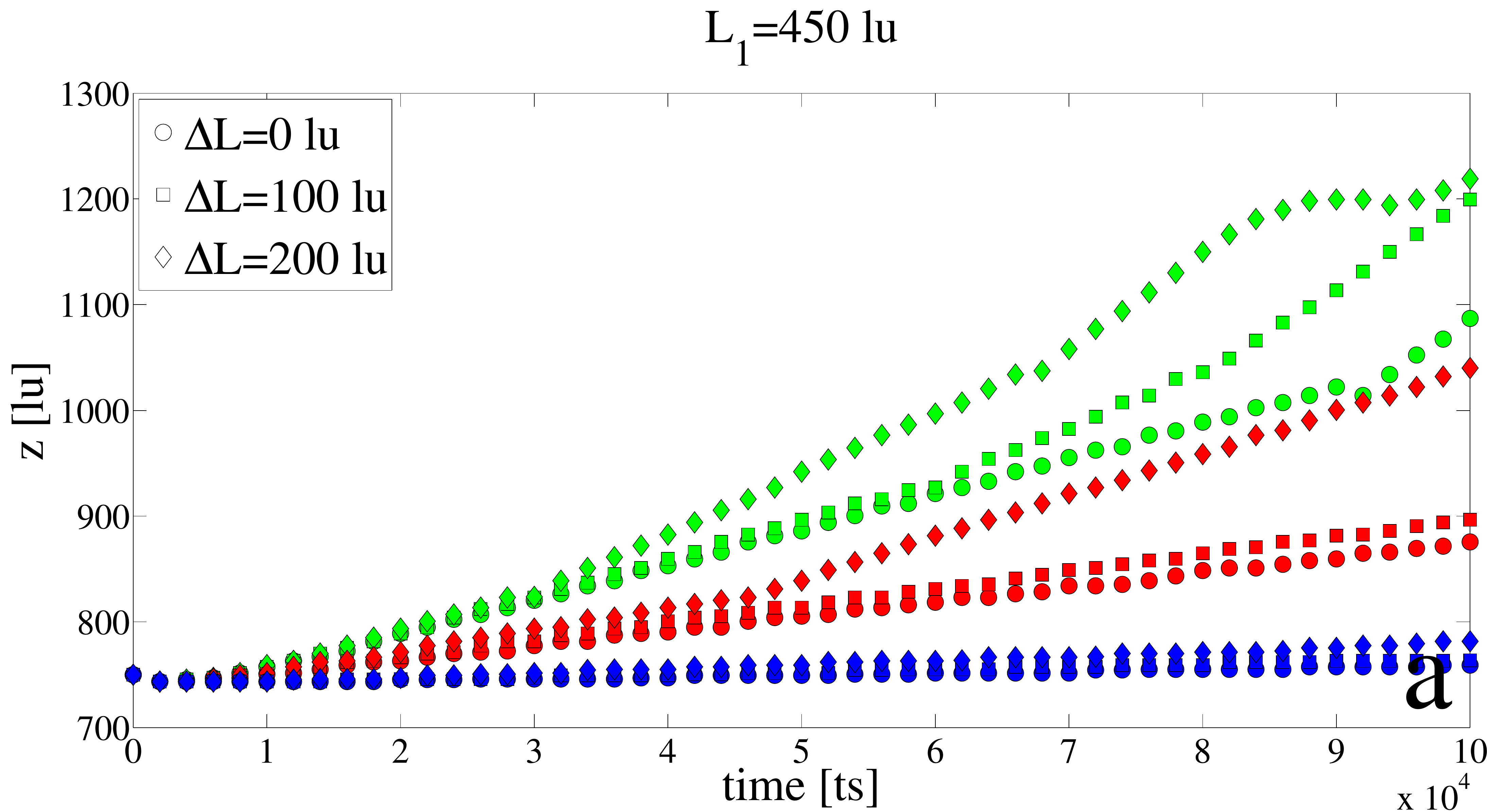}
\includegraphics[width=8.5cm]{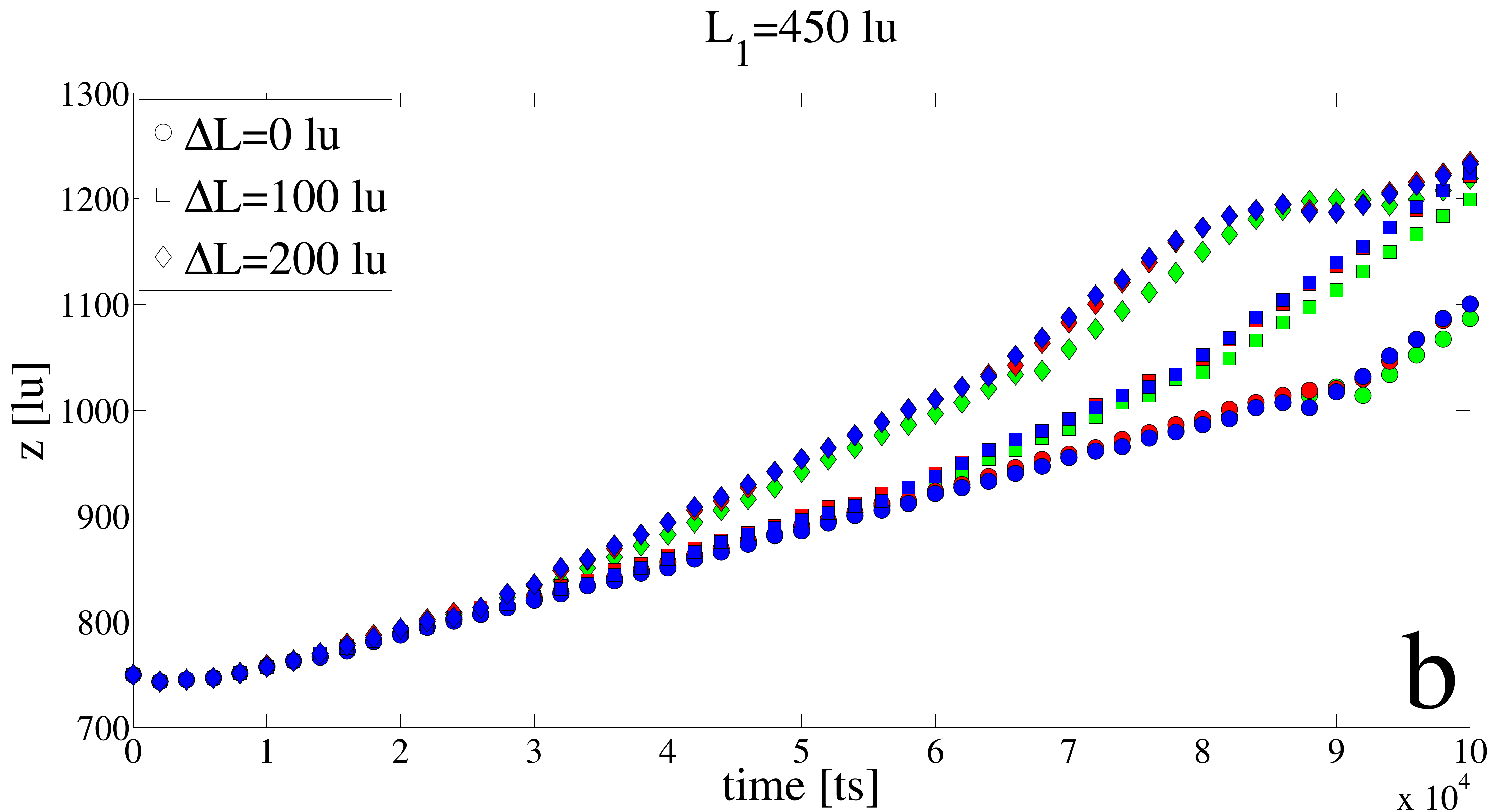}
\caption{\label{fig:constriction2}
Front dynamics for constrictions with inert solid surface. The length of the capillary is $L=750$ lu; the width of the 
simulation domain is $N_{y}=125$ lu. (a) Color code based on the width of the solid phase $\Delta H$: green, red and blue
for $\Delta H=25,38,50$ lu, leading to the minimum heights $H_{\mathrm{min}}=75,50,25$ lu. (b) The width of the solid phase
is $\Delta H=25$ lu (i.e., $H_{\mathrm{min}}=75$ lu). Color code based on the shape of the junction: green, red and blue for 
rectangular, convex and concave.}
\end{figure*}

\begin{figure*}[t]
\includegraphics[width=8.5cm]{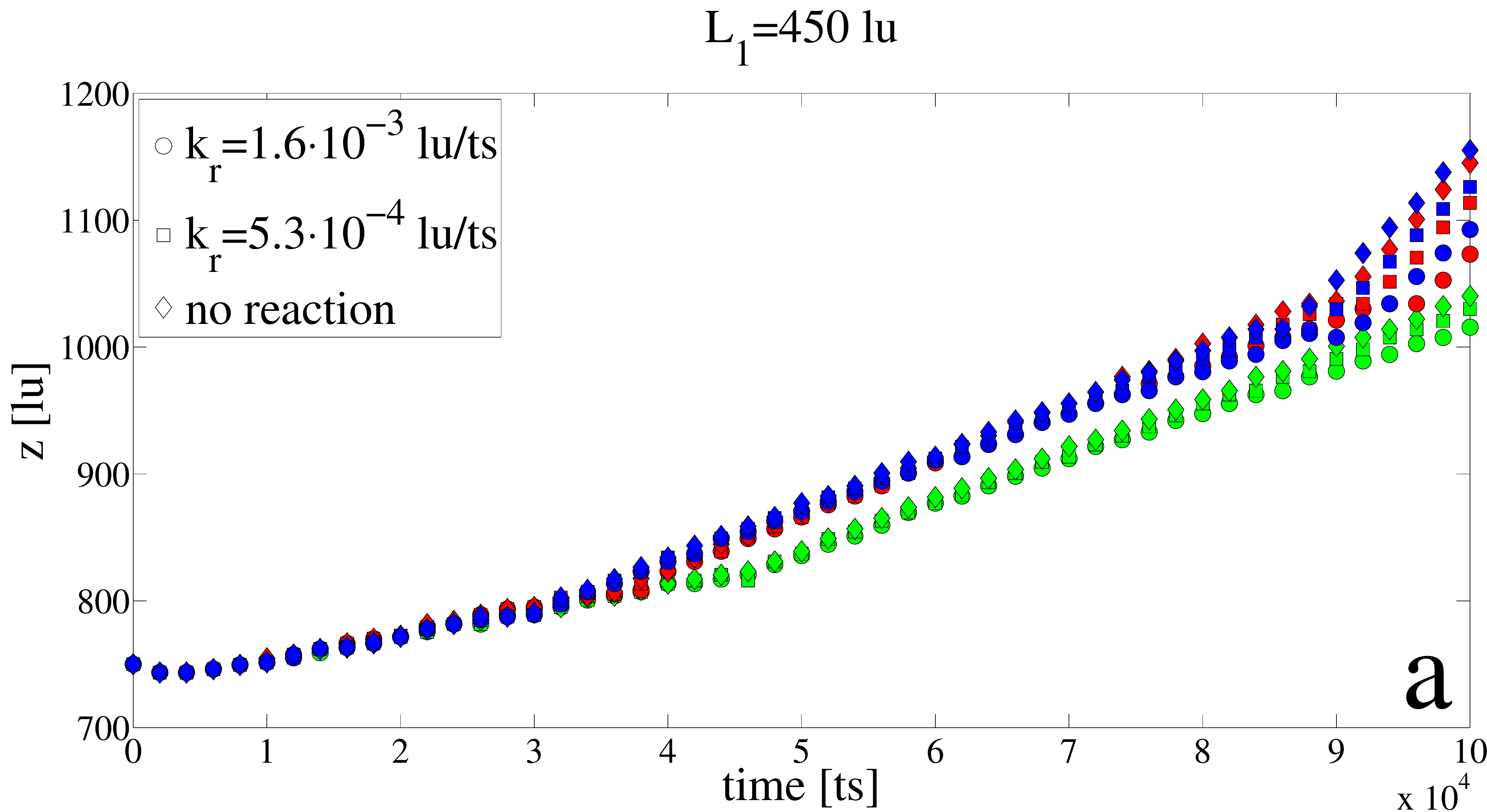}
\includegraphics[width=8.5cm]{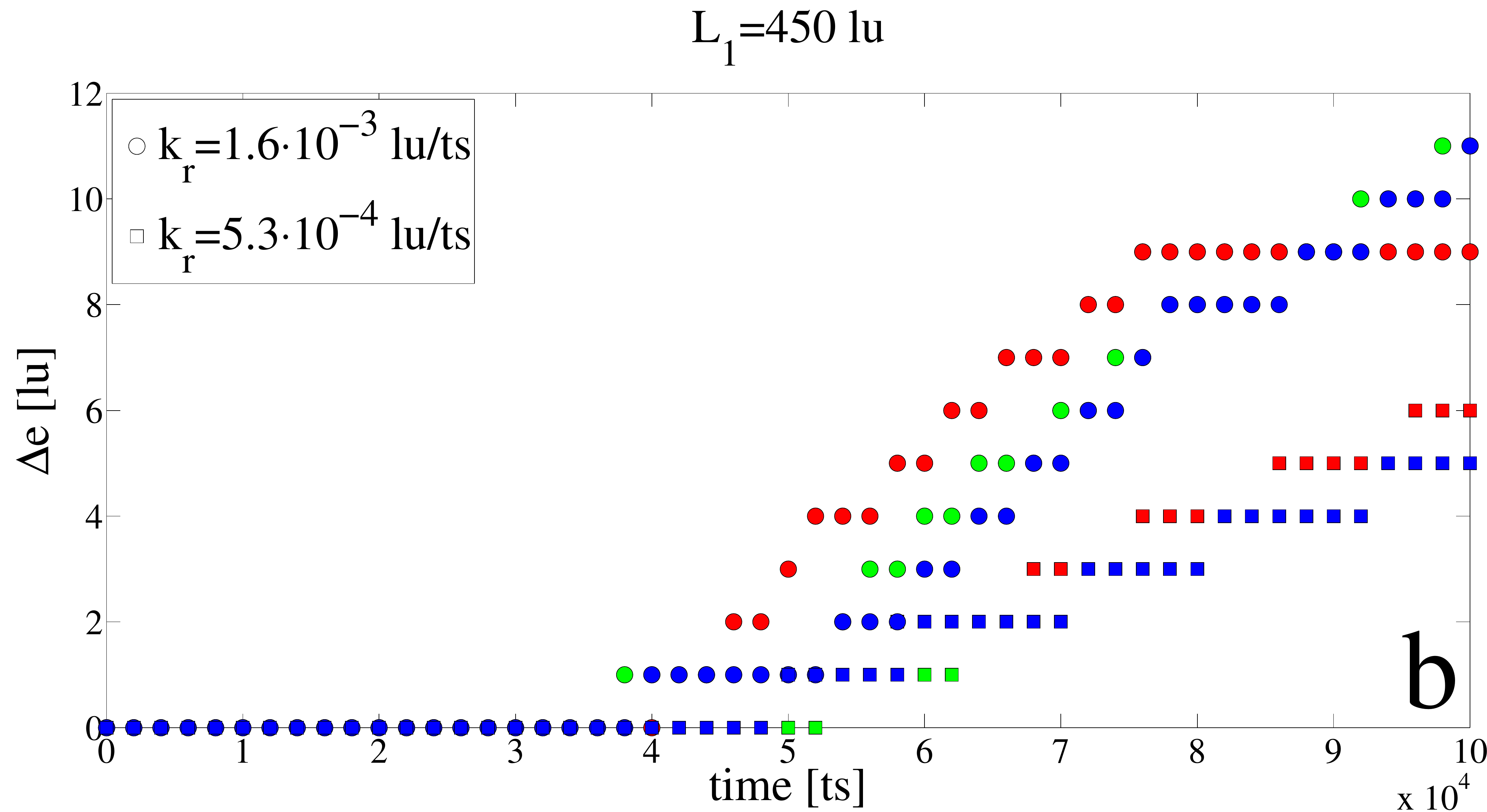}
\caption{\label{fig:constriction3}
Results for constrictions in the presence of surface reaction. The length of the capillaries is set to $L=750$ lu, the width 
of the simulation domain is $N_{y}=125$ lu and the misalignment is given by $\Delta L=200$ lu. The initial width of the solid 
surface is given by $\Delta H=38$ lu, corresponding to a minimum height of $H_{\mathrm{min}}=50$ lu. Color code based on the 
morphology of the junction: green, red and blue for rectangular, convex and concave. (a) Front position in the course of time. 
(b) Time dependence of the maximal thickening of the growing surface in the neighborhood of the junction.}
\end{figure*}
\begin{figure*}[t]
\includegraphics[width=8.5cm]{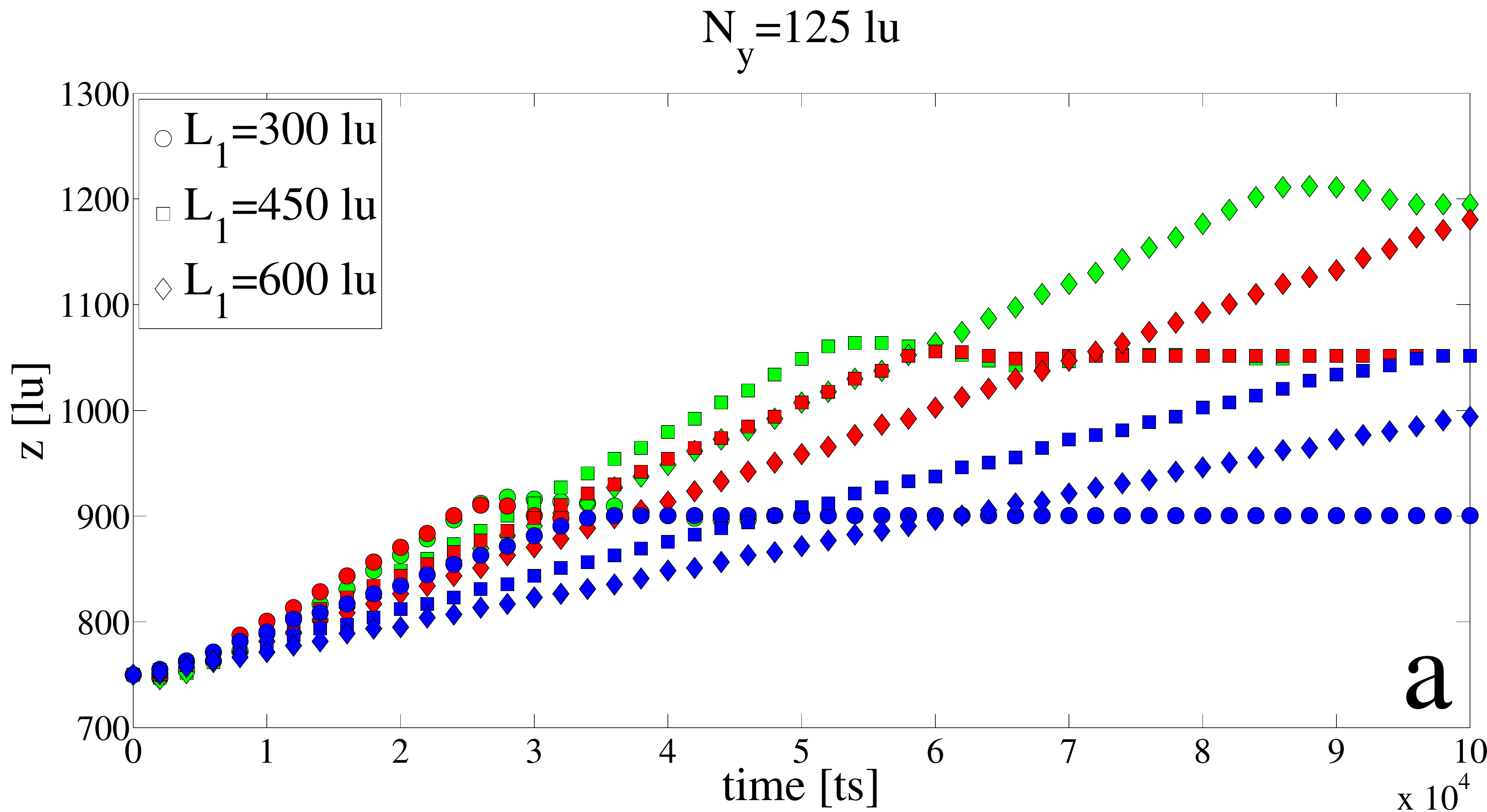}
\includegraphics[width=8.5cm]{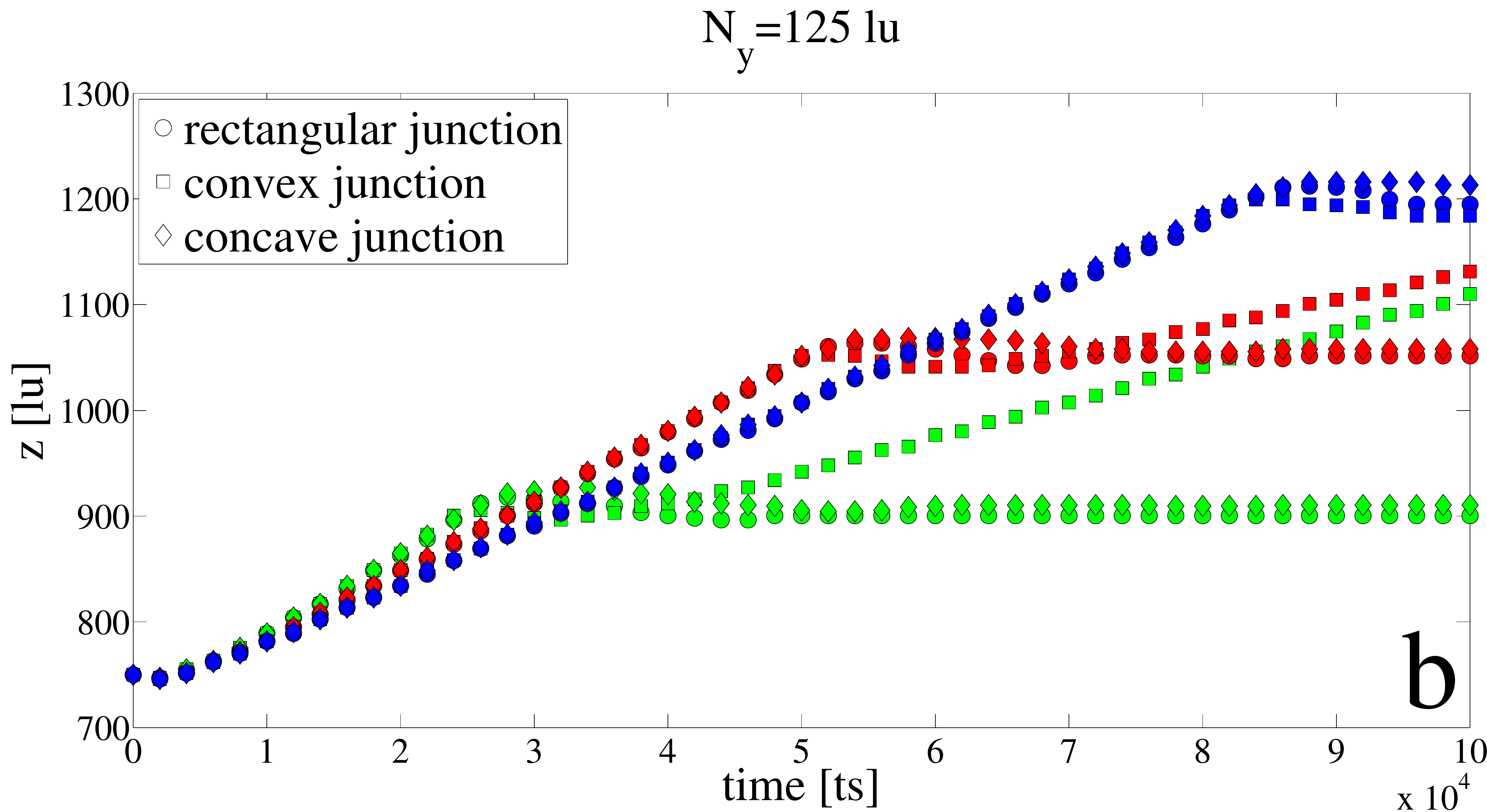}
\caption{\label{fig:expansion1}
Front position in the course of time for capillary systems with expansions. The surface reaction is disregarded. The length of the 
capillaries is $L=750$ lu. (a) The shape of the junction is rectangular. Color code based on the width of the solid phase 
$\Delta H=25,38,50$ lu for green, red and blue, respectively. These values determine the minimum heights $H_{\mathrm{min}}=75,50,25$ lu. 
(b) The width of the step is set to $\Delta H=25$ lu, i.e.~$H_{\mathrm{min}}=75$ lu. Color code based on the length of the narrower 
segment $L_{1}=300,450,600$ lu, corresponding to green, red and blue.}
\end{figure*}
\begin{figure*}[t]
\includegraphics[width=8.5cm]{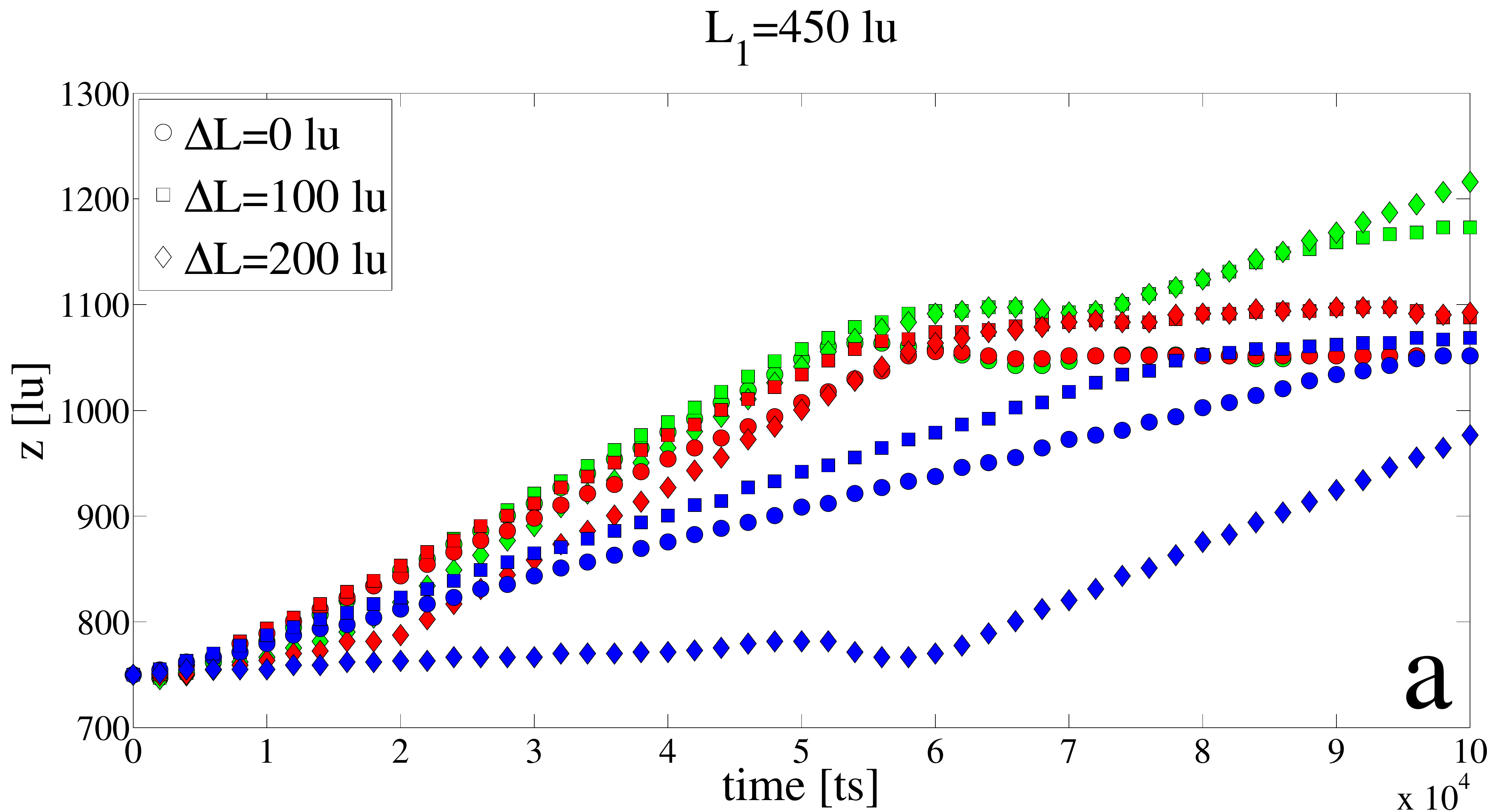}
\includegraphics[width=8.5cm]{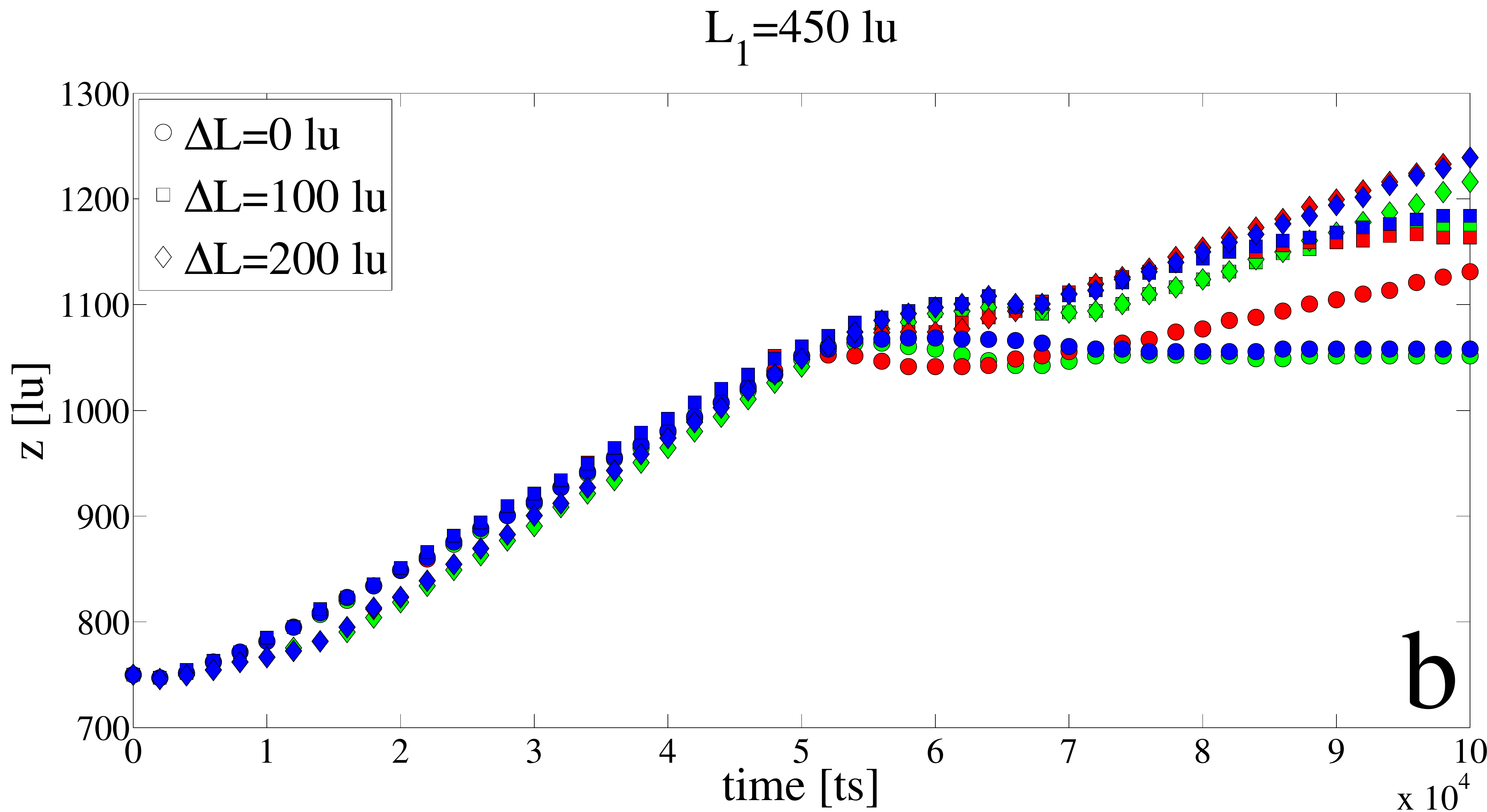}
\caption{\label{fig:expansion2}
Infiltration dynamics for capillary systems with expansions in the presence of misalignment between the walls. The surface reaction is not taken 
into account. The length of the capillaries is fixed to $L=750$ lu and the width of the simulation domain is $N_{y}=125$ lu. (a) The shape of
the junction is rectangular. Color code based on the width of the solid phase $\Delta H=25,38,50$ lu corresponding to green, red and blue. The minimum 
heights are thus $H_{\mathrm{min}}=75,50,25$ lu. (b) The width of the step is set to $\Delta H=25$, leading to $H_{\mathrm{min}}=75$ lu. Color code based 
on the shape of the junction: green, red and blue for rectangular, convex and concave, respectively.}
\end{figure*}
\begin{figure*}[t]
\includegraphics[width=8.5cm]{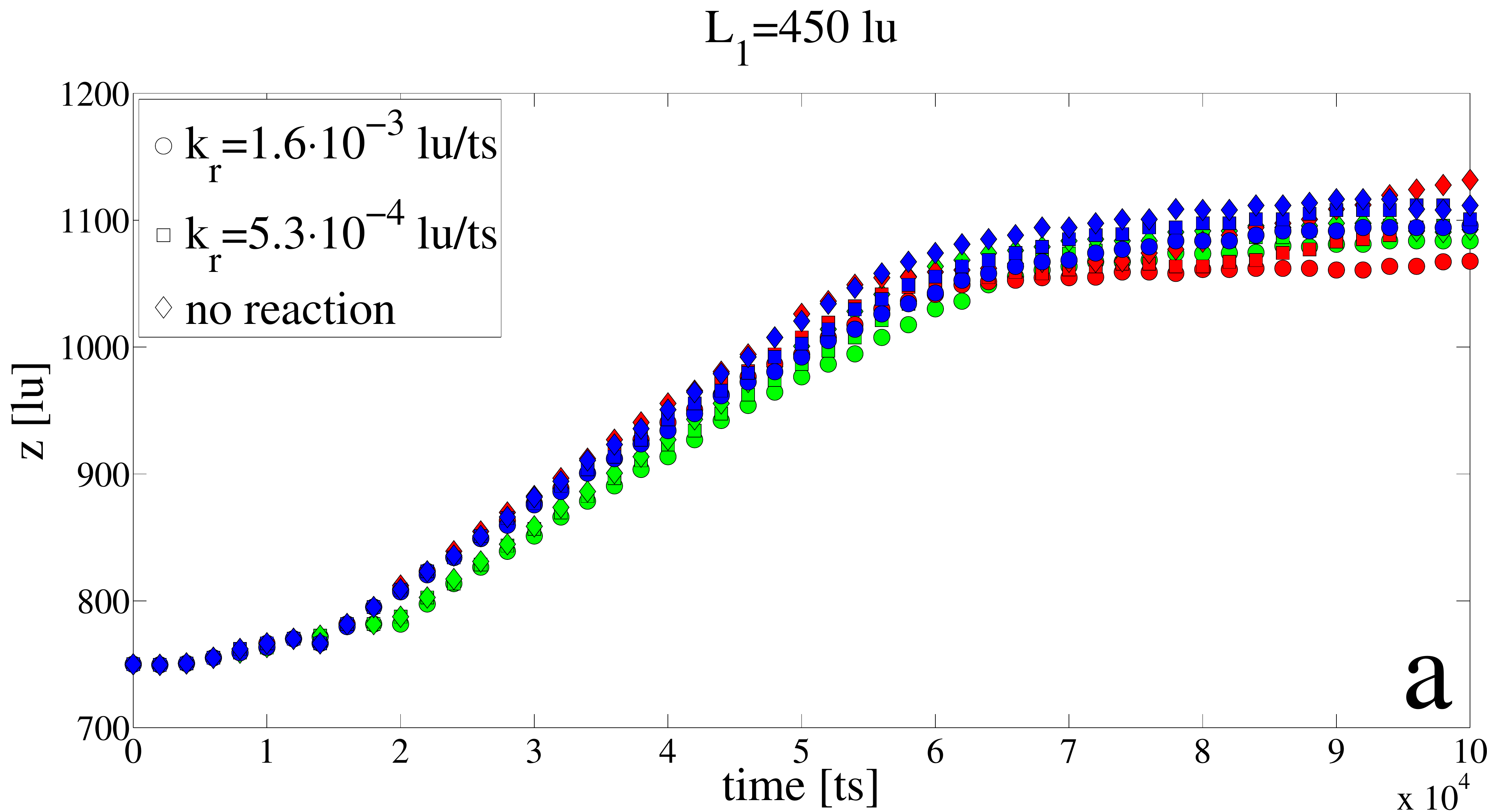}
\includegraphics[width=8.5cm]{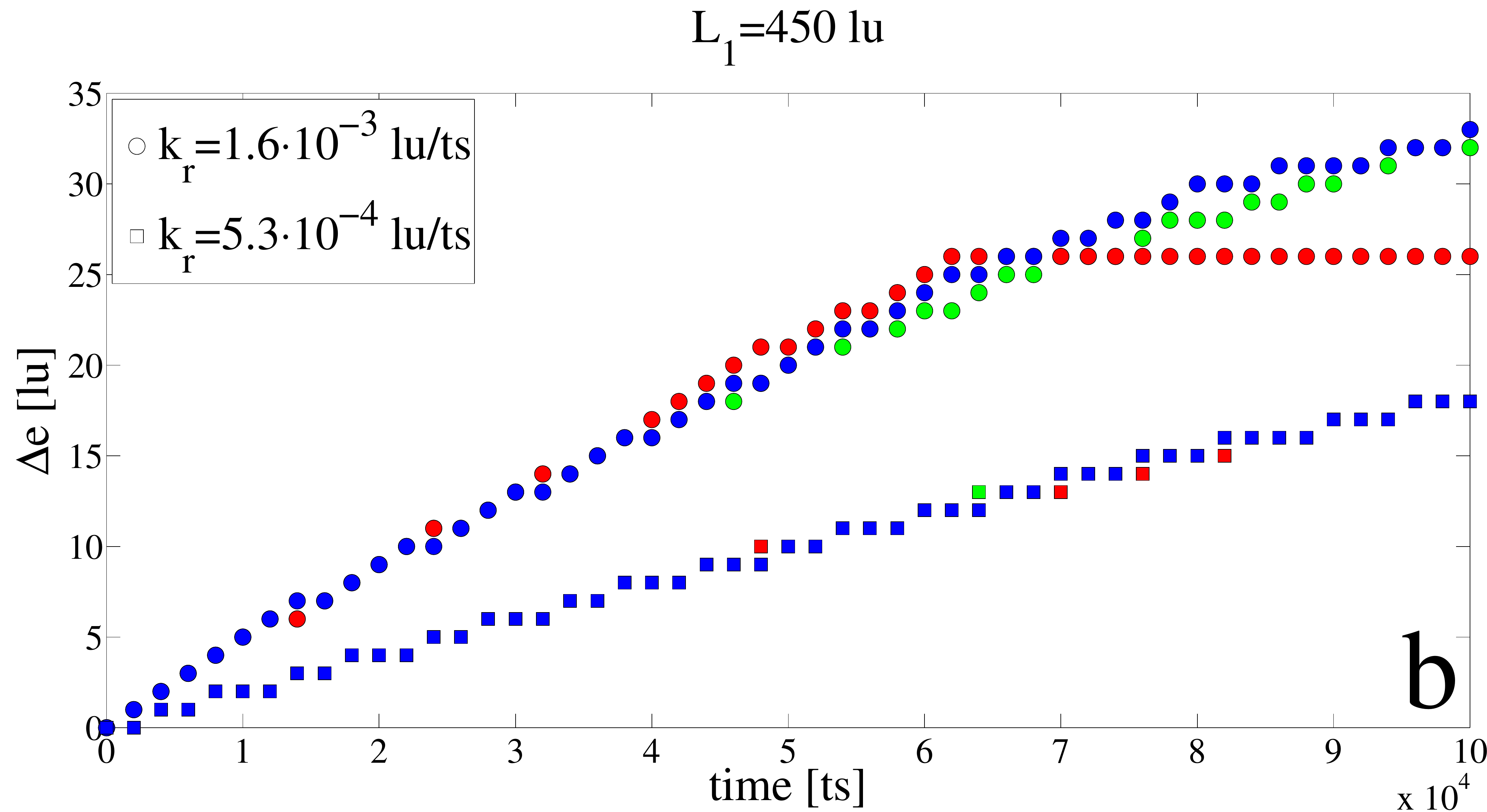}
\caption{\label{fig:expansion3}
Results for capillary systems with expansions in the presence of surface reaction. The length of the capillaries is $L=750$ lu, the width of
the simulation domain is $N_{y}=125$ lu and the misalignment between the walls is set to $\Delta L=200$ lu. The initial width of the solid
step is given by $\Delta H=38$ lu, determining the minimum height $H_{\mathrm{min}}=50$ lu. Color code based on the shape of the junction:
green, red and blue for a rectangular, convex and concave morphology, respectively. (a) Front position in the course of time. (b) Time
dependence of the maximal thickening of the growing surface.}
\end{figure*}


\section*{3.~~~CAPILLARY SYSTEMS AND SIMULATION SETTINGS}

Wetting and infiltration studies have highlighted the relevance of surface and pore microstructure (Alava et al., 2004; Grzelakowski et al., 2009; Qu\a'er\a'e, 2008).
The importance of pore characteristics has gained general acceptance also for the processing route of C/SiC composites 
(Gadow and Speicher, 2000; Israel et al., 2010; Paik et al., 2002; Salamone et al., 2008). One way to control the pore properties of the matrix is via the enrichment by 
ceramic powders. Our interest resides in the optimization of the infiltration process taking into account surface 
reactivity and subsequent thickening. We investigate basic, simplified capillary geometries that could arise from intra-particle 
porosity. In a general porous medium, the pores do not have a constant cross-section.
Furthermore, it is possible to identify larger void spaces. To start with, we thus consider capillaries
presenting the alternation between wide and narrow portions. Aligned periodic walls reproduce structures 
with such properties (Gern and Kochend\"{o}rfer, 1997; Patro et al., 2007). Our focus is on the sinusoidal profile. Other geometries are also considered, 
that is, rectangular steps and zig-zag walls. The degree of tortuosity is modified by introducing a misalignment 
between the upper and lower walls. Figure \ref{fig:profile1} illustrates some representative geometries for this type of systems.
Another complication arises from the fact that the pore size distribution is generally to some extent broad.
In order to undestand how this aspect can affect capillary impregnation, simulations are performed also for
capillaries with constrictions or expansions. The step at the junction can be rectangular, convex or concave.
Also in this case, the tortuosity of the systems is varied. Examples for this type of systems are shown in 
Fig.~\ref{fig:profile2}. Loosely speaking, a low degree of angularity is for example representative of directly synthesized powders,
while a higher degree of angularity can be observed for milled powders.

For the implementation of the various capillary systems we proceed as follows. The simulation domains are $N_{x}$
long and $N_{y}$ wide. The length of the capillaries is indicated by $L$ and their height by $H$. 
For all systems we have $L=N_{x}/2$. The left
extremity of the capillary coincides always with the point $x_{\mathrm{s}}=2N_{x}/5$. The periodic profiles are defined by the period $T$, 
the amplitude $A$ and the displacement $\Delta x$ controlling the misalignment between the walls. Since the minimum of the profiles 
corresponds to the starting point of the solid phase, the amplitude $A$ is equal to half the maximal initial width of the solid
phase (see Fig.~\ref{fig:profile1}). As an example, the lower wall of the sinusoidal capillary is described by the function 
$f(x)=A\sin(\omega \tilde{x}+3\pi/2)+A$ with $\omega=2\pi/T$ and $\tilde{x}=x-x_{\mathrm{s}}-\Delta x$. These parameters can take on 
the following values: $L=750,1'000,1'250$ lu, $N_{y}=100,125,150$ lu, $T=150,200,250$ lu, $A=(N_{y}-25)/8,3(N_{y}-25)/16,(N_{y}-25)/4$, 
$\Delta x=0,T/5,2T/5$. The choice for the amplitude $A$ guarantees that the width of the necks is at least of $25$ lu. 

For constricted channels and expansions, the overall length of the capillary is still denoted by
$L$. The length of the first segment, wider or narrower, is indicated by $L_{1}$ and that of the second one by
$L_{2}$; of course, $L=L_{1}+L_{2}$. The other parameters characterizing these systems are the width of the first segment
$H_{1}$, that of the second part $H_{2}$, the width of the step $\Delta H=(H_{1}-H_{2})/2$ (for constrictions)
and the displacement $\Delta L$ introducing a misalignment between the two walls (see Fig.~\ref{fig:profile2}). The 
convex and concave shapes for the step are defined using circles of center $(x_{\mathrm{c}},y_{\mathrm{c}})$ and radius 
$r=\Delta H$. For the lower wall, the coordinates of the center are given by $x_{\mathrm{c}}=2N_{x}/5+L_{1}\mp\Delta H/2-1$
and $y_{\mathrm{c}}=\Delta H$ (the signs refer to the convex and concave shapes respectively). The various parameters can vary as follows: 
$L=750,1'000,1'250$ lu, $H_{1}=100,125,150$ lu, $L_{1}=2L/5,3L/5,4L/5$, $\Delta H=(N_{y}-25)/4,3(N_{y}-25)/8,(N_{y}-25)/2$, 
$\Delta L=0,L_{2}/3,2L_{2}/3$ with $\Delta L<L_{1}$ (for constrictions). Furthermore, the step at the junction can have a 
rectangular, convex or concave shape (see Fig.~\ref{fig:profile2}). 

Our systems consist of binary mixtures (Chibbaro, 2008; Chibbaro et al., 2009b). In the initial condition, the first component always occupies 
half of the simulation domain. This means that the first fluid fills the capillary up to one-tenth of its length.
The density values are chosen so that $\rho_{0}=\rho_{1}+\rho_{2}=2$ mu/lu$^{2}$ and $\rho_{1}/\rho_{2}=2.5\%$, where
the first fluid is the main component; elsewhere the ratio between the densities of the two components is inversed.
Unless stated otherwise, the parameter for fluid-fluid interactions (cohesive forces) is $G_{\mathrm{c}}=0.9$ lu/mu/ts$^{2}$,
for which the surface tension turns out to be $\gamma=0.16403$ lu$\cdot$mu/ts$^{2}$ (Sergi et al., 2014). Solid-fluid interactions
are instead determined by the parameters $G_{\mathrm{ads},1}=-G_{\mathrm{ads},2}=-0.35$ lu/ts$^{2}$. In so doing, the first component
wets the solid phase while the second one is a non-wetting fluid. These settings (Sergi et al., 2014) reproduce an equilibrium 
contact angle around $30^{\circ}$ typical for droplets of molten Si on SiC substrates (Bougiouri, et al., 2006; Voytovych et al., 2008). Regarding
solute transport, we proceed as done by Sergi et al.~(2014). In the initial configuration, in the region filled with
the first fluid component, the wetting one, the solute concentration is $C_{1}=10^{-2}$ mu/lu$^{2}$; elsewhere the solute 
concentration is $C_{2}=2\cdot 10^{-3}$ mu/lu$^{2}$. The parameters for the function introducing an interface for solute are
assumed to be $G_{\mathrm{s}}=-4.875\cdot 10^{-3}$ mu/lu/ts$^{2}$, $\varphi_{0}=1$ and $C_{0}=4.9\cdot 10^{-3}$ mu/lu$^{2}$ (Sergi et al., 2014).
For the saturated concentration we choose $C_{\mathrm{s}}=5\cdot 10^{-3}$ mu/lu$^{2}$. At the beginning, the mass
deposited on solid boundaries is $b_{0}=2\cdot 10^{-3}$ mu; the threshold value determining surface growth
is $b_{\mathrm{max}}=10^{-2}$ mu. The relaxation time for solute transport is set to $\tau_{\mathrm{s}}=1$ ts, so
the diffusion coefficient $D$ is kept fixed. The reaction-rate constant is varied according to the rule 
$k_{\mathrm{r}}=(8/5)/I$ with $I=1'000,3'000$ ts. We recall that the Damkohler number is defined as $Da=k_{\mathrm{r}}N/D$
(Kang et al., 2003 and 2004; Lu et al., 2009), where $N$ is a characteristic length for the system. Last, the evolution
of every system amounts to $100'000$ ts. For the analysis, $50$ evenly-spaced frames are collected. The interfaces are
tracked as explained in the article by Sergi et al.~(2014). Here the interested reader can also find more details about
the motivations for the simulation settings.


\section*{4.~~~RESULTS AND DISCUSSION}

\subsection*{4.1~~~Periodic profiles}\label{sec:periodic}

Periodic profiles give rise to capillaries with a rich variety of morphology characteristics (see Fig.~\ref{fig:profile1}). Simply 
said, with $T$ variations the pore chambers and necks become longer, for increasing values. Increases in the amplitude $A$ lead 
to deeper pore chambers (wider stomaches) and narrower necks (closer tips). With increasing values of $\Delta x$ the pore 
chambers become less wide and the necks are less long. Moreover, the path connecting adjacent maxima and minima is shorter for 
the zig-zag and sinusoidal profiles while it is longer for the rectangular steps.

\subsubsection*{4.1.1~~~Sinusoidal profiles}\label{sec:sinus}

To start with, let us consider sinusoidal profiles in the absence of surface reaction. Figure \ref{fig:sin1} shows the behavior of 
the invading front in the course of time for varying channel width $N_{y}$, period $T$ and amplitude $A$. In general, it is possible 
to distinguish phases of pronounced acceleration followed by phases of weaker acceleration leading eventually to plateaus. This phenomenon 
is usually referred to as pinning of the interface 
(Blow et al., 2009; Chibbaro et al., 2009a and 2009c; Kusumaatmaja et al., 2008; Mognetti and Yeomans, 2009; Wiklund  and Uesaka, 2012 and 2013).
Basically, pinning occurs in the proximity of the necks, i.e.~close to narrow-to-wide structures. This phenomenon is of course more marked for
capillaries having narrower necks. The role of the period is not so clear. It seems that for narrower necks pinning occurs for the smaller period. 
Increasing periods tend mainly to weaken the capillary forces without enhancing the phenomenon of pinning (results not shown for brevity),
as we could verify with longer simulations up to $500'000$ timesteps.
It should be noted that for these systems the minimum and maximum radii are the same while the average and hydraulic radii are comparable.
The hydraulic radius is defined as $r_{\mathrm{h}}=(1/2)\mathrm{Area}/\mathrm{Perimeter}$ (Dullien, 1992).
The plots of Fig.~\ref{fig:sin1} also indicate that the various curves are associated with three distinct infiltration behaviors. It is important 
to note that the minimum radius is not the same within a group of capillaries. What seems to really matter is rather the ratio of the amplitude to 
the domain width $A/N_{y}$. It follows that to first approximation the infiltration process depends on the 
structure of the capillary. For the sake of clarity, we consider the quantity $R=<(V-<V>)^{2}>/<V>$, where $V$ is a variable and brackets 
indicate average. Let us consider the data for which $N_{y}$ varies: if $V=A/N_{y}$, it is found that $R$ is at least $1'000$ times smaller than for 
$V=r_{\mathrm{min}}$. Thus, $A/N_{y}$ characterizes better the infiltration behavior. It really has to be noted that if $T=150$ lu and $A=(N_{y}-25)/8$, 
the minimum heights are $H_{\mathrm{min}}=62,75$ and $87$ lu for $N_{y}=100,125$ and $150$ lu, respectively. For $N_{y}=150$ lu the minimum radius is 
bigger, but this is not sufficient in order to ease penetration, since the pore chamber becomes deeper (stronger drag force).

Figure \ref{fig:sin2} refers to capillary systems with varying channel width $N_{y}$ and period $T$. It clearly arises that, for a medium value
of the average minimum height $<H_{\mathrm{min}}>$, the phenomenon of pinning is accentuated with shorter periods and longer capillary lengths, but this 
does not result in a strong retardation for capillary infiltration. If we assume that the infiltration velocity is proportional to $1/L$, as for smooth 
capillaries (Chibbaro et al., 2009b), for the infiltration velocity we can write $v(L+\Delta L)/v(L)\approx (1-\Delta L/L)$. In passing from 
$L=750$ lu to $L=1'000$ lu, this ratio becomes $v(L+\Delta L)/v(L)=0.67$. From Fig.~\ref{fig:sin2} there appears that the invading fronts reach
almost the same depth. Precisely, on average we find that $v(L+\Delta L)/v(L)=0.82$; the infiltration velocities are determined by
applying the method of least squares. If we restrict our attention to the data for $T=200$ lu, it even turns out that $v(L+\Delta L)/v(L)=0.96$.
In this case, if we overcome the obstacle of estimating the velocity by considering $z$ as a function of $L$ for the time fixed at $t=100'000$ lu,
this phenomenon is clearer. It is found that $z(L+\Delta L)/z(L)=0.91$ for $L=750$ lu and $z(L+\Delta L)/z(L)=0.88$ for $L=1'000$ lu.
As a result, in the case of sinusoidal profiles, the length of the capillaries affects the infiltration process to a weaker extent. The discussion
on the results of Fig.~\ref{fig:sin1} is valid also for longer capillaries. Unless stated otherwise, in the sequel the conclusions can be
assumed to hold for all capillary lengths.

Shown in Fig.~\ref{fig:sin3} are the results for capillary systems with misalignment between the upper and lower walls (see Fig.~\ref{fig:profile1}).
The infiltration process is faster for increasing misalignment of the walls. In that respect, it is important to note that
for a given value of $\Delta x$ the minimum and maximum radii vary with $N_{y}$ within a group of systems displaying similar behavior.
In general, there appears that the effect of pinning is weaker for increasing misalignment of the walls. Furthermore, for higher misalignment, it 
is found that pinning is clearly stronger in the case of narrower necks and small periods (results not shown for brevity). We also consider the tortuosity 
(Duda et al., 2011; Matyka and Koza, 2012). The tortuosity is a measure of the departure of fluid flow from straight pathways. 
The calculations are made by means of the formula $\lambda=<u>/<u_{x}>$ (Duda et al., 2011; Matyka and Koza, 2012), $\bm{u}$ being the overall fluid velocity 
(Kang et al., 2002a; Shan and Chen, 1993; Shan and Doolen, 1995). It should be kept in mind that at the interface there are spurious currents (Wagner, 2003).
The results shown in Fig.~\ref{fig:sin3} indicate that the tortuosity is smaller for higher misalignment of the walls. This means that it is more important
to have wider pathways rather than only narrow, straight ones. The curves are grouped approximately according to the misalignment of the walls. 

Figure \ref{fig:sin4} has to do with sinusoidal capillaries in the presence of reaction. Without reaction, the increase in period $T$ results
in a slightly lower penetration depth for $H_{\mathrm{min}}=75$ and $50$ lu. When the reactivity is enabled, no drastic difference for the penetration
depth can be observed. In the absence of reaction, in passing from $T=150$ lu to $T=200$ lu, flow slows down significantly for $H_{\mathrm{min}}=25$ lu
(cf.~discussion on Fig.~\ref{fig:sin1}). Again, this points out that the structure of the capillary is more important than the minimum radius
(cf.~comments on Fig.~\ref{fig:sin1}). With surface reaction, the results for the maximal thickening $\Delta e$ of the solid surface (Sergi et al., 2014) 
indicates that the process of surface growth is not strongly affected by fluid behavior. The morphology presented in Fig.~\ref{fig:system1} supports
this thesis. This consequence is quite expected since we know that the infiltration 
velocity was proven to have a marginal role on the kinetics of pore closure (Sergi et al., 2014). There appears an exception for the system with 
$H_{\mathrm{min}}=50$ lu and $k_{\mathrm{r}}=1.6\cdot10^{-3}$ lu/ts. Visual inspection of the dynamics reveals that this difference is chiefly due to an 
asymmetry in the thickening of the surface near the neck. For other $N_{y}$ values the curves for $\Delta e$ have a similar behavior.

In Fig.~\ref{fig:sin5} we analyze the effects of pinning for surface growth. When the period $T$ varies, it is possible to distinguish different 
infiltration behaviors due to pinning. What is important to remark is that the maximal width of the growing surface $\Delta e$ is not affected by the 
hydrodynamic behavior. The evolution of $\Delta e$ depends essentially on the reaction-rate constant $k_{\mathrm{r}}$. It is observed that the infiltration 
process tends to be retarded to a larger extent by surface growth with smaller periods as the plateaus become longer (i.e., pinning more marked). Concerning 
the results for different tortuosities, from Fig.~\ref{fig:sin5} it can be seen that the various curves can be considered as grouped according to the degree 
of misalignment. On the other hand, the quantity $\Delta e$ depends mainly on the reaction-rate constant $k_{\mathrm{r}}$. These findings give further evidence 
for the prominence of the capillary structure, while the hydrodynamic behavior is of secondary importance for the occurrence of pore closure. The 
infiltration process is slowed down by surface growth more significantly for aligned walls.

We recall that for a uniform capillary, in the absence of surface reaction, the process of capillary infiltration is described by the 
following equation (Chibbaro et al., 2009b): 
\begin{equation}
z(t)=\frac{V_{\mathrm{cap}}H\cos\theta}{6L}t_{\mathrm{d}}[\exp(-t/t_{\mathrm{d}})+t/t_{\mathrm{d}}-1]+z_{0}\ ,
\label{eq:z}
\end{equation}
where $V_{\mathrm{cap}}=\gamma/\mu$ and $t_{\mathrm{d}}=H^{2}/12\mu$. By $z$ we designate the centerline position of the invading front, $\gamma$ is the surface 
tension, $\mu$ the dynamic viscosity, $\rho$ the density, $H$ the capillary height, $L$ the length, $\theta$ the contact angle. 
The quantity $z$ has a linear dependence with time. We want to use this formula in order to extract approximations for the effective radius $r_{\mathrm{eff}}$.
The well-known result for the capillary pressure $\Delta P=2\gamma\cos\theta/H$ allows to fit the simulation data to a straight line with
a slope given by $H^{2}\Delta P/12\mu L$. By using for $\Delta P$ a suitable average, an estimate for the effective radius can be obtained. From
this result, we then calculate the dynamic contact angle using the capillary pressure (Joos et al., 1990; Van Remoortere and Joos, 1993). Finally, by
taking an average for the contact angle it is possible to employ Eq.~\ref{eq:z}, as shown in Fig.~\ref{fig:sin6}. Deviations from the behavior
predicted by Eq.~\ref{eq:z} are more marked when the phenomenon of pinning occurs. The reactivity turns out to enhance the effect of pinning.
At the beginning of every plateau the dynamic contact angle $\theta$ reaches the local minimum value. When the infiltration accelerates, $\theta$
increases sharply. When pinning is more significant, $\theta$ remains quite large and then decreases more slowly. This phenomenon is clearer 
from the results with reaction, for which pinning is stronger. It is found that the effective radius is better approximated by the average
radius, with the exception when pinning becomes more significant. In this case $r_{\mathrm{min}}$ is closer to $r_{\mathrm{eff}}$. It is
important to recall that pinning depends mainly on the structure: $r_{\mathrm{min}}$ is a parameter among others defining it, as seen before
for the discussion related to Fig.~\ref{fig:sin1}.


\subsubsection*{4.1.2~~~Step-shaped profiles}\label{sec:step}

For periodic step-shaped walls, we limit ourselves to capillaries of length $L=750$ lu. From Fig.~\ref{fig:step1} it emerges that the fluid advances no more when the 
meniscus reaches the zones of enlargement at $x=865$ lu. This occurs because at the corners the contact line remains pinned. This behavior was ascertained by repeating 
the simulations for $N_{y}=100,125$ lu with $500'000$ timesteps. The same conclusions hold also for longer periods. As seen before, an increase in capillary
forces could be obtained by considering smaller ratios $A/N_{y}$ (Mognetti and Yeomans, 2009). A detailed analysis of the conditions for depinning is outside the scope
of the present study: the interested reader is addressed to the existing literature
(Blow et al., 2009; Chibbaro et al., 2009a and 2009c; Kusumaatmaja et al., 2008; Mognetti and Yeomans, 2009).

Figure \ref{fig:step2} presents the results in the case of misalignment between the upper and lower walls. The effect of misalignment is that the pinning barrier can
be overcome and the meniscus does not remain stuck. The curves are grouped as for sinusoidal profiles (see Fig.~\ref{fig:sin1}). Different degrees of misalignment do not 
separate the infiltration behaviors neatly. Moreover, the dependence on the period is not so marked. The infiltration is slightly faster for $T=250$ lu, the largest period. The 
results for the other capillary systems lead to the same conclusions. Of course, the infiltration process is slower for step-shaped walls than for sinusoidal profiles. For the 
considered number of timesteps, the difference amounts to about $100$ lu. Furthermore, the systems with period $T=150$ lu have the same minimum radius of the sinusoidal 
capillaries of Fig.~\ref{fig:sin6}. It thus follows that the presence of corners quite significantly lowers the effective radius.

The results with reactivity are also in line with previous findings. A typical morphology is shown in Fig.~\ref{fig:system1}. Again, there appears clearly that the quantity 
$\Delta e$, controlling pore closure, does not depend on the flow velocity, as it comes out from Fig.~\ref{fig:step3}. The surface growth opposite to fluid flow seems to 
have a marginal influence on the infiltration process. The remarks presented in the previous paragraph are material to this statement. The curves of Fig.~\ref{fig:step3} 
give further support to this thesis since their 
behavior is related. Similar results are obtained for the other capillary systems. From a comparison with the equivalent sinusoidal capillary systems it turns out 
that the process of pore closure occurs sooner with step-shaped walls. Corners are thus more detrimental. Interestingly, it follows that the phenomenon of pore 
closure is actually sensitive to the wall structure. 


\subsubsection*{4.1.3~~~Zig-zag profiles}\label{sec:zigzag}

Also for zig-zag profiles we restrict the simulations to the case of capillaries of length $750$ lu. In general, the effect of pinning appears to be weaker
for this geometry. As it can be seen in Fig.~\ref{fig:zigzag1}, the dependence of $z$ on time is almost linear, with the exception for narrower necks. The
curves are grouped as for sinusoidal profiles. But the infiltration process turns out to be faster for zig-zag profiles. The period $T$ does not seem to
have a prominent role. For narrower necks, the same conclusions reached for sinusoidal profiles hold (see Fig.~\ref{fig:sin1}), as it was verified up to
$500'000$ timesteps also in this case. Namely, the infiltration depth is almost the same but for larger periods there are less evident plateaus.

The misalignment between the walls has the effect to speed up capillary infiltration especially for narrower necks, as shown in Fig.~\ref{fig:zigzag2}.
For wider necks, the advantage of misalignment is comparable to that for the other geometries, that is within $150$ lu (see Fig.~\ref{fig:sin3}). It is 
found that the dynamics is slightly faster for larger periods (not shown for brevity); the corresponding characteristic radii (e.g., minimum and 
hydraulic) are almost the same. The other capillary systems lead to the same conclusions.

From Fig.~\ref{fig:zigzag3} it can be seen that, for the same reaction conditions, $\Delta e$ does not display approximately the same behavior until
pore closure occurs, as for the other profiles (see Figs.~\ref{fig:sin4}, \ref{fig:sin5} and \ref{fig:step3}). The difference can be ascribed to the
fact that the surface does not grow principally near the maximum, i.e.~the peak (see Fig.~\ref{fig:system1}). By varying the width of the channels, this 
phenomenon takes place for other combinations of the other parameters. It is thus difficult to identify the causes with precision. In any case, we can conclude 
that the process of pore closure results to be slower for zig-zag profiles and this is due to their structural properties (sharp peaks). 

Since the infiltration process appears to be slightly faster for zig-zag profiles than for sinusoidal ones, it turns out that the effective radius for
this geometry is higher (see Fig.~\ref{fig:sin6}), as we could verify. Presumably, the reason resides
in the fact that the role of pinning is less significant. However, in general this phenomenon is not captured exhaustively by characteristic radii. 
Although all periodic profiles have the same minimum radius, in the absence of misalignment, for the step-shaped structure the flow can not proceed.
Of course, the tortuosity is expected to be smaller for zig-zag profiles. As an example, for the equivalent systems of Fig.~\ref{fig:sin3}, it is
found that the first maxima are slightly lower. For completeness, for the equivalent systems with step-shaped structures, the first maxima are around $3$
for the highest misalignment of the walls.


\subsection*{4.2~~~Two-pore systems}\label{sec:dual}
                           
Constrictions and expansions are characterized by constant radii (see Fig.~\ref{fig:profile2}). In our systems, $L_{1}$ 
determines the relative length of the two segments. $\Delta H$ controls the width of the narrower segment. As $\Delta L$ 
increases, the necks become shorter while the main cavity becomes less large. For the rectangular step, at the junction there 
are two corners. For the convex and concave shapes, there is only one corner, but the profile of the junction is longer.

\subsubsection*{4.2.1~~~Constrictions}\label{sec:shrink}

Also for constrictions the length of the capillary is fixed to $L=750$ lu. Furthermore, we consider for the width of the 
simulation domain only the case $N_{y}=125$ lu. From Fig.~\ref{fig:constriction1} it is interesting to see that for the
channels with higher minimum radius, when the capillary shrinks, the flow accelerates. Otherwise, the infiltration is clearly 
faster for the capillaries with longer segments of large width. The shape of the junction has no noticeable effect. The same 
conclusions can be drawn from the other capillary systems. Comparison with the periodic profiles indicates that the infiltration 
process is slower (see Figs.~\ref{fig:sin1} and \ref{fig:zigzag1}), despite the absence of pinning.

Figure \ref{fig:constriction2} shows the results in the presence of misalignment between the upper and lower walls.
Misalignment can speed up the infiltration process. As the minimum radius increases, that is, the step $\Delta H$ decreases,
the effect of misalignment becomes more significant. For smaller segments of the wider part, it is found that misalignment
is associated with a weaker resistance to flow (results not shown for brevity). For higher misalignment, when the channels 
widen, the front remains pinned, creating a plateau. Again, the shape of the junction has no appreciable consequence. We start
seeing a clear separation of the curves for smaller minimum radii and higher misalignment: the rectangular shape leads
to a slower dynamics. For increasing $L_{1}$, it also turns out that the convex shape is associated with a faster infiltration
process (not shown for brevity). For periodic profiles, the misalignment of the walls leads to a slightly faster 
dynamics (see Figs.~\ref{fig:sin3} and \ref{fig:zigzag2}).

For these capillary systems the reaction has a less marked effect. In Fig.~\ref{fig:constriction3} it is instructive to notice
that the process of surface growth starts becoming detrimental especially when the front reaches the minimum radius.
This observation holds also for the other structures. As a result, the reaction has a higher impact for smaller segments
with large radius. The analysis proposed for sinusoidal profiles in order to estimate the effective radius was also
performed for the systems of Figs.~\ref{fig:constriction2} and \ref{fig:constriction3}. It arises that the average radius
characterizes the infiltration process better for large minimum radii and high misalignment. Departure from these
conditions leads to effective radii closer to the minimum one. For the equivalent sinusoidal profiles the dynamics is
faster. Their effective radius is higher. Also the results for the tortuosity indicates that the resistance 
to flow tends to be stronger for constrictions (results not shown for brevity). 


\subsubsection*{4.2.2~~~Expansions}\label{sec:expand}

Also in this case we restrict ourselves to lengths of $750$ lu and widths of $N_{y}=125$ lu for the channels. The results for 
the channels presenting expansions are very interesting for the problem of pinning because it might be presumed that the drag force 
is weaker while the capillary forces increase because of longer flat walls. From Fig.~\ref{fig:expansion1} it can be seen that, 
for the rectangular shape of the junction, the pinning barrier can not be overcome in any case. For the largest minimum height 
$H_{\mathrm{min}}=75$ lu, the front can proceed for the convex geometry. This is easier for the smaller value of $L_{1}$.  
It might appear surprising that the concave geometry can be less favorable. We point out that, 
for the convex shape, the front has higher momentum when it encounters the sharp corner. For the data of Fig.~\ref{fig:expansion1}, 
with $L_{1}=450$ lu, the average over the last three frames before a negative momentum is recorded yields $724$ mu$\cdot$lu/ts for the 
convex shape and $333$ mu$\cdot$lu/ts for the concave shape. Furthermore, because of the convex geometry, the contact line can slide along 
the border of the junction and reach again the uniform part of the channel. The penetration for the convex geometry is slightly slower
than for constrictions (see Fig.~\ref{fig:constriction1}). For the sake of completeness, simulations up to $500'000$ ts proves
that the pinning barrier is overcome only for the convex geometry with $L_{1}=300,450$ lu and $\Delta H=25$ lu, corresponding
to $H_{\mathrm{min}}=75$ lu.

Figure \ref{fig:expansion2} shows the results in the presence of misalignment between the walls. For a rectangular shape
of the junction, it is interesting to remark that the misalignment seems to have no effect for $H_{\mathrm{min}}=50$ lu. The centerline
position of the front remains almost unchanged, but the contact line can proceed inside the widening part of the channel.
We observe this behavior during a shorter time interval for smaller $L_{1}$ (results not shown for brevity). In this case, the capillary 
forces are expected to be stronger. In the light also of the considerations of the previous paragraph, this configuration can be assumed 
to be that easing to a larger extent the infiltration. At the beginning, the infiltration process is particularly slow for $H_{\mathrm{min}}=25$ lu 
and higher misalignment because the orifice is not behind the contact line. From Fig.~\ref{fig:expansion2} we also see that the shape of 
the junction has no particular influence for $H_{\mathrm{min}}=75$ lu with misalignment. Also for smaller minimum heights we distinguish no 
privileged geometry. In general, the effect of misalignment is weaker for longer narrow segments and larger minimum heights. Comparison 
with constrictions indicates that for wider channels the results are similar (see Fig.~\ref{fig:constriction2}). Otherwise, constrictions 
are more interesting for the absence of the phenomenon of pinning. In any case, constrictions can be regarded as expansions for flow in 
the other sense. So, in a random porous medium, both types of channel are present.

The results with reaction provide a compelling example for the independence of surface growth from the infiltration velocity (see 
Fig.~\ref{fig:expansion3}). Comparison with the outcome for constrictions allows to make this statement more precise (see 
Fig.~\ref{fig:constriction3}). We see that this is true in particular when the structural features responsible for the retardation
of the flow are already behind the contact line. For an extended porous medium, this means that the flow is interrupted because
of pore obstruction at the surface of the solid with a relation linear with time. Now, the infiltration velocity is controlled
by the pore characteristics. Our finding does not imply that the porous structure is irrelevant. Indeed, a larger porosity should
require more time before the flow stops and, on the basis of previous experiments, it might be possible to make predictions.
We verified that the curves for $z$ separate more neatly with reactivity for the other values of $L_{1}$ and $\Delta L$, since at the
beginning both solid steps are behind the contact line for these configurations. Furthermore, the effect of surface growth is more
marked for decreasing minimum radii. In general, we recognize no particular influence of the shape of the junction for surface 
growth.

The infiltration process starts with a strong acceleration near the converging part of the channel, followed by a phase where the front 
(centerline position) is almost at rest (see Fig.~\ref{fig:expansion3}). As a consequence, it is more difficult to fit the data to a linear 
relation. In any case, it follows that the results for expansions are in part comparable with those for constrictions 
and periodic profiles (sinusoidal and zig-zag) and even slightly faster (see Figs.~\ref{fig:sin6} and \ref{fig:constriction3}). The 
analysis for tortuosity reveals that expansions are less advantageous than constrictions. It also appears that the concave geometry 
favors infiltration more than the rectangular shape (results not shown for brevity). As an example, for higher misalignment and 
$H_{\mathrm{min}}=50$ lu, it is found that the tortuosity is relatively low, around $1.35$, with a peak in the range $2$-$3$, when the 
flow meets the narrowing part of the channel. With reaction, the peaks can reach $5$ with stronger reactivity.


\section*{5.~~~CONCLUSIONS}

Our work deals with LB simulations in 2D for capillary penetration into single channels with walls characterized by
different geometric properties. The focus is on the effects for fluid flow of surface growth from a supersaturated
solution (Kang et al., 2007, 2002b, 2003, 2004; Lu et al., 2009). The study is motivated by the problem of Si infiltration 
into C preforms (Bougiouri et al., 2006; Dezellus and Eustathopoulos, 2010; Dezellus et al., 2003; Einset, 1996 and 1998; Eustathopoulos et al., 1999; 
Hillig et al., 1975; Israel et al., 2010; Liu et al., 2010; Messner and Chiang, 1990; Mortensen et al., 1997; Voytovych et al., 2008). 
Several inconsistencies exist between the experiments and the models, as well as the simulation conditions. For example, 
in the LB models the liquid and vapor phases have the same density. But this choice allows to reproduce a linear dynamics 
for the Washburn infiltration (Chibbaro, 2008; Chibbaro et al., 2009b), as observed in experiments (Israel et al., 2010; Voytovych et al., 2008). 
However, the accordance with experimental results is poor since in our simulations the Reynolds number is overestimated 
(Sergi et al., 2014). This shortcoming could be limited by considering longer capillaries with more iterations. This is not 
necessary for our purposes as the process of surface growth exhibits a relatively weak dependence on the infiltration 
velocity. This is expected to be the case especially for smaller Reynolds numbers. Furthermore, for interconnected porous 
systems smaller Reynolds numbers could be attained. Another point is 
that in our simulations the reaction-rate constant $k_{\mathrm{r}}$ varies, while in real systems this quantity is fixed 
to a single value. 
From experimental data the values of the parameter $k_{\mathrm{r}}$ could be chosen more accurately on the basis 
of the morphology of the growing solid phase at the surface. 
This kind of analysis is capital because a proper comparison with the
experimental results should be based on dimensionless, characteristic numbers
involving the time set by the process of pore closure and the corresponding
infiltrated distance. Indeed, the flow behavior is expected to depend on the
relative effects between the intervening mechanical forces and the speed of
reactivity. In this way, it would be possible to have good accordance for the
full infiltration dynamics resulting from simulations and experiments. This is
easy to understand because otherwise it is imposed only the condition to realize
the last stage of the process.
Finally, it is worth noticing that other phenomena are even completely 
disregarded in our description, as thermal effects (Sangsuwan et al., 1999) and the transition to wetting 
(i.e., the reaction at the contact line). As a consequence, our investigation provides
no explanation for the emergence of the linear Washburn behavior (Bougiouri et al., 2006;
Israel et al., 2010; Voytovych et al., 2008).

Our modeling approach consists in decomposing intra-particle porosity into
basic, simplified structures. Their function for capillary infiltration and
surface growth is assessed. Inputs for preform preparation can be singled out.
In general, it is found that the structure of the channels is more important
than the minimum radius. More specifically, for periodic profiles,
the simulations indicate that the ratio of the amplitude to the channel width, $A/N_{y}$, mainly
affect the infiltration speed. Smaller ratios are more advantageous. The misalignment between the walls further speeds up
the infiltration process. The period does not appear as a prominent parameter. Of course, smaller periods are associated
with a higher drag force. For the sinusoidal profiles, the length of the capillaries is not so important, as opposed to uniform
channels. Faster infiltration occurs for zig-zag profiles. The results for the sinusoidal profiles are similar. Step-walled
capillaries lead to a slower dynamics. In this case, the phenomenon of pinning is more marked.

These considerations have interesting consequences for the selection of ceramic powders and the preform preparation.
Small ratios $A/N_{y}$ can be realized by wide pathways with radius variations weakly pronounced. It follows that
powders containing larger particles are good candidates. The results for the misalignment of the walls suggest that
an excessive compaction of the grains should be avoided, especially in the direction normal to the infiltration.
The use of particles with rounded edges is highly recommended. In particular, the results for step-walled capillaries 
show that powders with fine grains are detrimental.

Constrictions and expansions can arise by the arrangement of quite coarse particles. By inverting the sense of fluid motion
the other configuration is obtained. The danger of pinning associated with expansions should be minor in random porous
media given the typical size of the grains. It is interesting to notice that the effect of reaction is weaker for
constrictions and expansions. An important consequence is that the flow paths allowing to guide the fluid into the porous
structures should be created by means of large aggregates. Small, faceted grains could instead be added in order to
enhance the process of surface growth. We recall that in C/SiC composites the SiC formation is desired for the mechanical
properties.

To summarize, in order to ensure the impregnation by Si it would thus be advisable to start with preforms presenting pore pathways as
wide as possible, straight and with round morphologies. Preferably, this could be achieved by using large ceramic particles.
The residual Si in the ceramics could be limited by adding carbon powders containing smaller grains. Particles of small size are likely
to slow down the infiltration significantly, in particular if the surface is characterized by corners. Previous investigations have 
already highlighted the importance of the particle size, the packing properties and the microstructure for the preform to be impregnated
(Gadow and Speicher, 2000; Israel et al., 2010; Paik et al., 2002; Salamone et al., 2008). Another finding of relevance is that, especially 
at the solid surface of the preform, the growth kinetics is expected to depend marginally on the infiltration velocity. As explained, this 
finding can have practical consequences for the processing conditions (see Sec.~4.2.2). 
As suggested clearly by the analysis for misalignment, the porosity alone does not characterize
exhaustively the whole infiltration process. Our investigation still leaves the open question on
the role of the flow pattern inside real porous media for reactive infiltration. Further progress
can be made by considering more complex models for the microstructure including interconnected pore
channels. Needless to say that this problem is critical for a better determination of the
optimal configuration of the porous preform for the manufacturing of ceramic components devised
to advanced applications. In that respect, the LB method seems to have an advantage over other
numerical schemes for hydrodynamics.


\acknowledgments

The research leading to these results has received funding from the European
Union Seventh Framework Programme (FP7/2007-2013) under grant agreement
n$^{\circ}$ 280464, project "High-frequency ELectro-Magnetic technologies
for advanced processing of ceramic matrix composites and graphite expansion''
(HELM).


\end{document}